\documentclass[preprint,amsmath,amssymb,aps]{revtex4-2}

\usepackage{graphicx}
\usepackage{dcolumn}
\usepackage{bm}
\usepackage{epsfig}
\usepackage{epstopdf}
\usepackage{subfig}
\usepackage{caption}
\usepackage{enumerate}
\usepackage{comment}
\usepackage{physics}

\usepackage{geometry}
\usepackage{amsmath}
\usepackage{slashed}
\usepackage{ulem}
\usepackage{color}
\usepackage{hyperref}

\begin{document}

\title{Cancellation of Infrared Divergences in $e^{+}e^{-}\rightarrow q\bar{q}g$ in Light Front Coherent State Formalism}
\author{Deepesh Bhamre}
  \email{deepesh.bhamre@cs.cruzeirodosul.edu.br}
  \affiliation{Laborat\'{o}rio de F\'{i}sica Te\'{o}rica e Computacional-LFTC, Universidade Cruzeiro do Sul and Universidade Cidade de S\~{a}o Paulo (UNICID), Rua Galv\~{a}o Bueno, 01506-000, S\~{a}o Paulo, Brasil}
 \author{Shrey Gogia}
  \email{shreygogiaphys@kgpian.iitkgp.ac.in}
  \affiliation{Department of Physics, Indian Institute of Technology Kharagpur, Kharagpur-721302, India}
 \author{Anuradha Misra}
  \email{misra@physics.mu.ac.in}
  \affiliation{Centre for Excellence in Basic Sciences (UMDAE-CEBS), University of Mumbai, Santacruz (East), Mumbai-400098, India}

 \date{\today}

\begin{abstract}

We address the issue of  cancellation of infrared (IR) divergences at the amplitude level in Light Front Quantum Chromodynamics (LFQCD) using the coherent state formalism. We consider the process $e^{+}e^{-}\rightarrow q\bar{q}g$ upto $\mathcal{O}(g^3)$ in light-cone-time-ordered Hamiltonian perturbation theory and show that IR divergences in S-matrix elements appear due to vanishing energy denominators. We construct the coherent state formalism for LFQCD and explicitly show that these divergences are cancelled when a coherent state basis is used for calculating the S-matrix elements.

\begin{description}

\item[Keywords]
{Infrared Divergences, Coherent State Formalism, Light Front QCD}

\end{description}

\end{abstract}

\maketitle

\section{Introduction}\label{sec:intro}

\paragraph*{}
Infrared (IR) divergences in gauge field theories, and methods to deal with them have been a subject of study since a very long time. The origin of logarithmic divergences that appear in the probability of scattering of an electron, accompanied by emission of light quanta, in a Coulomb field was studied as far back as 1937 by Bloch and Nordsieck \citep{blochnordsieck}. A cure to this problem of `infrared catastrophe' was proposed, at least in perturbation theory, by Kinoshita, Lee and Nauenberg in the form of what is known today as the KLN theorem \citep{kinoshitaIR, leenauenbergIR}. It was shown that such divergences, although appearing in partial transition probabilities, always cancel when the total probability is calculated by considering an appropriate ensemble of degenerate states.

\paragraph*{}
Subsequently, Chung showed the absence of IR divergences in QED perturbation theory when the matrix elements were calculated between appropriately chosen external photon states \citep{chung}. Again in the context of QED, a space of asymptotic states was proposed and a definition of an IR-finite S-matrix was provided by Kulish and Faddeev using the method of asymptotic dynamics \citep{kulishfaddeev}. This approach by Kulish and Faddeev (KF), referred to as the `coherent state approach' has led to numerous studies later. A few of these are mentioned below.

\paragraph*{}
A detailed study consisting of a series of articles by Kibble discusses the coherent state approach in QED \cite{kibble1, kibble2, kibble3, kibble4}. Butler and Nelson extended the KF approach to the non-Abelian case and constructed states that resulted in an IR-finite matrix element for a quark scattering from an external color-singlet potential \citep{butlernelson}. QCD brings in an additional dimension of complexity to the subject. The divergences mentioned uptil now arise due to the vanishing momentum of a particle (soft divergences). In the presence of theories involving a massless vertex i.e. an interaction involving all massless particles, there is also a presence of collinear divergences. The factorization of collinear and soft divergences was studied in Ref.\citep{delduca}. The cancellation of IR divergences for the process $qq \rightarrow qqg$ was shown in Ref.\citep{nelson} using the coherent state approach in QCD. Frenkel \textit{et al} proposed splitting the interaction Hamiltonian into a soft and a hard part and demonstrated that asymptotic states constructed using the soft evolution operator lead to IR-finite S-matrix elements in non-abelian gauge theories \citep{frenkel}. A coherent state operator for soft gluon emission was constructed by Catani and Ciafaloni \citep{catanigeneralizedcohst, ciafalonicohstasydyn} and some theoretical/technical aspects of the QCD coherent states were also studied by them \citep{catanigaugecovariance}. Catani \textit{et al} have discussed the violation of the Bloch-Nordsieck theorem in QCD and have shown that the non-cancelling terms are of higher twist \citep{cataninoncancellingir}. A short review concerning the coherent state formalism can be found in Ref.\citep{ciafaloniirsingcohst}. The IR-finiteness of S-matrix using the QCD coherent states was shown by Giavarini and Marchesini \citep{giavarinimarchesini}. Contopanagos and Einhorn have discussed the properties of asymptotic S-matrix in field theories with massless particles and have established the relationship between the coherent state method and the scattering matrix in the space of asymptotic states \citep{contopanagos}. The relation between asymptotic symmetry and the KF approach applied to QCD has been brought out in Refs.\citep{furugorinojiri, gonzo}. Using an approach similar to the coherent state formalism, Forde and Signer have constructed IR-finte amplitudes for $e^{+}e^{-} \rightarrow$ 2 jets at NLO and have recovered the known standard result for its total cross-section \citep{fordesigner}. KF method has been discussed in detail recently in Ref.{\citep{prabhu}} by Prabhu \textit{et al} who have proposed a reformulation of KF construction in case of massless QED, Yang-Mills theories and quantum gravity. A recent review on the topic of infrared structure of perturbative gauge theories can be found in \citep{magneairreview}.

\paragraph*{}
As in the case of conventional equal-time formalism, Light Front Field Theories (LFFTs) too face the problem of infrared divergences when massless fields are involved. LFFTs are quantum field theories wherein the quantization surface is a light-front. This formalism has been one of the most important first-principles approaches in the study of the non-perturbative regime of strong interactions. A comprehensive review is provided in Ref.\citep{brodskyreview}. The problem of IR divergences in LFFTs, which are based on a Hamiltonian approach, needs to be addressed. A coherent state formalism for LFFTs was developed by one of us \citep{misraqed, misradlcq, misraqcd} and was used to demonstrate cancellation of IR divergences to all orders in fermion self-energy corrections in QED \citep{jaimisra, jaimisrafeynmangauge, jaimisraallorder}. For Light Front QCD (LFQCD) too, similar studies have been initiated by us \citep{misraqcd, misrafbs, bhamremisra2jet}. We propose to take these studies further in the current and future work. In this article, we calculate the transition amplitude for $e^{+}e^{-} \rightarrow q\bar{q}g$ in the LF coherent state formalism at $\mathcal{O}(g^{3})$. We firstly show that it contains IR divergences (both soft and collinear) in the Fock basis. We then calculate the transition matrix elements in coherent state basis and show that the IR divergences are cancelled in the coherent state basis.

\paragraph*{}
Conventionally, parton shower algorithms used in Monte Carlo event generators have relied on the cancellation of infrared divergences at the cross-section level in partonic (perturbative) calculations. The subtraction method developed by Catani, Seymour and others \citep{cataniseymour, cataniseymourNLO} is a prime example of how this cancellation is implemented. There has been much progress in the recent past, both on the theoretical and computational front, in the development of improved parton shower algorithms. One aim of these developments is the construction of all-order amplitude-level parton shower algorithms \citep{martinez, forshaw2019}. We are hopeful that our studies will add a new dimension on the theoretical front of construction of better parton shower algorithms.

\paragraph*{}
In sec.\ref{sec:basics} we provide, for completeness, a very brief introduction of the LFQCD Hamiltonian, IR divergences in LFQCD and the coherent state formalism. More details can be found in Ref.\citep{bhamremisra2jet}. In sec.\ref{sec:coh st construct}, we construct the $\ket{qqg}$ coherent state which we will use for arriving at an IR finte $e^{+}e^{-} \rightarrow q\bar{q}g$ amplitude. Next we calculate, in sec.\ref{sec:Fock diag}, the $e^{+}e^{-} \rightarrow q\bar{q}g$ amplitude to $\mathcal{O}(g^{3})$ in the Fock basis. This consists of a number of time-ordered diagrams involving various combinations of the LFQCD interactions. We illustrate a few of these in sec.\ref{sec:Fock diag} and present the remaining ones in Appendix \ref{app:A}. In sec.\ref{sec:coh components} we calculate the transition amplitude in coherent state basis. We show in sec.\ref{sec:cancel} that the IR divergences present in the Fock basis are cancelled by the higher order terms in the coherent state expansion. We conclude in sec.\ref{sec:conclusion} with an outlook and open questions for the near future.

\section{IR Divergences and Coherent state formalism in LFQCD}\label{sec:basics}
\subsection{Light-Front Quantum Chromodynamics}

\begin{figure}[h]
    \centering
    \includegraphics[width=0.2\linewidth]{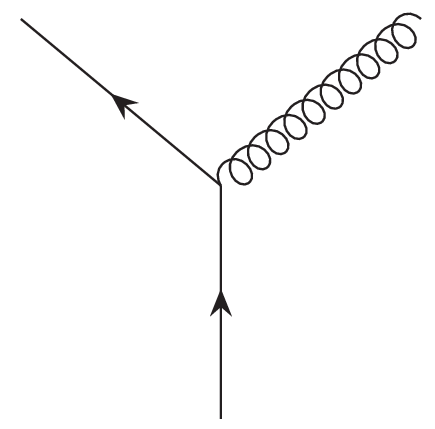}
    \includegraphics[width=0.2\linewidth]{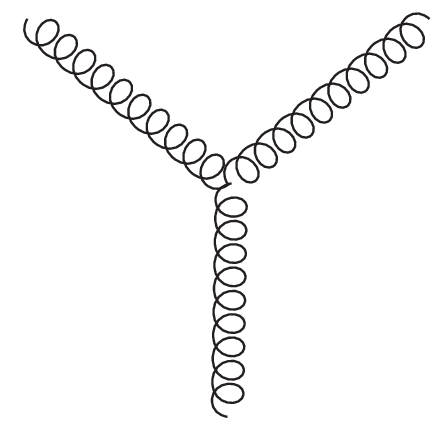}
    \includegraphics[width=0.2\linewidth]{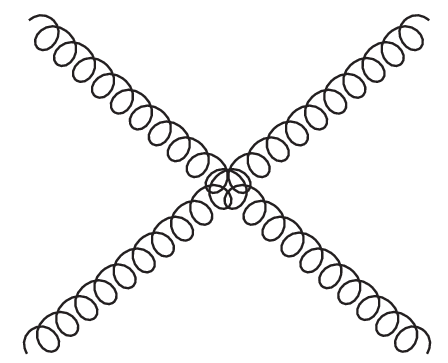}\\
    (a) $V_1$ \hspace{0.14\linewidth} (b) $V_2$ \hspace{0.14\linewidth} (c) $V_3$\\
    \includegraphics[width=0.33\linewidth]{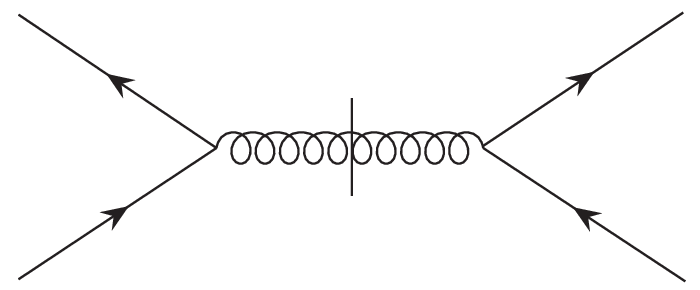}
    \includegraphics[width=0.33\linewidth]{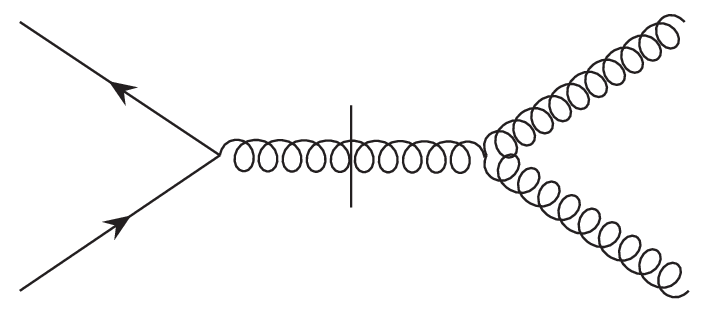}\\
    (d) $W_1$ \hspace{0.265\linewidth} (e) $W_2$ \\
    \includegraphics[width=0.33\linewidth]{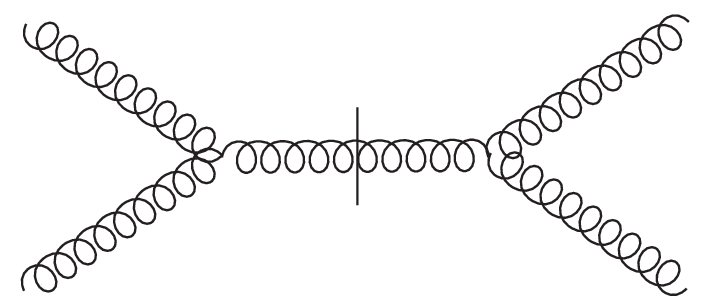}
    \includegraphics[width=0.33\linewidth]{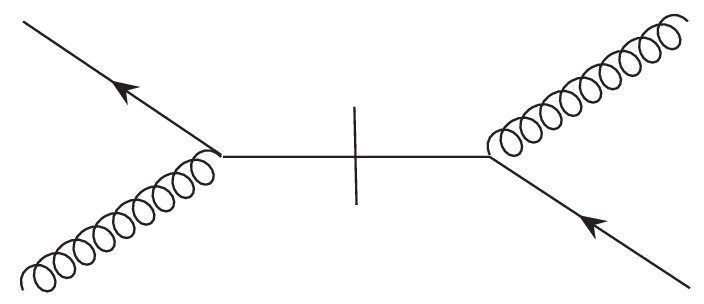}\\
    (f) $W_3$ \hspace{0.265\linewidth} (g) $W_4$ \\
    \caption{Interaction Vertices in Light-Front QCD}
    \label{fig:vertices}
\end{figure}

On the light cone, the Hamiltonian is defined in terms of the energy-momentum tensor as {\cite{harindranath}}
\begin{equation}
    P^-=P_+ =\int d^2 \mathbf{x}_\perp d x^-T^{+-}
\end{equation}
In the light-front gauge $A^+_a=0$, the Light-Front QCD Hamiltonian is \cite{bhamremisra2jet}
\begin{equation}\label{lfham}
    \begin{split}
       P^-=\int d^2 \mathbf{x}_\perp d x^- \Biggl[
            \frac{1}{2}\Bar{\Psi}\gamma^+\frac{m^2+(i\grad_\perp)^2}{(i\partial^+)}\Psi+\frac{1}{2}A_a^\mu(i\grad_\perp)^2 A^a_\mu
            &+gJ^\mu_a A_\mu^a+\frac{g^4}{4}B^{\mu\nu}_a B_{\mu\nu}^a\\+\frac{g^2}{2}J_a^+\frac{1}{(i\partial^+)^2}J_a^+
            &+\frac{g^2}{2}\left(\Bar{\Psi}\gamma^\mu T^a A_\mu^a\right)\frac{\gamma^+}{(i\partial^+)}(\gamma^\nu T^b A_\nu^b{\Psi})\Biggr]
    \end{split}
\end{equation}
where $J^\mu_a=\Bar{\Psi}_c\gamma^\mu T^a_{cc'} \Psi_{c'}+f^{abd}\partial^{\mu}A^{\rho}_{b} A^d_\rho$  and $B^{\mu\nu}_a=f^{abc}A^\mu_bA^\nu_c$. A detailed derivation can be found in \cite{bhamremisra2jet}.
The interaction Hamiltonian $H_\text{int}$, in Eq. \eqref{lfham}, has the form 
\begin{equation}
    H_\text{int}=V_1+V_2+V_3+W_1+W_2+W_3+W_4
\end{equation}
where $V_i$ $(i=1,2,3)$ are the standard QCD interactions given by
\begin{equation}
    V_1\equiv\int d^2 \mathbf{x}_\perp dx^- g \Bar{\Psi}\gamma^\mu T^a \Psi A^a_\mu
\end{equation}

\begin{equation}
    V_2\equiv\int d^2 \mathbf{x}_\perp dx^- g f^{abd} (\partial^\mu A^\nu_b) A^d_\nu A^a_\mu
\end{equation}

\begin{equation}
    V_3\equiv\int d^2 \mathbf{x}_\perp dx^- \frac{g^2}{4} f^{abd}f^{aef}  A^\mu_b A^\nu_d A^e_\mu A^f_\nu
\end{equation}
In addition to the usual QCD interaction vertices $V_i$, the LFQCD Hamiltonian contains \textit{instantaneous} interaction vertices $W_i$ $(i=1,2,3,4)$ which are non-local interactions resulting from the elimination of dependent degrees of freedom from the theory. 

\begin{equation}
    W_1\equiv\int d^2 \mathbf{x}_\perp dx^- \frac{g^2}{2} (\Bar{\Psi}\gamma^+T^a\Psi)\frac{1}{(i\partial^+)^2}  (\Bar{\Psi}\gamma^+T^a\Psi)
\end{equation}

\begin{equation}
\begin{split}
    W_2\equiv\int d^2 \mathbf{x}_\perp dx^- \frac{g^2}{2} \Biggl[(\Bar{\Psi}\gamma^+T^a\Psi)\frac{1}{(i\partial^+)^2} f^{abd} (\partial^+ A^\mu_b) A^d_\mu + f^{abd} (\partial^+ A^\mu_b) A^d_\mu \frac{1}{(i\partial^+)^2}  (\Bar{\Psi}\gamma^+T^a\Psi)\Biggr]
\end{split}
\end{equation}

\begin{equation}
\begin{split}
    W_3\equiv\int d^2 \mathbf{x}_\perp dx^- \frac{g^2}{2}  f^{abd}  f^{aef} (\partial^+ A^\mu_b) A^d_\mu \frac{1}{(i\partial^+)^2}  (\partial^+ A^\nu_e) A^f_\nu  
\end{split}
\end{equation}

\begin{equation}
    W_4\equiv\int d^2 \mathbf{x}_\perp dx^- \frac{g^2}{2} (\Bar{\Psi}\gamma^\mu T^a A^a_\mu)\frac{\gamma^{+}}{(i\partial^{+})} (\gamma^\nu T^b A^b_\nu \Psi)
\end{equation}
The Feynman diagrams for all the interactions are shown in Figure \ref{fig:vertices}.

\subsection{Infrared Divergences in LFQCD}

Infrared divergences occur in field theories when massless propagators go on shell. In light-front formulation, the IR limit $\mathbf{k}=(k^+,\mathbf{k}_\perp)\rightarrow \mathbf{0}$ leads to IR divergences which may be 'true' or 'spurious' IR divergences \cite{misraqed}. Spurious divergences are a manifestation of UV divergences of the covariant formulation and are, therefore, not considered here. This work addresses the cancellation of true IR divergences at $\mathcal{O}(g^3)$ in the process $e^{+}e^{-}\rightarrow q\bar{q}g$ using the Light-Front Coherent State formalism.

\subsection{Coherent State Formalism in LFFT}
In Light-Front Field Theory, the time evolution operator is given by 
\begin{equation}
    U(x^+,x_0^+)=T^+ \text{exp}\biggl(-i\int_{x_0^+}^{x^+} dx'^+ H_{\text{int}}(x'^+)\biggr)
\end{equation}
where $T^+$ denotes the light-cone time $(x^+)$ ordering.\\
The M\o{}ller operators are 
\begin{equation}
    \begin{split}
        \Omega_{\pm}&\equiv U (0, \pm \infty)\\
                    &=T^+ \text{exp}\biggl(-i\int_{\pm\infty}^{0} dx'^+ H_{\text{int}}(x'^+)\biggr)
    \end{split}
\end{equation}
and asymptotic states are
\begin{equation}
    \ket{s(t)}=e^{-ix^+H_0}\ket{s(0)}
\end{equation}
governed by the free Hamiltonian $H_0$.

In the coherent state formalism, the evolution of the asymptotic states is governed by the asymptotic Hamiltonian $H_{A}(\Delta)$ \cite{fordesigner}
\begin{equation}
    \ket{s(x^+)}=e^{-ix^+H_{A}(\Delta)}\ket{s(0)}
\end{equation}
where $H_{A}(\Delta)$ is obtained by taking the asymptotic limit $|x^+|\rightarrow \infty$ of the full interaction Hamiltonian. The parameter $\Delta$ is a cutoff introduced for separation of long-range interactions from $H_\text{int}$.
The asymptotic M\o{}ller operators are defined as
\begin{equation}
    \Omega_{A\pm}(\Delta)=T^+ \text{exp}\biggl(-i\int_{\pm \infty}^0 dx'^{+}H_{\Delta}(x'^{+})\biggr)
\end{equation}
where, and now onwards, $H_{A}(\Delta)$ is written as $H_{\Delta}$ for brevity. The incoming and outgoing coherent states are given by
\begin{equation}
    \begin{split}
        \ket{s:\text{coh}} & \equiv \Omega^\dagger_{A+}(\Delta)\ket{s} \\
        \bra{s:\text{coh}} & \equiv \bra{s}\Omega_{A}^-(\Delta) 
    \end{split}
\end{equation}
respectively\citep{bhamremisra2jet}.

Thus, the $q\Bar{q}g$ coherent state takes the form
\begin{equation}\label{cohstate}
\begin{split}
    \ket{q\bar{q}g:\text{coh}}&=\Biggl(1+(-i) \int_0^\infty dx_1^+ \Theta_\Delta H_\text{int}(x_1^+) + \cdots \Biggr)\ket{q\bar{q}g}\\ &=\Biggl(1+(-i) \int_0^\infty dx_1^+ \Theta_\Delta V_1(x_1^+) +\cdots\Biggr)\ket{q\bar{q}g}
\end{split}
\end{equation}
where $\Theta_\Delta$ is a product of Heaviside step functions and ensures that only those terms in the potential are included that survive in the asymptotic limit $|{x^+}|\rightarrow \infty$.

\section{Construction of Coherent States}\label{sec:coh st construct}

Consider the term in the asympotic M\o{}ller operator resulting from 3-point interaction $V_1$:
\vspace{0.1cm}
\begin{multline}
    \Omega^\dagger_{\Delta(V_1)}\ket{q_{p_3}\Bar{q}_{p_4}g_{p_5}}=-ig\int_0^\infty d x^+\int d^2 \mathbf{x}_\perp d x^-\int \Pi[d k_i] \Theta_\Delta (b_{k_1}^\dagger \Bar{u}_{k_1}e^{ik_1\cdot x}+d_{k_1} \Bar{v}_{k_1}e^{-ik_1\cdot x})\gamma^\mu T^a \\
    \times (b_{k_2} {u}_{k_2}e^{-ik_2\cdot x}+d_{k_2}^\dagger {v}_{k_1}e^{ik_2\cdot x})(a_{k_3} {\epsilon}^a_{\mu{k_3}}e^{-ik_2\cdot x}+a_{k_3}^\dagger {\epsilon}^{\ast a}_{\mu{k_3}}e^{ik_2\cdot x})\ket{q_{p_3}\Bar{q}_{p_4}g_{p_5}}
\end{multline}
at $\mathcal{O}(g)$.\\
In the asymptotic limit, only terms satisfying $\sum_i k_i^- \rightarrow 0$ survive due to the presence of $\Theta_\Delta$, and hence the contribution of $V_1$ to the coherent state at $\mathcal{O}(g)$ is given by
\begin{multline}\label{cohV1}
     \Omega^\dagger_{\Delta(V_1)}\ket{q_{p_3}\Bar{q}_{p_4}g_{p_5}}=-ig\int_0^\infty d x^+\int d^2 \mathbf{x}_\perp d x^-\int \Pi[d k_i] \Theta_\Delta (
     \Bar{u}_{k_1}\gamma^\mu T^a {u}_{k_2} {\epsilon}^{a}_{\mu{k_3}} b_{k_1}^\dagger b_{k_2} {{a}_{\mu}}_{k_3} e^{i(k_1-k_2-k_3)x} \\ 
     +\Bar{u}_{k_1}\gamma^\mu T^a {u}_{k_2} {\epsilon}^{\ast a}_{\mu{k_3}} b_{k_1}^\dagger b_{k_2} {a}_{\mu{k_3}}^\dagger e^{i(k_1-k_2+k_3)x}+ \Bar{u}_{k_1}\gamma^\mu T^a {v}_{k_2} {\epsilon}^{a}_{\mu{k_3}} b_{k_1}^\dagger d^\dagger_{k_2} {a}_{\mu{k_3}} e^{i(k_1+k_2-k_3)x}\\     +\Bar{v}_{k_1}\gamma^\mu T^a {u}_{k_2} {\epsilon}^{\ast a}_{\mu{k_3}} d_{k_1} b_{k_2} {a}_{\mu{k_3}}^\dagger e^{i(-k_1-k_2+k_3)x} +\Bar{v}_{k_1}\gamma^\mu T^a {v}_{k_2} {\epsilon}^{a}_{\mu{k_3}} d_{k_1} d_{k_2}^\dagger {a}_{\mu{k_3}}e^{i(-k_1+k_2-k_3)x}\\   +\Bar{v}_{k_1}\gamma^\mu T^a {v}_{k_2} {\epsilon}^{\ast a}_{\mu{k_3}}d_{k_1} d_{k_2}^\dagger {a}^\dagger_{\mu{k_3}}e^{i(-k_1+k_2+k_3)x})b^\dagger_{p_3}d^\dagger_{p_4}a^\dagger_{p_5}\ket{0}
 \end{multline}
Integrating over the spatial coordinates, we get
\begin{multline}\label{cohV1_2}
\Omega^\dagger_{\Delta(V_1)}\ket{q_{p_3}\Bar{q}_{p_4}g_{p_5}}= \frac{g}{(2\pi)^6}\int [dk] \Theta_\Delta \Biggl[ \frac{\bar{u}_{p_5-k}\slashed{\epsilon}^a_{p_5}T^a{v}_{k}b^\dagger_{p_5-k}d^\dagger_{k}b^\dagger_{p_3}d^\dagger_{p_4}}{\sqrt{2(p_5-k)^+}\sqrt{2p_5^+}((p_5-k)^-+k^--p_5^-)} \\ +\  \frac{\bar{u}_{p_3-k}\slashed{\epsilon}^{\ast a}_{k}T^a{u}_{p_3}b^\dagger_{p_3-k}a^\dagger_{k}a^\dagger_{p_5}d^\dagger_{p_4}}{\sqrt{2(p_3-k)^+}\sqrt{2p_3^+}((p_3-k)^-+k^--p_3^-)}
-  \frac{\bar{v}_{p_4}\slashed{\epsilon}^{\ast a}_{k}T^a{v}_{p_4-k}d^\dagger_{p_4-k}a^\dagger_{k}a^\dagger_{p_5}b^\dagger_{p_3}}{\sqrt{2(p_3-k)^+}\sqrt{2p_3^+}((p_3-k)^-+k^--p_3^-)}\Biggr]\ket{0}
\end{multline}
$\Theta_\Delta$ is a function of $k_i$ and limits the region of integration by imposing a condition on the sum of the energy components of these momenta $(\sum_i k_i^-)$. The argument of $\Theta_\Delta$ is in general different for each term of the coherent state. For example in the first term of Eq. \eqref{cohV1_2}, $\Theta_\Delta$ is a product of step functions which ensure that $(p_5-k)^- +k^--p_5^-<\Delta$, where $\Delta$ is a cutoff. \\
Similarly, the component of the asymptotic coherent state corresponding to the $V_2$ interaction is
\vspace{0.1cm}
\begin{multline}
    \Omega^\dagger_{\Delta(V_2)}\ket{q_{p_3}\Bar{q}_{p_4}g_{p_5}}=(-i) \int_0^\infty dx_1^+ \Theta_\Delta V_2(x_1^+)\ket{q\bar{q}g}\\= \frac{-gf^{abd}}{(2\pi)^6}\int [dk] \Theta_\Delta \Biggl[ \frac{-ip_5^\mu\epsilon^{\nu}_{b_{p_5}} \epsilon^{\ast d}_{\nu_k} \epsilon^{\ast a}_{\mu_{p_5-k}} + ik^\mu\epsilon^{\ast \nu}_{b_{k}} \epsilon^{\ast d}_{\nu_{p_5-k}} \epsilon^{a}_{\mu_{p_5}}+ i(p_5-k)^\mu\epsilon^{\ast \nu}_{b_{p_5-k}} \epsilon^{d}_{\nu_{p_5}} \epsilon^{\ast a}_{\mu_{k}}}{\sqrt{2(p_5-k)^+}\sqrt{2p_5^+}(k^-+(p_5-k)^--p_5^-)}\Biggr]a^\dagger_{k}a^\dagger_{p_5-k}b^\dagger_{p_3}d^\dagger_{p_4}\ket{0}
\end{multline}

Transition amplitude in the coherent state basis is given by 
\begin{equation}\label{eq:transition-amplitude}
\begin{split}
    T_{fi}&=\bra{q_{p_3}\Bar{q}_{p_4}g_{p_5}:\text{coh}}V\frac{1}{p_i^- -H_0}V\cdots\frac{1}{p_i^- -H_0}V_{em}\frac{1}{p_i^- -H_0}V_{em}\ket{e^-_{p_1}e^+_{p_2}}\\
    &=\bra{q_{p_3}\Bar{q}_{p_4}g_{p_5}}V\frac{1}{p_i^- -H_0}V\cdots\frac{1}{p_i^- -H_0}V_{em}\frac{1}{p_i^- -H_0}V_{em}\ket{e^-_{p_1}e^+_{p_2}}\\
    &+ \bra{q_{p_3}\Bar{q}_{p_4}g_{p_5}}\Omega_{\Delta(V_1)}V\frac{1}{p_i^- -H_0}V\cdots\frac{1}{p_i^- -H_0}V_{em}\frac{1}{p_i^- -H_0}V_{em}\ket{e^-_{p_1}e^+_{p_2}}\\
    &+\bra{q_{p_3}\Bar{q}_{p_4}g_{p_5}}\Omega_{\Delta(V_2)}V\frac{1}{p_i^- -H_0}V\cdots\frac{1}{p_i^- -H_0}V_{em}\frac{1}{p_i^- -H_0}V_{em}\ket{e^-_{p_1}e^+_{p_2}}\\
    &+ \cdots
\end{split}
\end{equation}
where $V\in\{V_i,W_j\}$. The first term in this expansion is simply the transition amplitude in the Fock basis. This will be evaluated in the next section. The corrections due to the redefinition of the asymptotic states as coherent states will be evaluated in Section \ref{sec:coh components}.

\section{$e^{+}e^{-}\rightarrow q\bar{q}g$ at $\mathcal{O}(g^{3})$ in Fock Basis}\label{sec:Fock diag}

The amplitude for the process $e^{+}e^{-}\rightarrow q\bar{q}g$ at $\mathcal{O}(g^{3})$ has contributions from several diagrams involving various interaction potentials in different combinations and time-orderings. Consider the diagram $T_{1c}$ of Fig. \ref{fig:T1}. It is given by:
\vspace{0.1cm}
\begin{multline}\label{eq:ir-finite}
    T_{1c}=\frac{-e^2g^3}{(2\pi)^{15/2}\prod_i\sqrt{2p_i^+}}\int [dk]\Biggl[\frac{\Bar{u}_{p_3}\slashed{\epsilon}^{\ast^a}_{p_5}T^a{u}_{p_3+p_5}\Bar{v}_{p_4-k}\slashed{\epsilon}^b_{k}T^b{v}_{p_4}\Bar{u}_{p_3+p_5}\slashed{\epsilon}^{\ast c}_{k}T^c{u}_{p_3+p_5+k}}{(2(p_3+p_5+k)^+)(2(p_3+k)^+)(p_3^-+p_5^--(p_3+p_5)^-)(2(p_4-k)^+)(2(p_1+p_2)^+)}\\ \times \frac{\Bar{u}_{p_3+p_5+k}\slashed{\epsilon}_{p_1+p_2}v_{p_4-k}\Bar{v}_{p_2}\slashed{\epsilon}^\ast_{p_1+p_2}u_{p_1}}{(2k^+)(p_1^-+p_2^--(p_4-k)^--(p_3+p_5+k)^-)(p_1^-+p_2^--k^--(p_3+p_5)^--(p_4-k)^-)(p_1^-+p_2^--(p_1+p_2)^-)}\Biggr]
\end{multline}
IR divergences in the LF Hamiltonian formalism appear due to vanishing denominators. In the limit $\mathbf{k}=(k^+,\mathbf{k}_\perp)\rightarrow \mathbf{0}$ the energy denominators in Eq. \eqref{eq:ir-finite} remain finite, as can be easily verified. Thus, diagram $T_{1c}$ is IR-finte.

In the following subsections, all the IR divergent terms corresponding to various combinations of ${V_i}$ and ${W_i}$ contributing to the first term of the transition amplitude series of Eq. \eqref{eq:transition-amplitude} are presented, along with the corresponding Feynman diagrams. Another set of possible combinations which can be obtained by performing the $(q\leftrightarrow\bar{q})$ operation can be dealt with in a similar manner and are not presented here.

Contributions to transition amplitude have been categorized into classes according to the interaction vertices involved at $\mathcal{O}(e^2g^3)$. We present the diagrams and expressions for each of the types below:

\subsection{Amplitudes involving $V_1V_1V_1$ vertices}
We briefly sketch the calculation of the transition amplitude for $V_1V_1V_1$ type terms here. The same procedure is followed for all the other amplitudes. The $V_1V_1V_1$ transition amplitude is given by
 \begin{equation}
    T_1=\bra{q_{p_3}\Bar{q}_{p_4}g_{p_5}}V_1\frac{1}{p_i^- -H_0}V_1\frac{1}{p_i^- -H_0}V_1\frac{1}{p_i^- -H_0}V_{em}\frac{1}{p_i^- -H_0}V_{em}\ket{e^-_{p_1}e^+_{p_2}} 
 \end{equation}
 Following the standard procedure of inserting identity operators $\int [dq_i]\ket{q_i}\bra{q_i}$ \cite{bhamremisra2jet} we obtain
 \vspace{0.1cm}
 \begin{multline}
     T_1=\int d^3k d^3k'\prod_i^7[dq_i]\prod_i^7[dq'_i]\biggl[\bra{q_{p_3}\Bar{q}_{p_4}g_{p_5}}V_1\ket{q_{q_6}\Bar{q}_{q_7}}\bra{q_{q_6}\Bar{q}_{q_7}}\frac{1}{p_i^- -H_0}\ket{q_{q'_6}\Bar{q}_{q'_7}}\\ \times\bra{q_{q'_6}\Bar{q}_{q'_7}}V_1\ket{q_{q_3}\Bar{q}_{q_4}g_{q_5}} \bra{q_{q_3}\Bar{q}_{q_4}g_{q_5}}\frac{1}{p_i^- -H_0}\ket{q_{q'_3}\Bar{q}_{q'_4}g_{q'_5}}\bra{q_{q'_3}\Bar{q}_{q'_4}g_{q'_5}}V_1\ket{q_{q_1}\Bar{q}_{q_2}}\\ \times \bra{q_{q_1}\Bar{q}_{q_2}}\frac{1}{p_i^- -H_0}\ket{q_{q'_1}\Bar{q}_{q'_2}}\bra{q_{q'_1}\Bar{q}_{q'_2}}V_{em}\ket{\gamma_k}\bra{\gamma_k}\frac{1}{p_i^- -H_0}\ket{\gamma_{k'}}\bra{\gamma_{k'}}V_{em}\ket{e^-_{p_1}e^+_{p_2}} \biggr]
 \end{multline}
 which leads to
 \begin{multline}\label{eq:T1}
     T_1=\int d^3k \prod_i^7[dq_i]\biggl[\frac{\bra{q_{p_3}\Bar{q}_{p_4}g_{p_5}}V_1\ket{q_{q_6}\Bar{q}_{q_7}}\bra{q_{q_6}\Bar{q}_{q_7}}V_1\ket{q_{q_3}\Bar{q}_{q_4}g_{q_5}} \bra{q_{q_3}\Bar{q}_{q_4}g_{q_5}}V_1\ket{q_{q_1}\Bar{q}_{q_2}}}{(p_1^-+p_2^- -q_6^--q_7^-)(p_1^-+p_2^- -q_3^--q_4^--q_5^-)}\\ \times \frac{\bra{q_{q_1}\Bar{q}_{q_2}}V_{em}\ket{\gamma_k}\bra{\gamma_k}V_{em}\ket{e^-_{p_1}e^+_{p_2}}}{(p_1^-+p_2^- -q_1^--q_2^-)(p_1^-+p_2^- -k^-)} \biggr]
 \end{multline}
Each matrix element of the form $\bra{q_{q_i}\cdots}V_1\ket{q_{q_j}\cdots}$ is then evaluated separately.

\begin{figure}[h]
    \centering
    \includegraphics[width=0.23\linewidth]{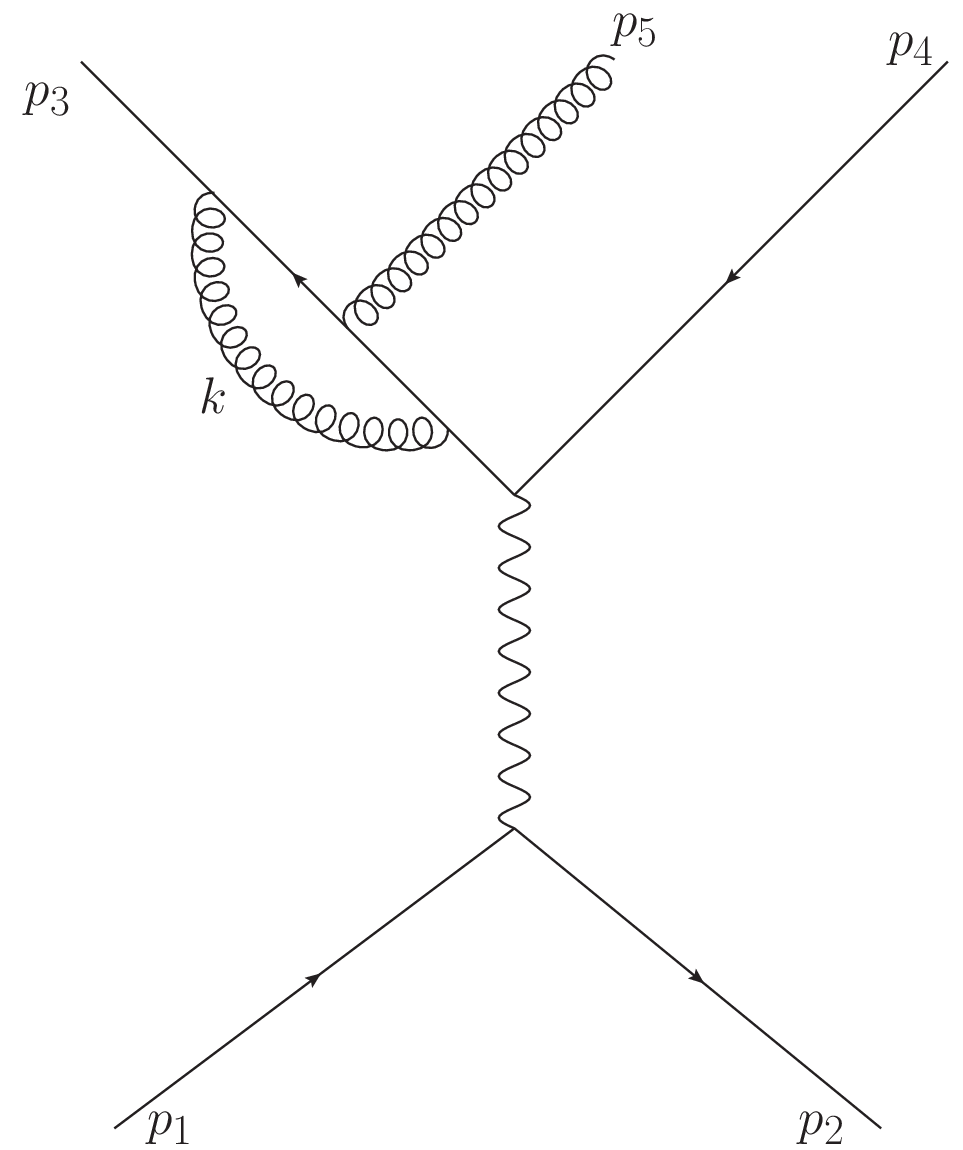} \includegraphics[width=0.23\linewidth]{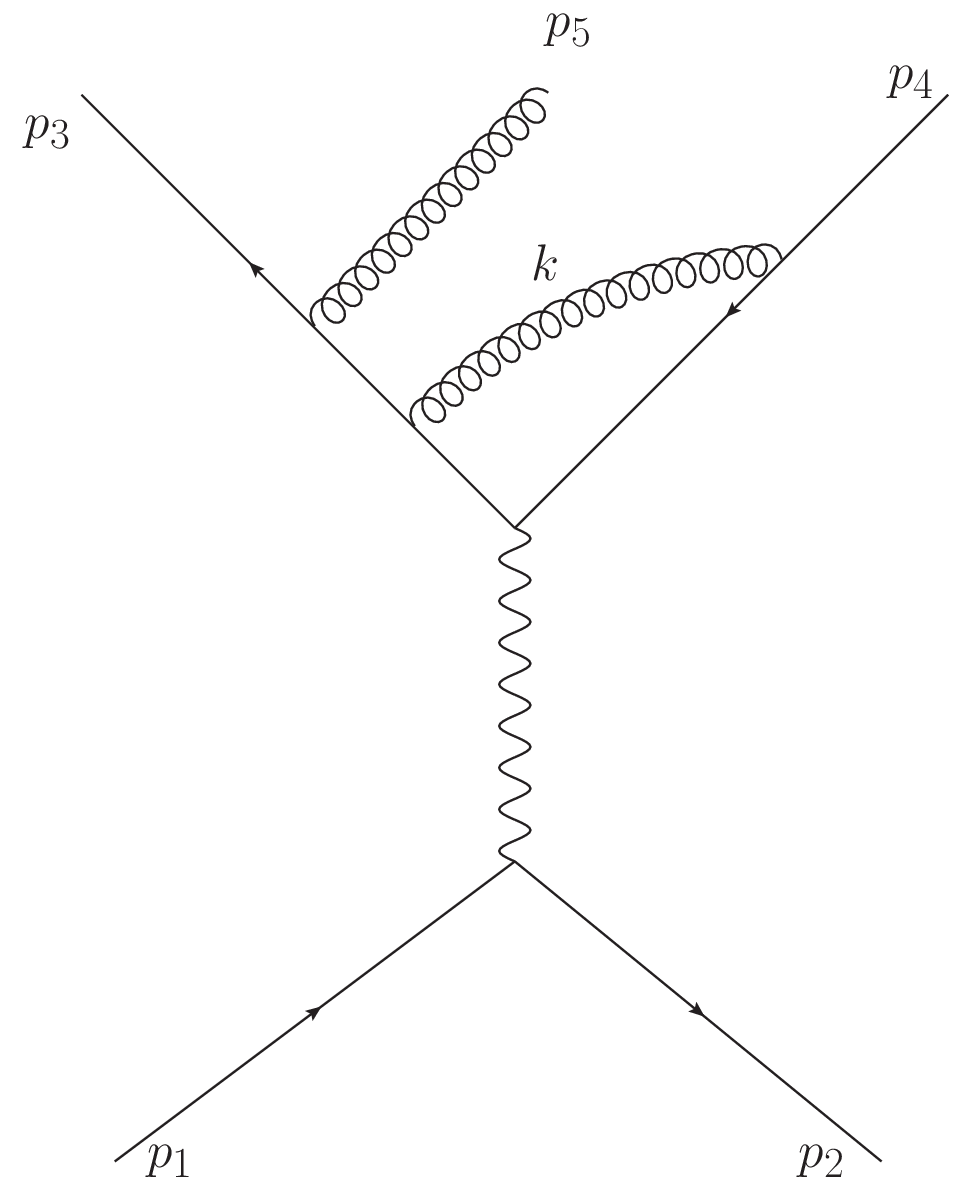} \includegraphics[width=0.23\linewidth]{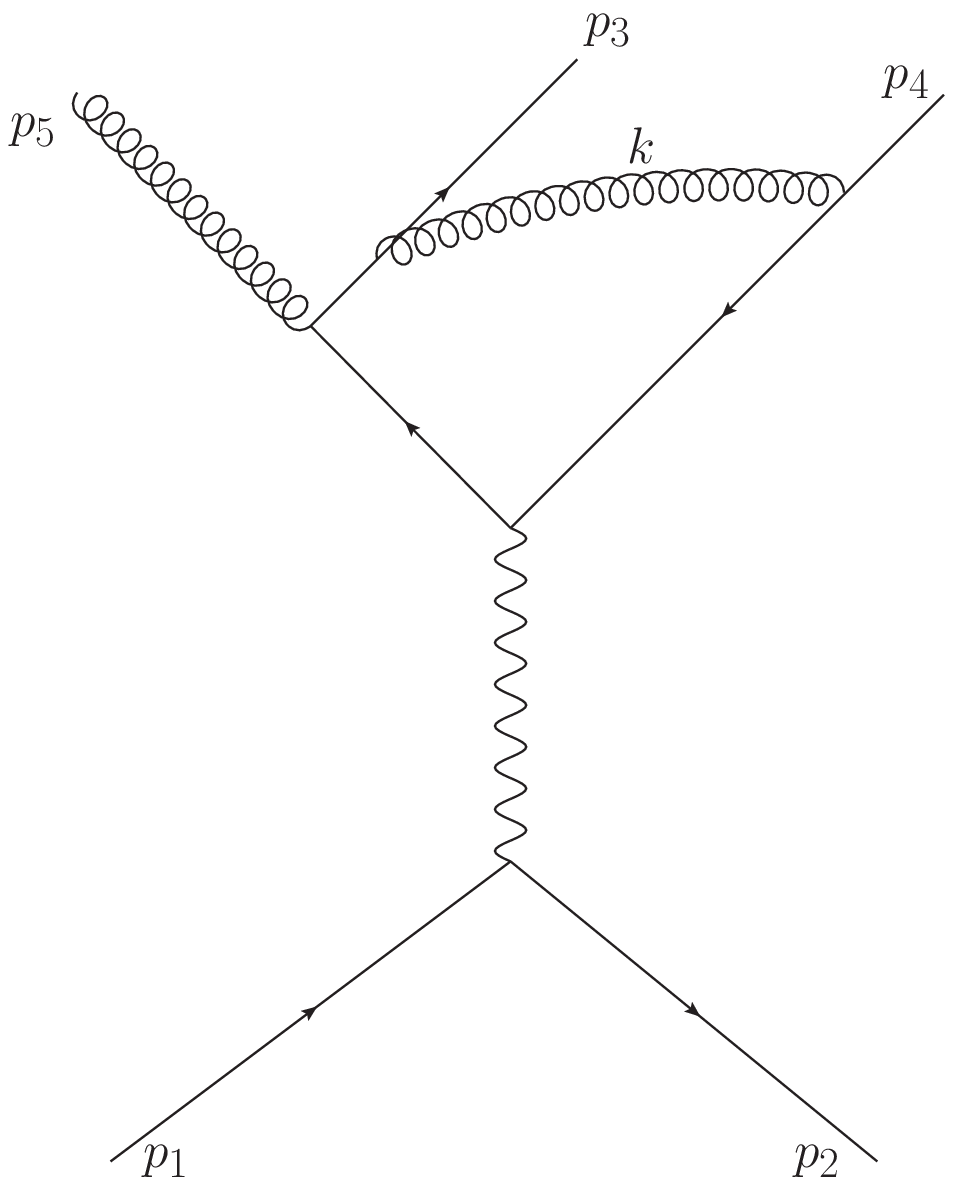}
    \includegraphics[width=0.23\linewidth]{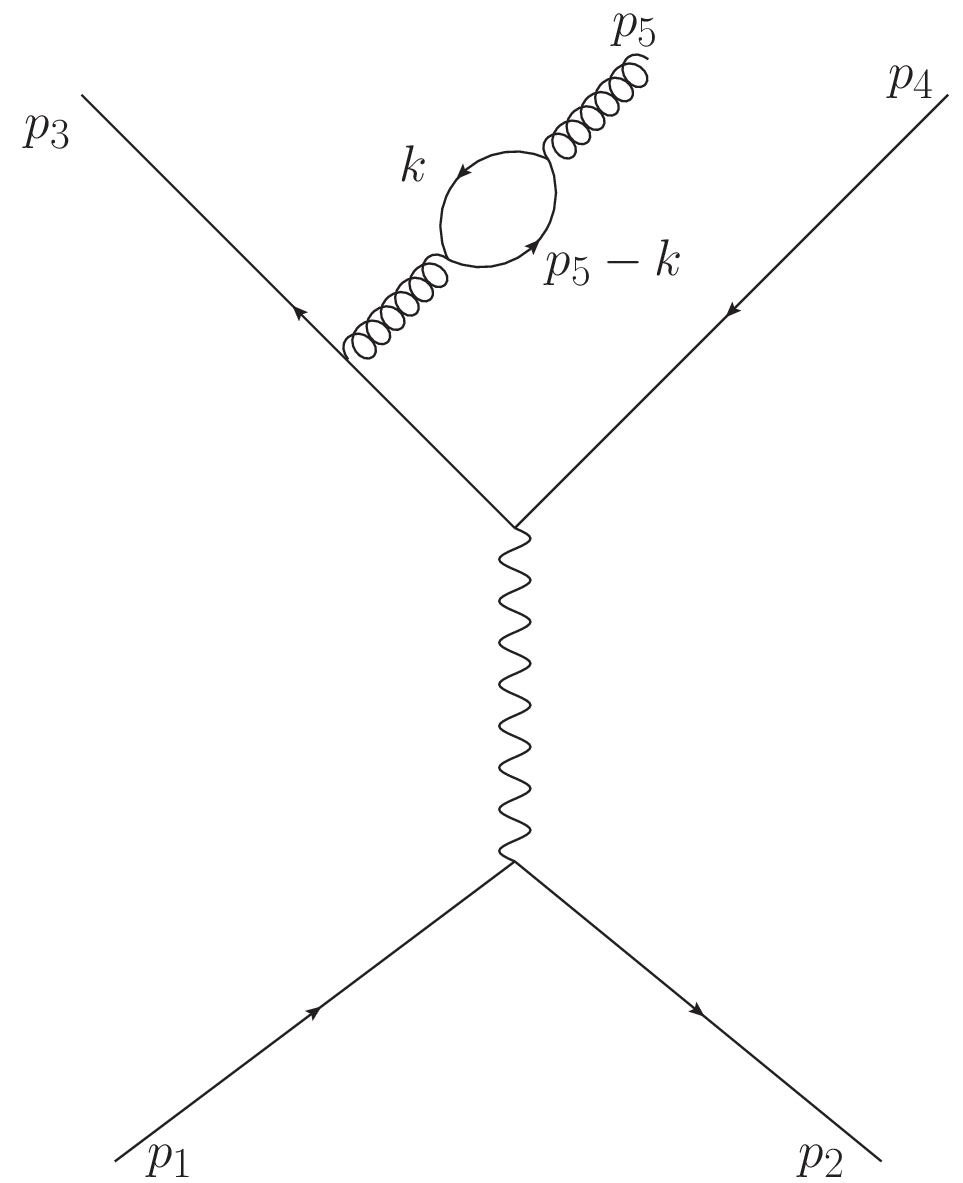}\\
    \hspace{0.02\linewidth} $T_{1a}$ \hspace{0.21\linewidth} $T_{1b}$\hspace{0.21\linewidth}$T_{1c}$\hspace{0.21\linewidth}$T_{1d}$\\
    \includegraphics[width=0.23\linewidth]{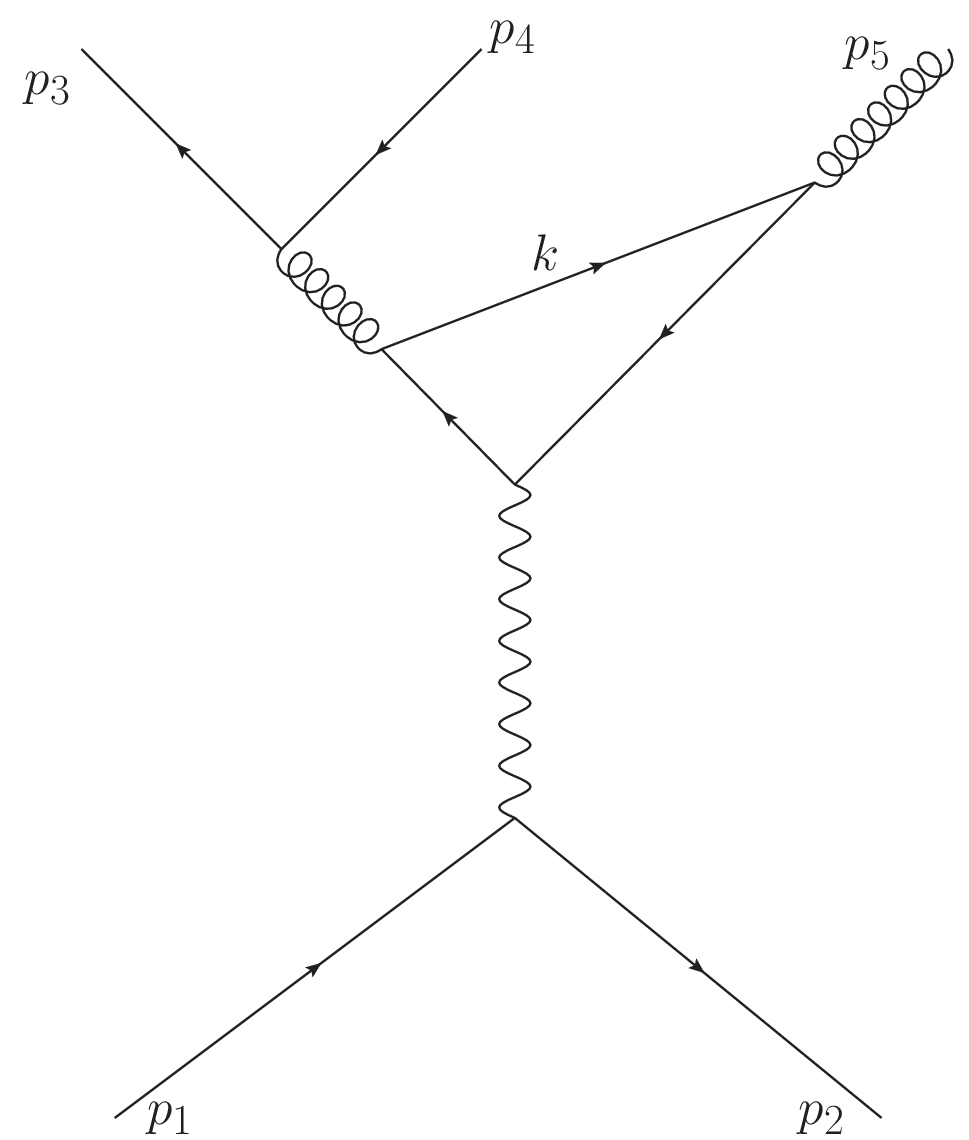}
    \includegraphics[width=0.23\linewidth]{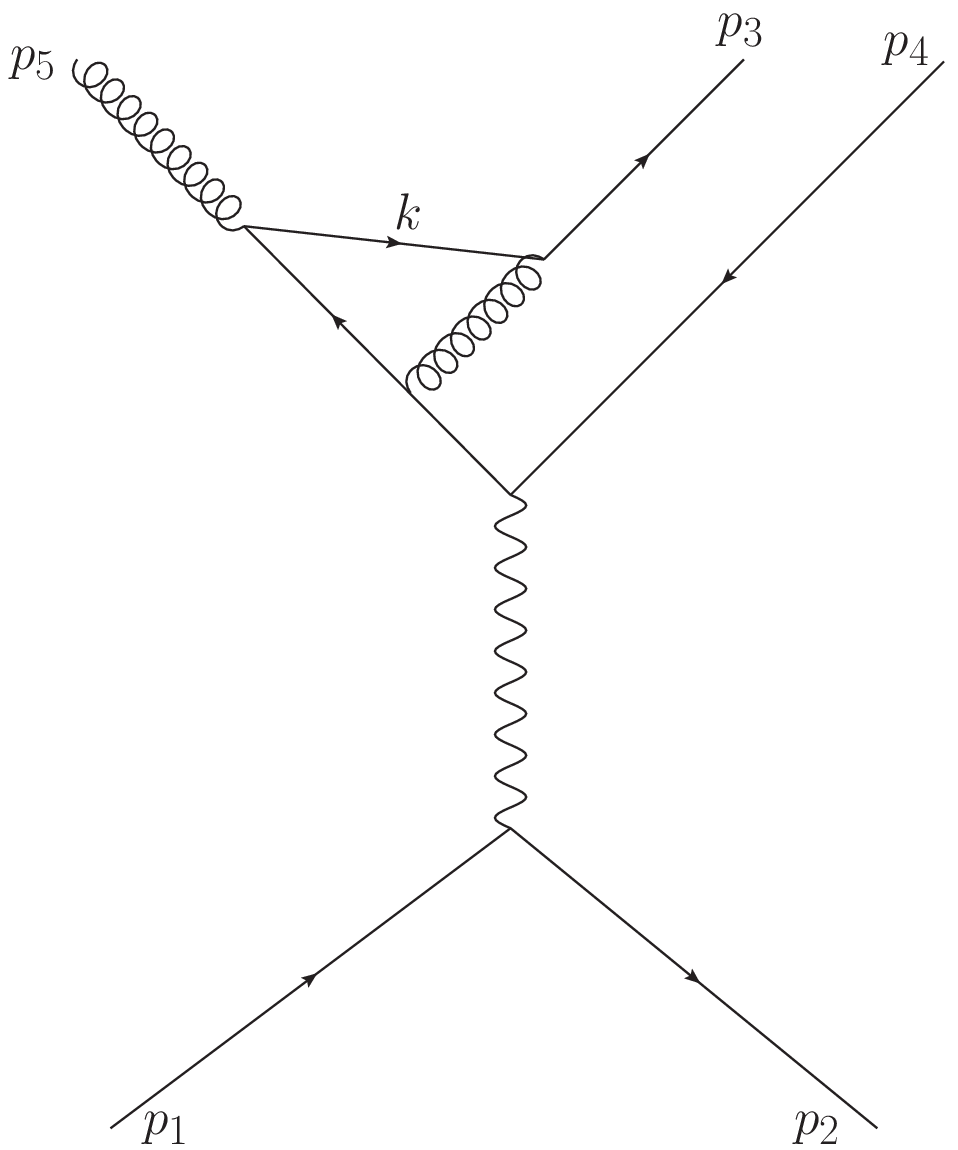}
    \includegraphics[width=0.23\linewidth]{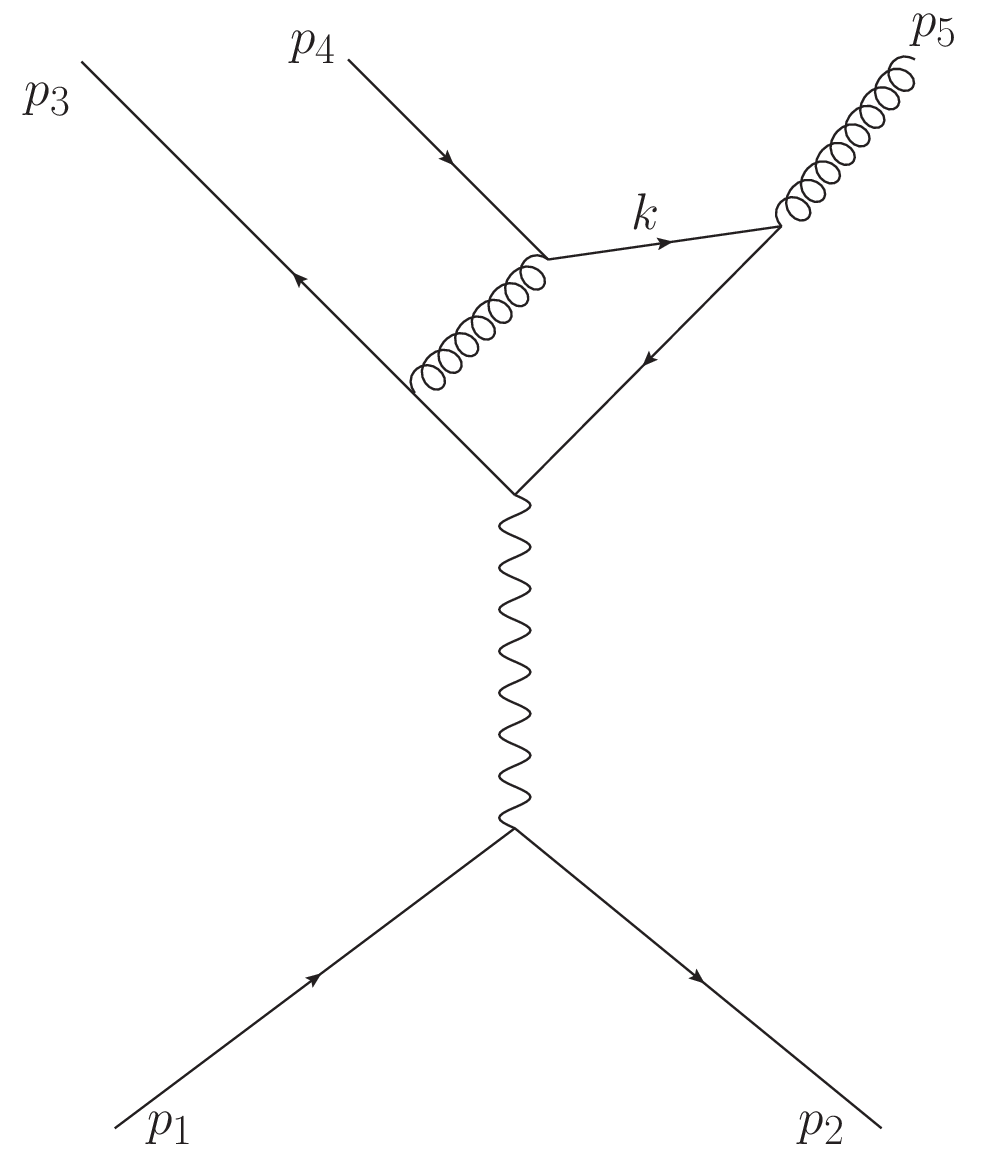}\\
    \hspace{0.02\linewidth} $T_{1e}$ \hspace{0.2\linewidth} $T_{1f}$\hspace{0.2\linewidth}$T_{1g}$\\
    \caption{Diagrams involving $V_1V_1V_1$ combination of interaction vertices}
    \label{fig:T1}
\end{figure}

The resulting diagrams are given in Fig. \ref{fig:T1}.

The amplitudes corresponding to diagrams $T_{1a}$ and $T_{1e}$ are presented here. The amplitudes for the other diagrams can be found in Appendix \ref{app:A}.
\vspace{0.1cm}
\begin{multline}
    T_{1a}=\frac{e^2g^3}{(2\pi)^{15/2}\prod_i\sqrt{2p_i^+}}\int [dk]\Biggl[\frac{\Bar{u}_{p_3}\slashed{\epsilon}^a_{k}T^a{u}_{p_3-k}\Bar{u}_{p_3-k}\slashed{\epsilon}^{\ast b}_{p_5}T^b{u}_{p_3+p_5-k}}{(2(p_3-k)^+)(2(p_3+p_5)^+)(2(p_3+p_5-k)^+)(2k^+)(p_3^--(p_3-k)^--k^-)}\\ \times \frac{\Bar{u}_{p_3+p_5-k}\slashed{\epsilon}^{\ast c}_{k}T^c{u}_{p_3+p_5}\Bar{u}_{p_3+p_5}\slashed{\epsilon}_{p_1+p_2}v_{p_4}\Bar{v}_{p_2}\slashed{\epsilon}^\ast_{p_1+p_2}u_{p_1}}{(2(p_1+p_2)^+)(p_3^-+p_5^--(p_3+p_5)^-)(p_1^-+p_2^--(p_1+p_2)^-)(p_3^-+p_5^--k^--(p_3+p_5-k)^-)}\Biggr]
\end{multline}
which is IR divergent as $(p_3^--(p_3-k)^--k^-)\rightarrow0$ in the limit $k\rightarrow0$.


Similarly,
\vspace{0.1cm}
\begin{multline}
    T_{1e}=\frac{e^2g^3}{(2\pi)^{15/2}\prod_i\sqrt{2p_i^+}}\int [dk]\Biggl[\frac{\Bar{v}_{p_5-k}\slashed{\epsilon}^{\ast a}_{p_5}T^a{u}_{k}\Bar{u}_{p_3}\slashed{\epsilon}^b_{p_3+p_4}T^b{v}_{p_4}\Bar{u}_{k}\slashed{\epsilon}^{\ast c}_{p_3+p_4}T^c{u}_{p_3+p_4+k}}{(2(p_3+p_4+k)^+)(2(p_5-k)^+)(2(p_1+p_2)^+)(2(p_3+p_4)^+)(p_5^--(p_5-k)^--k^-)}\\ \times \frac{\Bar{u}_{p_3+p_4+k}\slashed{\epsilon}_{p_1+p_2}v_{p_5-k}\Bar{v}_{p_2}\slashed{\epsilon}^\ast_{p_1+p_2}u_{p_1}}{(2k^+)(p_1^-+p_2^--(p_1+p_2)^-)(p_1^-+p_1^--(p_3+p_4)^--(p_5-k)^--k^-)(p_1^-+p_2^--(p_3+p_4+k)^--(p_5-k)^-)}\Biggr]
\end{multline}
which is IR divergent as $(p_5^--(p_5-k)^--k^-)\rightarrow0$ in the limit $k\rightarrow0$.

\subsection{Amplitudes involving $V_1V_2V_1$ vertices}

\begin{figure}[h]
    \centering
    \includegraphics[width=0.23\linewidth]{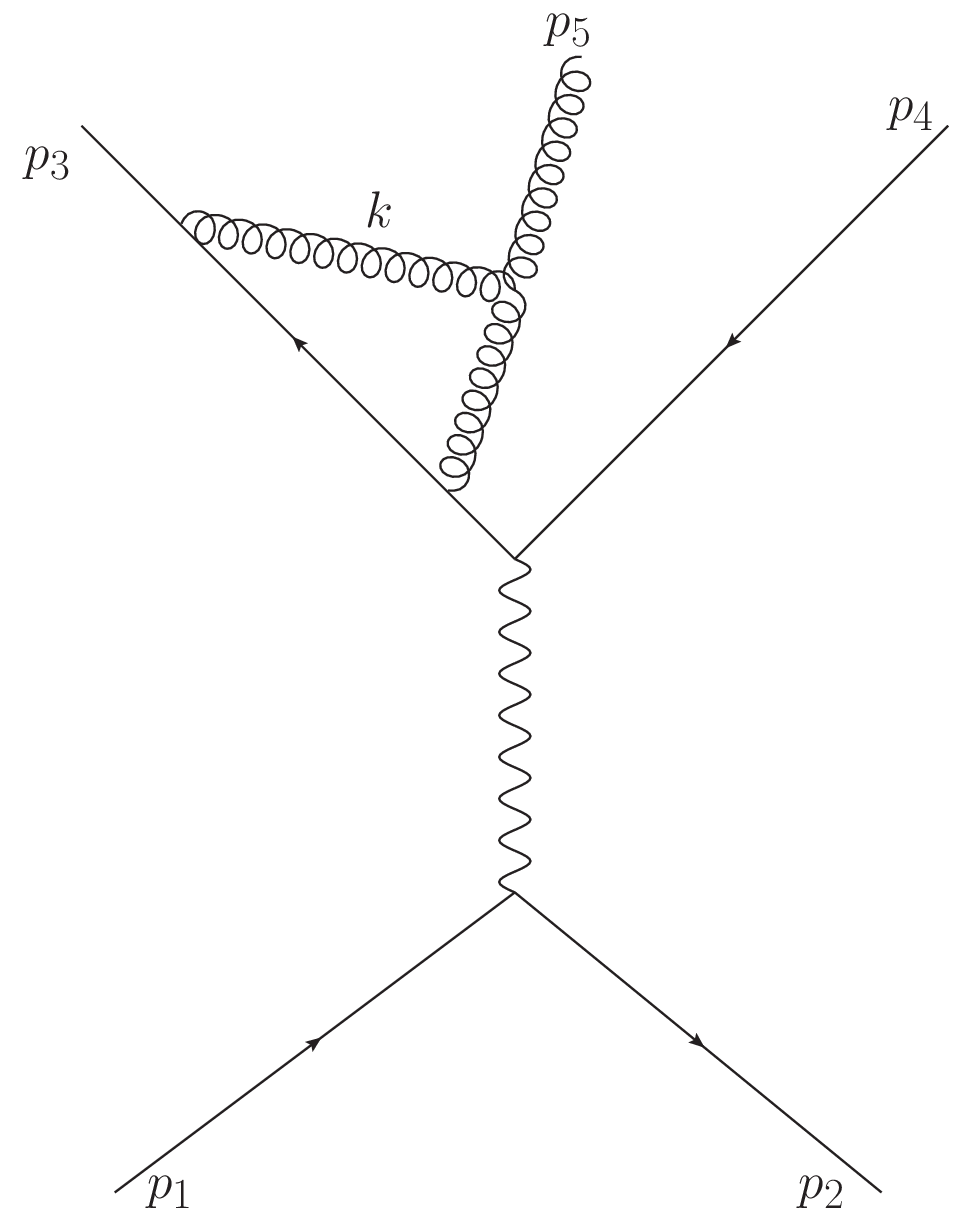} \includegraphics[width=0.23\linewidth]{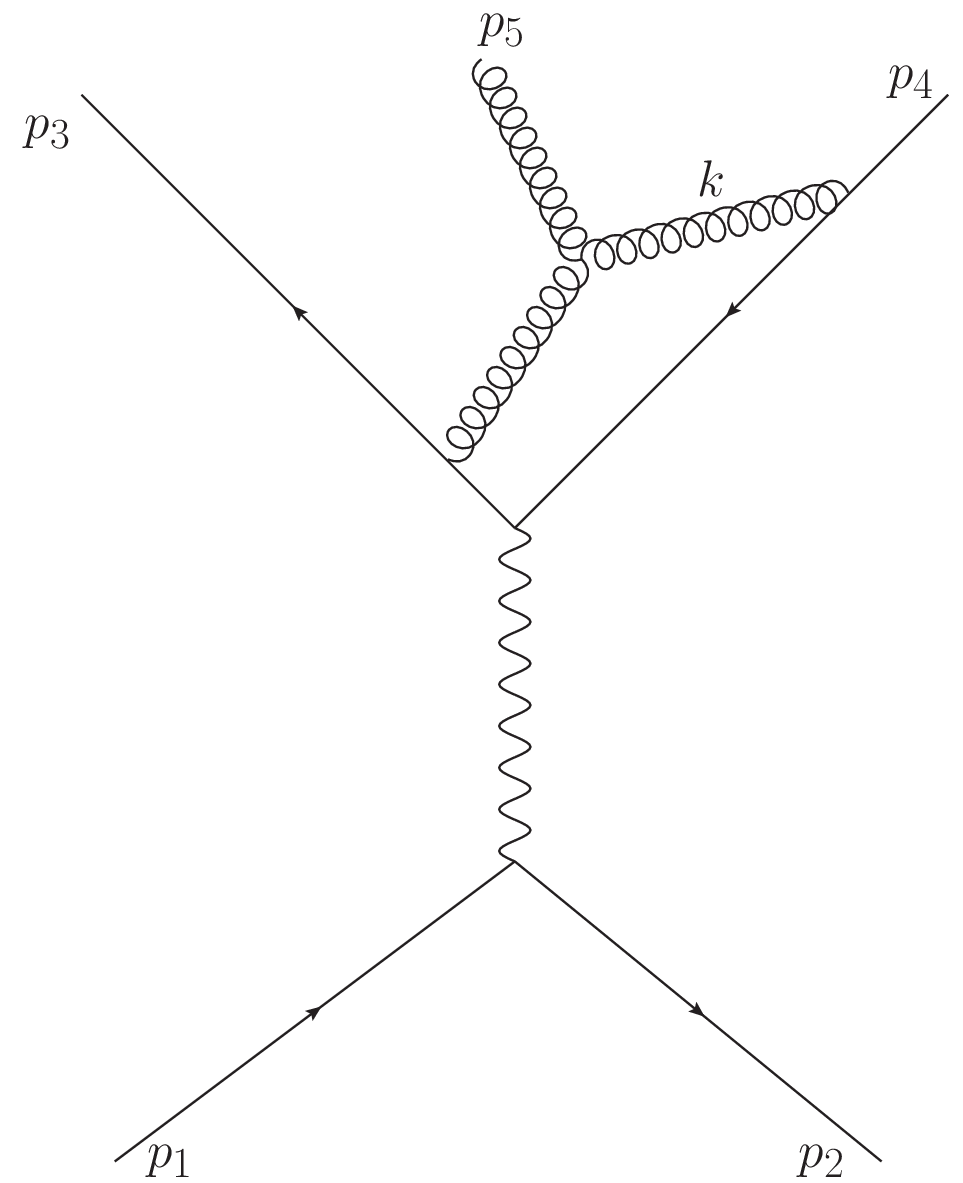}\\
    \hspace{0.02\linewidth} $T_{2a}$ \hspace{0.21\linewidth} $T_{2b}$\\
    \caption{Diagrams involving $V_1V_2V_1$ combination of interaction vertices}
    \label{fig:T2}
\end{figure}

Contributions involving $V_1V_2V_1$ vertices to the transition amplitude at $\mathcal{O}(g^3)$ can be represented by the diagrams in Fig. \ref{fig:T2}

and the corresponding amplitudes are
\begin{multline}
    T_{2a}=\frac{e^2g^3}{(2\pi)^{15/2}\prod_i\sqrt{2p_i^+}}\int [dk]\Biggl[\frac{\Bar{u}_{p_3}\slashed{\epsilon }^e_{k}T^e{u}_{p_3-k}f^{abd}(ik^\mu \epsilon^{\ast \nu}_{b k}(\epsilon^{\ast d}_{\nu p_5}\epsilon^{a}_{\mu p_5+k}+\epsilon^{d}_{\nu p_5+k}\epsilon^{\ast a}_{\mu p_5})+ip_5^\mu \epsilon^{\ast \nu}_{{p_5}b}(\epsilon^{\ast d}_{\nu k}\epsilon^{a}_{\mu p_5+k}}{(2k^+)(2(p_3-k)^+)(2(p_3+p_5)^+)(2(p_1+p_2)^+)(p_1^-+p_2^--(p_1+p_2)^-)}\\  \frac{+\epsilon^{d}_{\nu p_5+k}\epsilon^{\ast a}_{\mu k})-i(p_5+k)^\mu \epsilon^{\nu}_{b{p_5+k}}(\epsilon^{\ast d}_{\nu p_5}\epsilon^{\ast a}_{\mu k}+\epsilon^{\ast d}_{\nu k}\epsilon^{\ast a}_{\mu p_5}))  \Bar{u}_{p_3-k}\slashed{\epsilon}^{\ast c}_{p_5+k}T^c{u}_{p_3+p_5}\Bar{u}_{p_3+p_5} \slashed{\epsilon}_{p_1+p_2}v_{p_4}\Bar{v}_{p_2}\slashed{\epsilon}^{\ast}_{p_1+p_2}u_{p_1}}{(p_3^-+p_5^--(p_3+p_5)^-)(p_3^-+p_5^--(p_3-k)^--(p_5+k)^-)(p_3^--(p_3-k)^--k^-)}\Biggr]
\end{multline}

and

\begin{multline}
    T_{2b}=\frac{e^2g^3}{(2\pi)^{15/2}\prod_i\sqrt{2p_i^+}}\int [dk]\Biggl[\frac{\Bar{v}_{p_4-k}\slashed{\epsilon }^e_{k}T^e{v}_{p_4}f^{abd}(ik^\mu \epsilon^{\ast \nu}_{b{k}}(\epsilon^{\ast d}_{\nu p_5}\epsilon^{a}_{\mu p_5+k}+\epsilon^{d}_{\nu p_5+k}\epsilon^{\ast a}_{\mu p_5})+ip_5^\mu \epsilon^{\ast \nu}_{{p_5}b}(\epsilon^{\ast d}_{\nu k}\epsilon^{a}_{\mu p_5+k}}{(2k^+)(2(p_3+p_5+k)^+)(2(p_4-k)^+)(2(p_1+p_2)^+)(p_1^-+p_2^--(p_1+p_2)^-)}\\  \frac{+\epsilon^{d}_{\nu p_5+k}\epsilon^{\ast a}_{\mu k})-i(p_5+k)^\mu \epsilon^\nu_{{p_5+k}b}(\epsilon^{\ast d}_{\nu p_5}\epsilon^{\ast a}_{\mu k}+\epsilon^{\ast d}_{\nu k}\epsilon^{\ast a}_{\mu p_5}))  \Bar{u}_{p_3}\slashed{\epsilon}^{\ast c}_{p_5+k}T^c{u}_{p_3+p_5+k}\Bar{u}_{p_3+p_5+k} \slashed{\epsilon}_{p_1+p_2}v_{p_4-k}\Bar{v}_{p_2}\slashed{\epsilon}^\ast_{p_1+p_2}u_{p_1}}{(p_1^-+p_2^--(p_3+p_5+k)^--(p_4-k)^-)(p_4^-+p_5^--(p_4-k)^--(p_5+k)^-)(p_4^--(p_4-k)^--k^-)}\Biggr]
\end{multline}

$T_{2a}$ and $T_{2b}$ have IR divergences as $(p_3^--(p_3-k)^--k^-)\rightarrow0$ and $(p_4^--(p_4-k)^--k^-)\rightarrow0$ respectively in the limit $k\rightarrow0$.

\subsection{Amplitudes involving $V_2V_1V_1$ vertices}

\begin{figure}[h]
    \centering
    \includegraphics[width=0.23\linewidth]{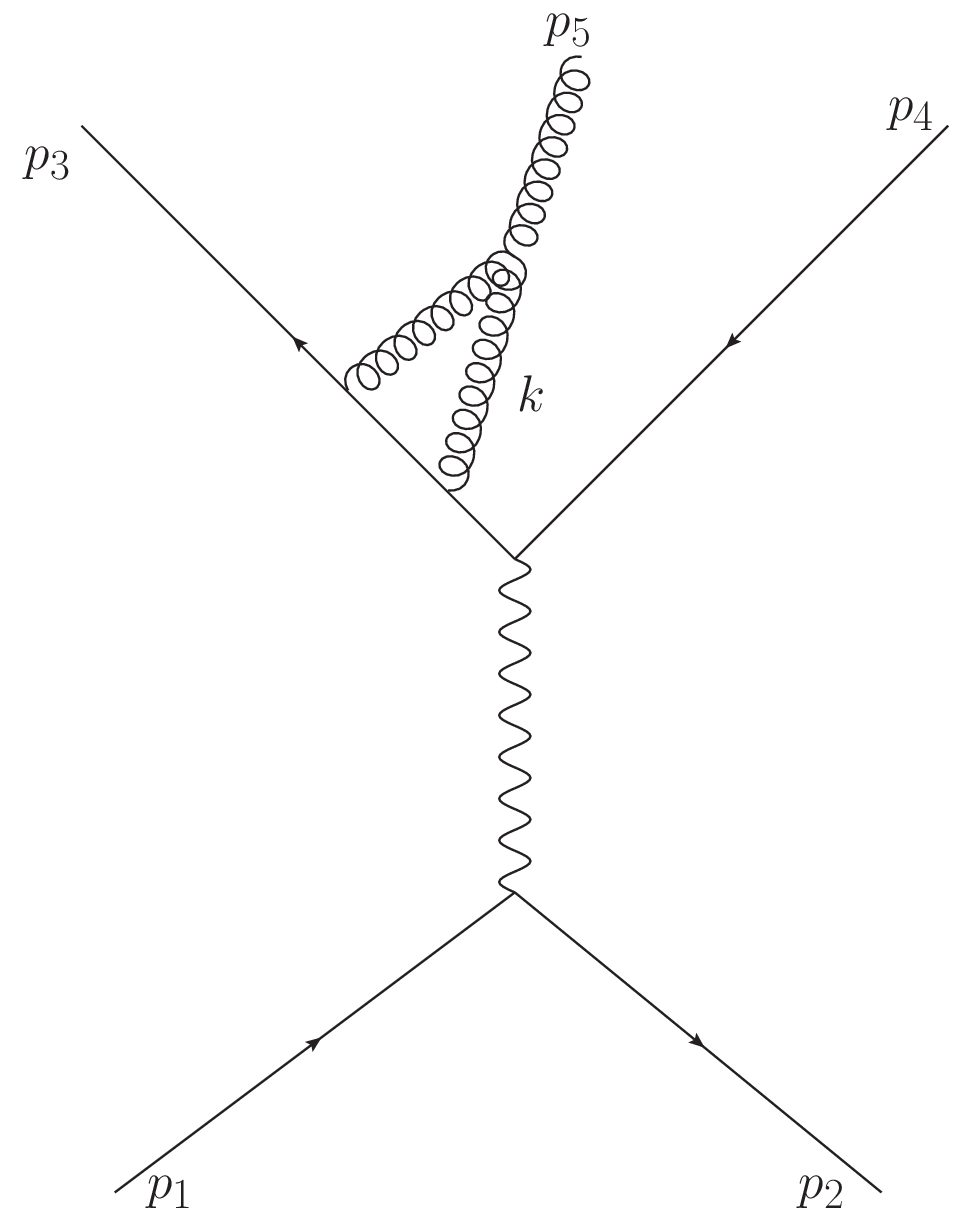} \includegraphics[width=0.23\linewidth]{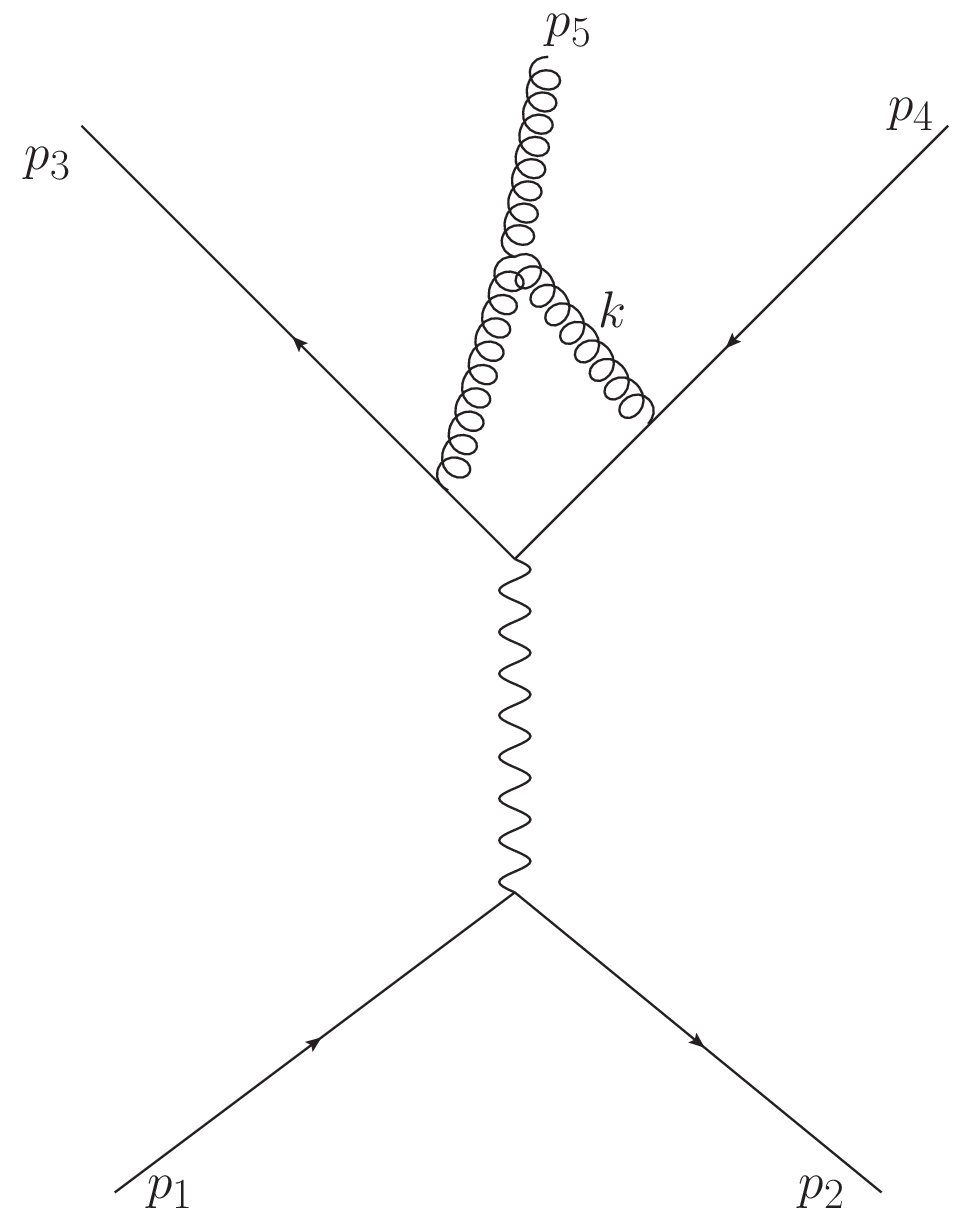}\\
    \hspace{0.02\linewidth} $T_{3a}$ \hspace{0.21\linewidth} $T_{3b}$\\
    \caption{Diagrams involving $V_2V_1V_1$ combination of interaction vertices}
    \label{fig:T3}
\end{figure}

Diagrams involving $V_2V_1V_1$ vertices and contributing to the transition amplitude at $\mathcal{O}(g^3)$ are represented in Fig. \ref{fig:T3}

and the corresponding amplitudes are
\begin{multline}
    T_{3a}=\frac{2e^2g^3}{(2\pi)^{15/2}\prod_i\sqrt{2p_i^+}}\int [dk]\Biggl[\frac{f^{abd}(-ik^\mu \epsilon^\nu_{b{k}}(\epsilon^{\ast d}_{\nu p_5}\epsilon^{a}_{\mu p_5-k}+\epsilon^{d}_{\nu p_5-k}\epsilon^{\ast a}_{\mu p_5})+ip_5^\mu \epsilon^{\ast \nu}_{{p_5}b}(\epsilon^{d}_{\nu k}\epsilon^{a}_{\mu p_5-k}+\epsilon^{d}_{\nu p_5-k}\epsilon^{a}_{\mu k})}{(2k^+)(2(p_3+p_5-k)^+)(2(p_3+p_5)^+)(2(p_1+p_2)^+)(p_1^-+p_2^--(p_1+p_2)^-)}\\  \frac{-i(p_5-k)^\mu \epsilon^{\nu}_{b{p_5-k}}(\epsilon^{\ast d}_{\nu p_5}\epsilon^{a}_{\mu k}+\epsilon^{d}_{\nu k}\epsilon^{\ast a}_{\mu p_5}))\Bar{u}_{p_3}\slashed{\epsilon}^{\ast e}_{p_5-k}T^e{u}_{p_3+p_5-k}\Bar{u}_{p_3+p_5-k}\slashed{\epsilon}^{\ast c}_{k}T^c{u}_{p_3+p_5}\Bar{u}_{p_3+p_5} \slashed{\epsilon}_{p_1+p_2}v_{p_4}\Bar{v}_{p_2}\slashed{\epsilon}^{\ast}_{p_1+p_2}u_{p_1}}{(p_3^-+p_5^--(p_3+p_5)^-)(p_3^-+p_5^--(p_3+p_5-k)^--k^-)(p_5^--(p_5-k)^--k^-)}\Biggr]
\end{multline}
and 
\begin{multline}
    T_{3b}=\frac{2e^2g^3}{(2\pi)^{15/2}\prod_i\sqrt{2p_i^+}}\int [dk]\Biggl[\frac{f^{abd}(-ik^\mu \epsilon^\nu_{b{k}}(\epsilon^{\ast d}_{\nu p_5}\epsilon^{a}_{\mu p_5-k}+\epsilon^{d}_{\nu p_5-k}\epsilon^{\ast a}_{\mu p_5})+ip_5^\mu \epsilon^{\ast \nu}_{{p_5}b}(\epsilon^{d}_{\nu k}\epsilon^{a}_{\mu p_5-k}+\epsilon^{d}_{\nu p_5-k}\epsilon^{a}_{\mu k})}{(2k^+)(2(p_3+p_5-k)^+)(2(p_4+k)^+)(2(p_1+p_2)^+)(p_1^-+p_2^--(p_1+p_2)^-)}\\  \frac{-i(p_5-k)^\mu \epsilon^{\nu}_{b{p_5-k}}(\epsilon^{\ast d}_{\nu p_5}\epsilon^{a}_{\mu k}+\epsilon^{d}_{\nu k}\epsilon^{\ast a}_{\mu p_5}))\Bar{v}_{p_4+k}\slashed{\epsilon}^{\ast e}_{k}T^e{v}_{p_4}\Bar{u}_{p_3}\slashed{\epsilon}^{\ast c}_{p_5-k}T^c{u}_{p_3+p_5-k}\Bar{u}_{p_3+p_5-k}\slashed{\epsilon}_{p_1+p_2}v_{p_4+k}\Bar{v}_{p_2}\slashed{\epsilon}^{\ast}_{p_1+p_2}u_{p_1}}{(p_1^-+p_2^--(p_4+k)^--(p_3+p_5-k)^-)(p_4^-+p_5^--(p_4+k)^--(p_5-k)^-)(p_5^--(p_5-k)^--k^-)}\Biggr]
\end{multline}

$T_{3a}$ and $T_{3b}$ have IR divergences as $(p_5^--(p_5-k)^--k^-)\rightarrow0$ in the limit $k\rightarrow0$.

\subsection{Amplitudes involving $V_1W_1$ vertices}

\begin{figure}[h]
    \centering
    \includegraphics[width=0.23\linewidth]{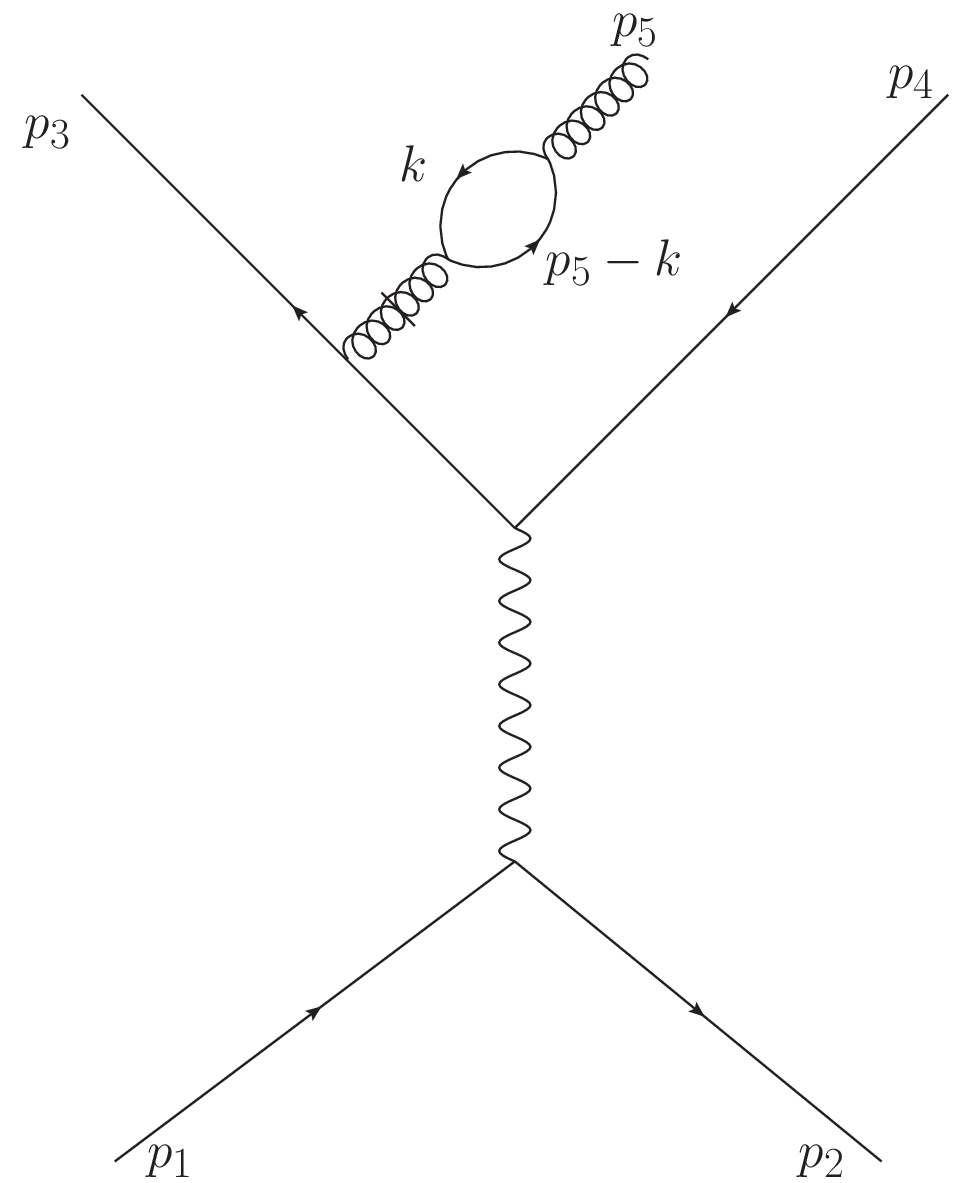} \includegraphics[width=0.23\linewidth]{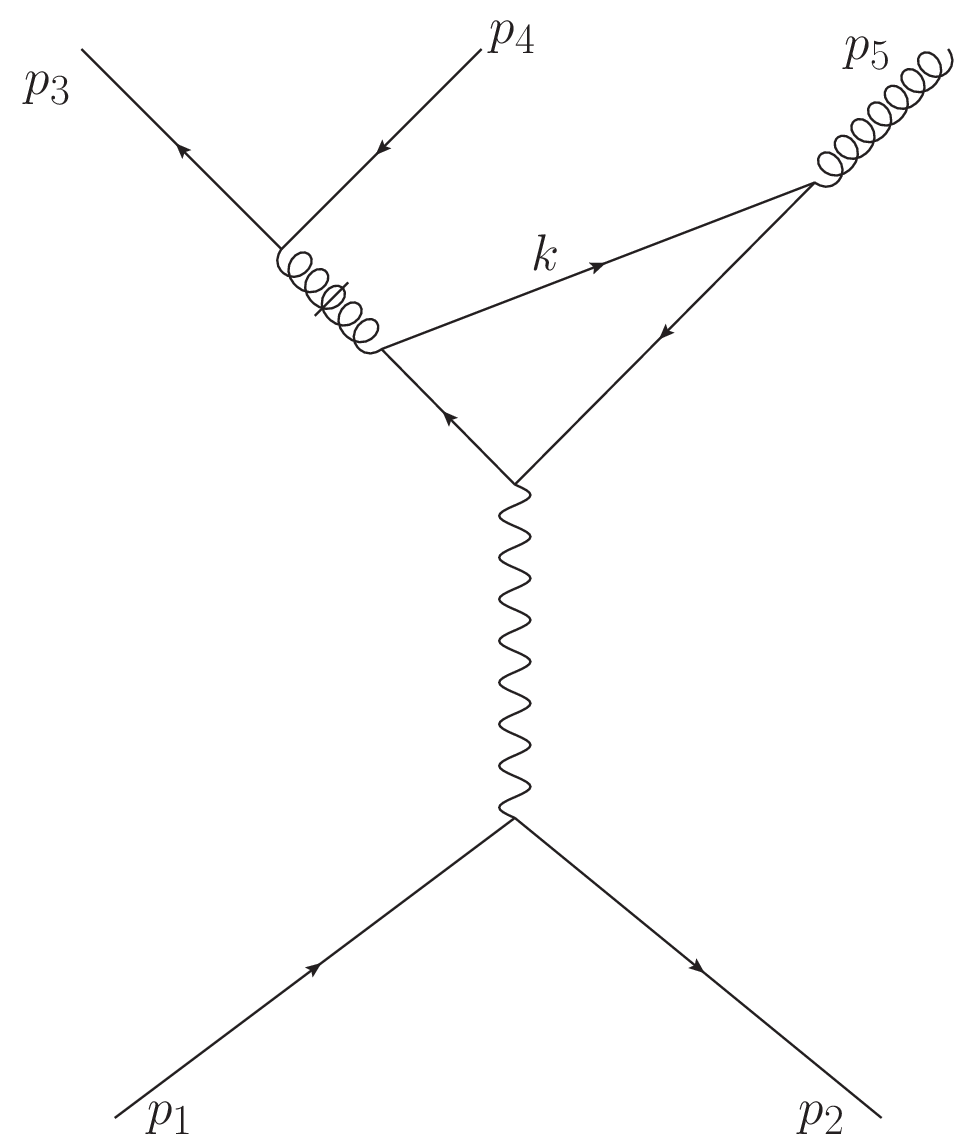} \includegraphics[width=0.23\linewidth]{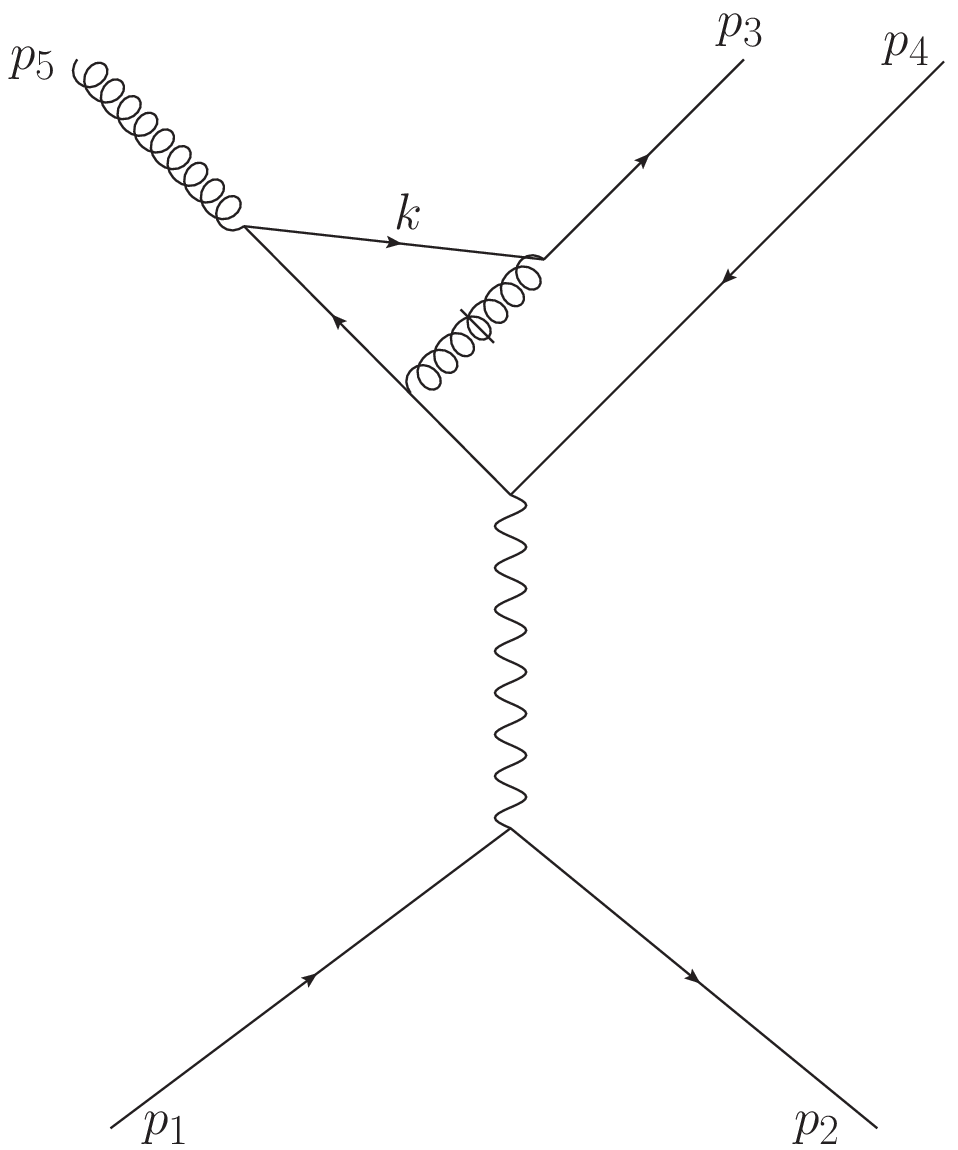}
    \includegraphics[width=0.23\linewidth]{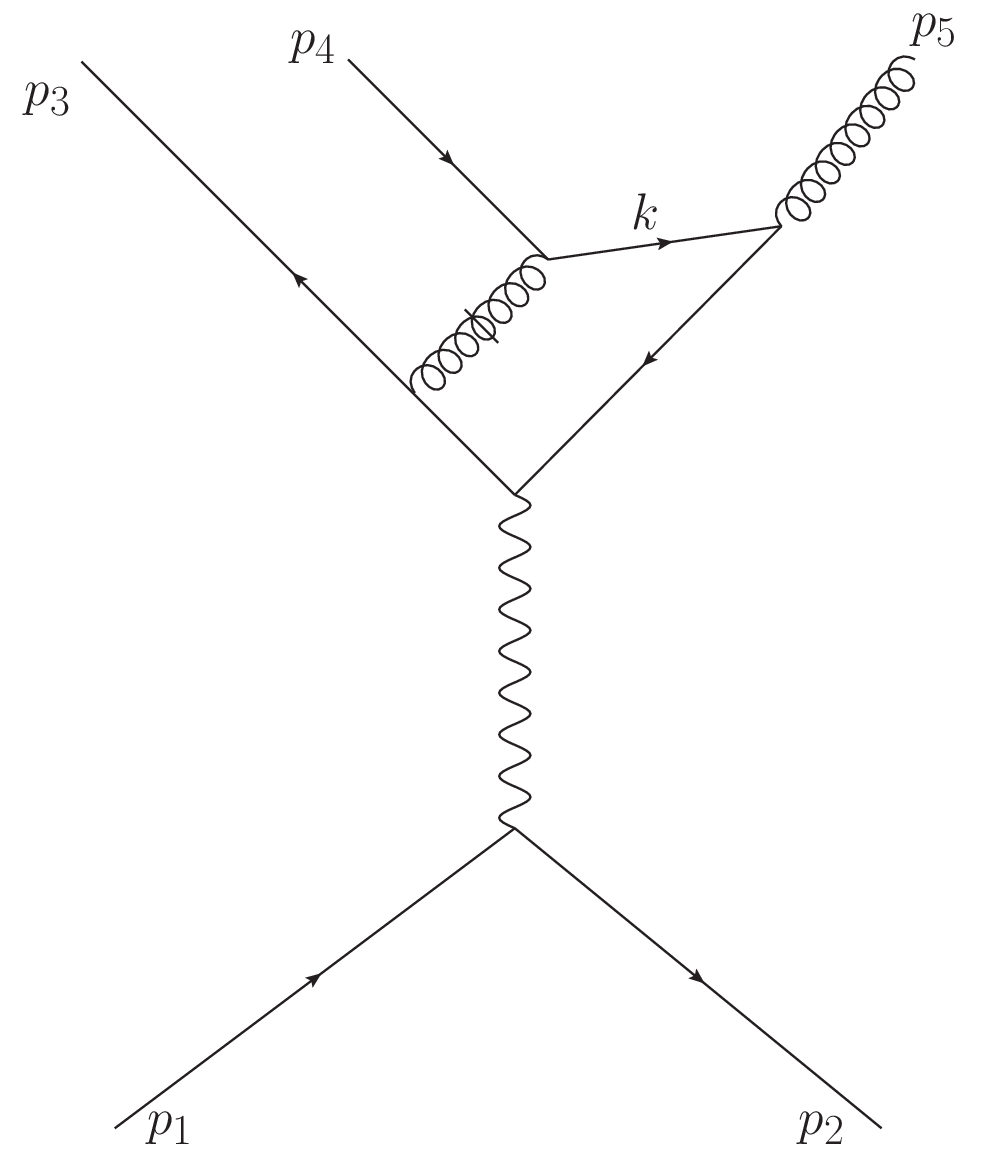}\\
    \hspace{0.02\linewidth} $T_{4a}$ \hspace{0.21\linewidth} $T_{4b}$\hspace{0.21\linewidth}$T_{4c}$\hspace{0.21\linewidth}$T_{4d}$
    \caption{Diagrams involving $V_1W_1$ combination of interaction vertices}
    \label{fig:T4}
\end{figure}

Diagrams involving $V_1W_1$ vertices and contributing to the transition amplitude at $\mathcal{O}(g^3)$ are illustrated in Fig. \ref{fig:T4}.

The amplitudes corresponding to diagrams $T_{4a}$ and $T_{4b}$ are presented here. The amplitudes for the other diagrams can be found in Appendix \ref{app:A}.	

\begin{multline}
    T_{4a}=\frac{-2e^2g^3}{(2\pi)^{21/2}\prod_i\sqrt{2p_i^+}}\int [dk]\frac{\Bar{v}_{p_5-k}\slashed{\epsilon }^{\ast c}_{p_5}T^c{u}_{k}\Bar{u}_{p_3}\gamma^+T^a{v}_{p_5-k}\Bar{u}_{k}\gamma^+T^a{u}_{p_3+p_5}}{(2(p_5-k)^+)(2(p_3+p_5)^+)(k^+ - (p_3+p_5)^+)^2(p_5^--(p_3-k)^--k^-)}\\  \times\frac{\Bar{u}_{p_3+p_5} \slashed{\epsilon}_{p_1+p_2}v_{p_4}\Bar{v}_{p_2}\slashed{\epsilon}^{\ast}_{p_1+p_2}u_{p_1}}{(2k^+)(2(p_1+p_2)^+)(p_1^-+p_2^--(p_1+p_2)^-)(p_3^-+p_5^--(p_3+p_5)^-)}
\end{multline}

\begin{multline}
    T_{4b}=\frac{-2e^2g^3}{(2\pi)^{21/2}\prod_i\sqrt{2p_i^+}}\int [dk]\frac{\Bar{v}_{p_5-k}\slashed{\epsilon }^{\ast c}_{p_5}T^c{u}_{k}\Bar{u}_{k}\gamma^+T^a{v}_{p_4}\Bar{u}_{p_3}\gamma^+T^a{u}_{p_3+p_4+k}}{(2(p_5-k)^+)(2(p_3+p_4+k)^+)(p_3^+ - (p_3+p_4+k)^+)^2(p_5^--(p_5-k)^--k^-)}\\  \times\frac{\Bar{u}_{p_3+p_4+k} \slashed{\epsilon}_{p_1+p_2}v_{p_5-k}\Bar{v}_{p_2}\slashed{\epsilon}^{\ast}_{p_1+p_2}u_{p_1}}{(2k^+)(2(p_1+p_2)^+)(p_1^-+p_2^--(p_1+p_2)^-)(p_1^-+p_2^--(p_3+p_4+k)^--(p_5-k)^-)}
\end{multline}

which have IR divergences as $(p_5^--(p_5-k)^--k^-)\rightarrow0$ in the limit $k\rightarrow0$.

\subsection{Amplitudes involving $V_1W_2$ vertices}

\begin{figure}[h]
    \centering
    \includegraphics[width=0.23\linewidth]{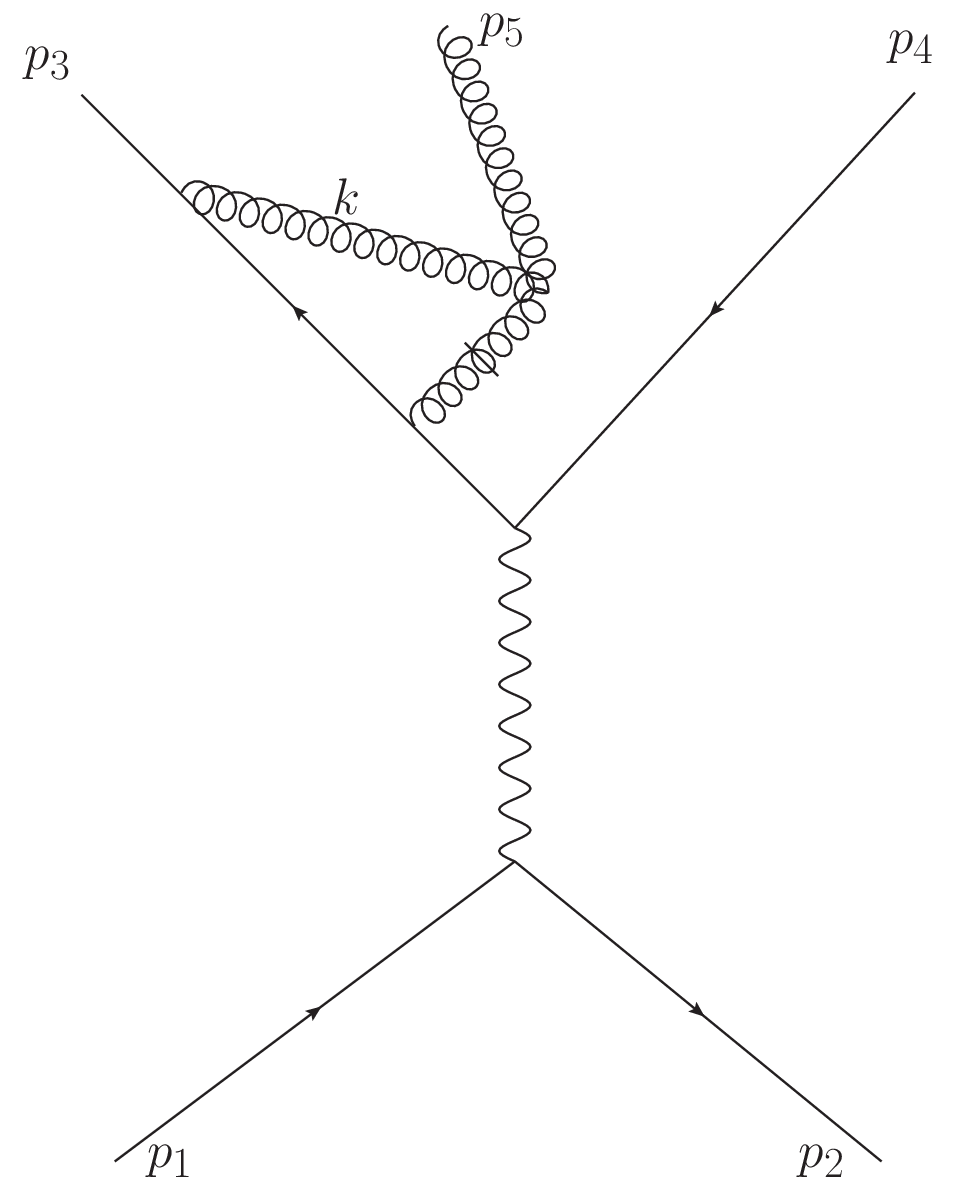} \includegraphics[width=0.23\linewidth]{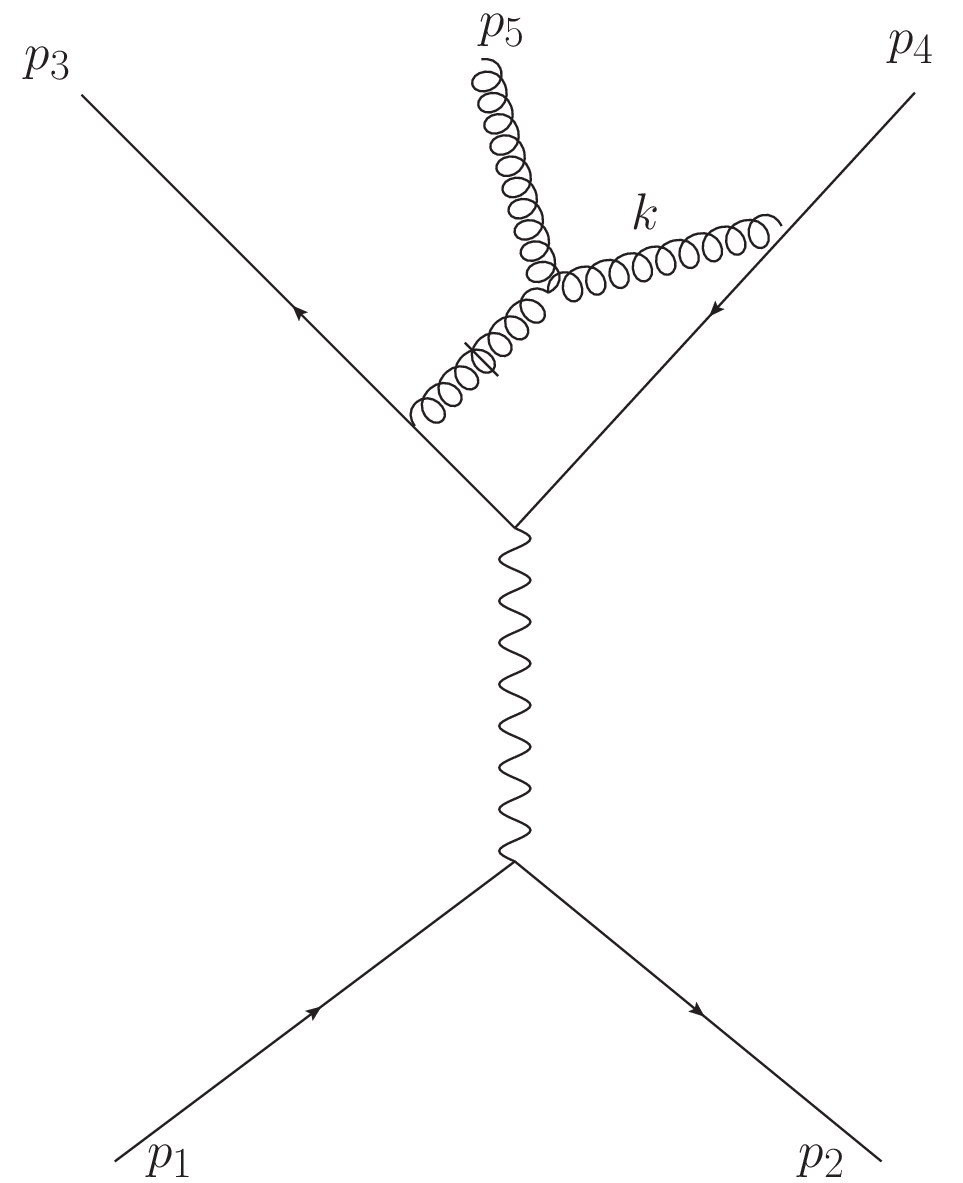}\\
    \hspace{0.02\linewidth} $T_{5a}$ \hspace{0.21\linewidth} $T_{5b}$\\
    \caption{Diagrams involving $V_1W_2$ combination of interaction vertices}
    \label{fig:T5}
\end{figure}

Diagrams with $V_1W_2$ vertices and contributing to the transition amplitude at $\mathcal{O}(g^3)$ are represented by in Fig. \ref{fig:T5}

and the corresponding amplitudes are
\begin{multline}
    T_{5a}=\frac{-e^2g^3}{(2\pi)^{21/2}\prod_i\sqrt{2p_i^+}}\int [dk]\frac{\Bar{u}_{p_3}\slashed{\epsilon }^e_{k}T^e{u}_{p_3-k}}{(2(p_1+p_2)^+)(p_1^-+p_2^--(p_1+p_2)^-)}\Biggl[\frac{\Bar{u}_{p_3-k}\gamma^+T^a_{c'c''}{u}_{p_3+p_5}f^{abd}(p_5^-\epsilon^{\ast \mu}_{p_5}\epsilon^{\ast}_{\mu k}+k^-\epsilon^{\ast \mu}_{k}\epsilon^{\ast}_{\mu p_5})}{(2(p_3-k)^+)(2(p_3+p_5)^+)(p_5^+-k^+)^2}\\  +\frac{f^{abd}(p_5^-\epsilon^{\ast \mu}_{p_5}\epsilon^{\ast}_{\mu k}+k^-\epsilon^{\ast \mu}_{k}\epsilon^{\ast}_{\mu p_5})\Bar{u}_{p_3-k}\gamma^+T^a_{c'c''}{u}_{p_3+p_5}}{(2(p_3-k)^+)(2(p_3+p_5)^+)((p_3-k)^+-(p_3+p_5)^+)^2}\Biggr]\frac{\Bar{u}_{p_3+p_5} \slashed{\epsilon}_{p_1+p_2}v_{p_4}\Bar{v}_{p_2}\slashed{\epsilon}^{\ast}_{p_1+p_2}u_{p_1}}{(2k^+)(p_3^-+p_5^--(p_3+p_5)^-)(p_3^--(p_3-k)^--k^-)}
\end{multline}

\begin{multline}
    T_{5b}=\frac{e^2g^3}{(2\pi)^{21/2}\prod_i\sqrt{2p_i^+}}\int [dk]\frac{\Bar{v}_{p_4-k}\slashed{\epsilon }^{c}_{k}T^c{v}_{p_4}}{(2(p_1+p_2)^+)(p_1^-+p_2^--(p_1+p_2)^-)}\Biggl[\frac{\Bar{u}_{p_3}\gamma^+T^a_{c'c''}{u}_{p_3+p_5+k}f^{abd}(p_5^-\epsilon^{\ast \mu}_{p_5}\epsilon^{\ast}_{\mu k}+k^-\epsilon^{\ast \mu}_{k}\epsilon^{\ast}_{\mu p_5})}{(2(p_4-k)^+)(2(p_3+p_5+k)^+)(p_5^+-k^+)^2}\\  +\frac{f^{abd}(p_5^-\epsilon^{\ast \mu}_{p_5}\epsilon^{\ast}_{\mu k}+k^-\epsilon^{\ast \mu}_{k}\epsilon^{\ast}_{\mu p_5})\Bar{u}_{p_3}\gamma^+T^a_{c'c''}{u}_{p_3+p_5+k}}{(2(p_4-k)^+)(2(p_3+p_5+k)^+)(p_3^+-(p_3+p_5+k)^+)^2}\Biggr]\frac{\Bar{u}_{p_3+p_5+k} \slashed{\epsilon}_{p_1+p_2}v_{p_4-k}\Bar{v}_{p_2}\slashed{\epsilon}^\ast_{p_1+p_2}u_{p_1}}{(2k^+)(p_1^-+p_2^--(p_3+p_5+k)^--(p_4-k)^-)(p_4^--(p_4-k)^--k^-)}
\end{multline}

$T_{5a}$ and $T_{5b}$ have IR divergences as $(p_3^--(p_3-k)^--k^-)\rightarrow0$ and $(p_4^--(p_4-k)^--k^-)\rightarrow0$ respectively in the limit $k\rightarrow0$.

\subsection{Amplitudes involving $V_2W_2$ vertices}

\begin{figure}[h]
    \centering
    \includegraphics[width=0.23\linewidth]{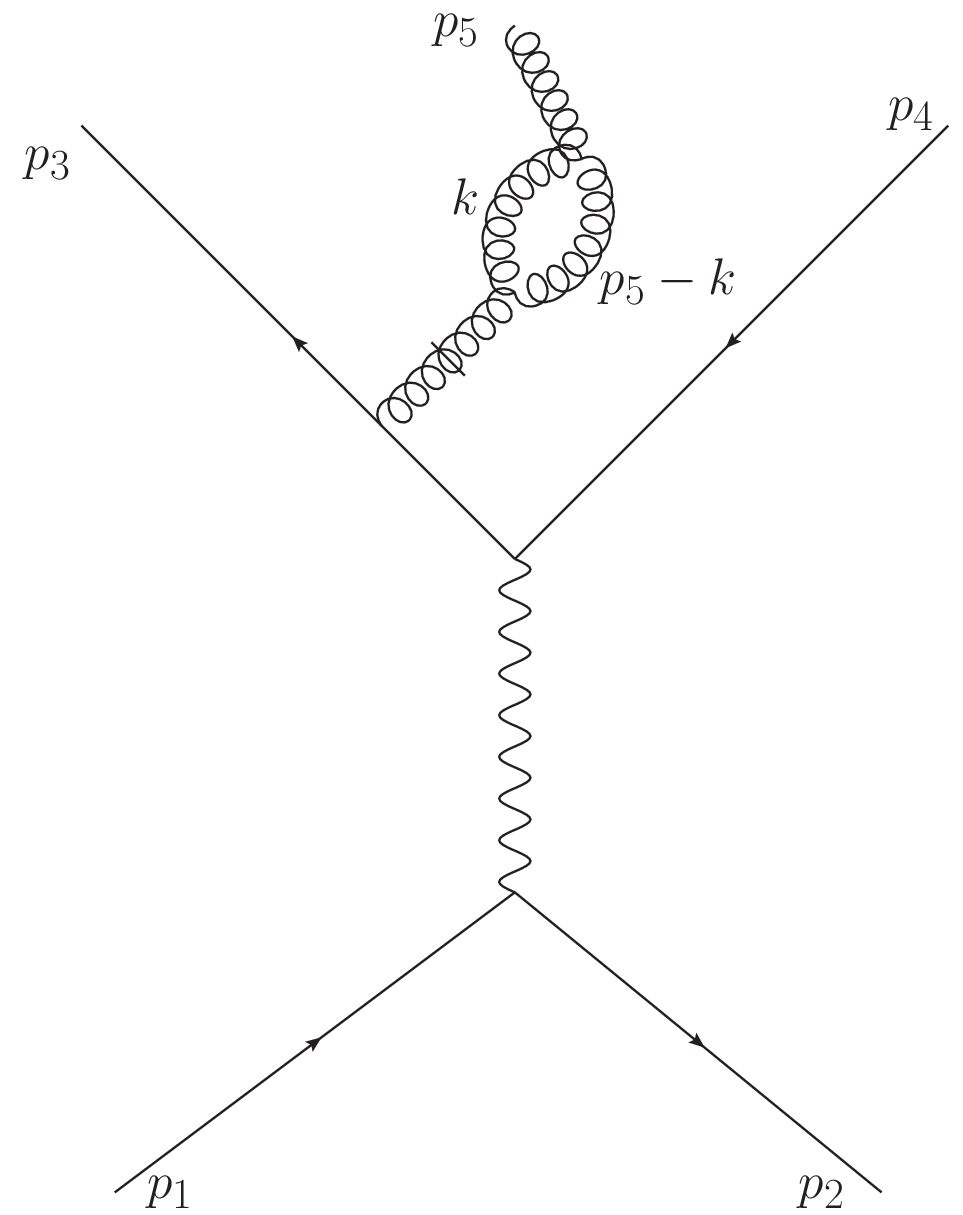} \\
    $T_{6a}$
    \caption{Diagrams involving $V_2W_2$ combination of interaction vertices}
    \label{fig:T6}
\end{figure}

Diagrams involving $V_2W_2$ vertices and contributing to the transition amplitude are represented in Fig. \ref{fig:T6}
and the corresponding amplitude is
\begin{multline}
    T_{6a}=\frac{e^2g^3}{2(2\pi)^{12}\prod_i\sqrt{2p_i^+}}\int [dk]\Biggl\{\frac{f^{abd}(p_5^{\mu} \epsilon^{\ast \nu}_{{p_5}b}(\epsilon^{d}_{\nu k}\epsilon^{a}_{\mu p_5-k}+\epsilon^{d}_{\nu p_5-k}\epsilon^{a}_{\mu k})-k^\mu \epsilon^\nu_{b{k}}(\epsilon^{\ast d}_{\nu p_5}\epsilon^{a}_{\mu p_5-k}+\epsilon^{d}_{\nu p_5-k}\epsilon^{\ast a}_{\mu p_5})}{(2k^+)(2(p_5-k)^+)(2(p_3+p_5)^+)(2(p_1+p_2)^+)(p_1^-+p_2^--(p_1+p_2)^-)}\\  \frac{-(p_5-k)^{\mu} \epsilon^{\nu}_{{p_5-k}b}(\epsilon^{\ast d}_{\nu p_5}\epsilon^{a}_{\mu k}+\epsilon^{d}_{\nu k}\epsilon^{\ast a}_{\mu p_5}))}{(p_3^-+p_5^--(p_3+p_5)^-)(p_5^--(p_5-k)^--k^-)}\Biggl[\frac{\Bar{u}_{p_3}\gamma^+T^a u_{p_3+p_5}f^{abd}(k^-\epsilon^{\ast \mu}_{b_k}\epsilon^{\ast d}_{\mu_{p_5-k}}+({p_5-k})^-\epsilon^{\ast \mu}_{b_{p_5-k}}\epsilon^{\ast d}_{\mu_k})}{((p_4+p_5)^+-p_4^+)^2}\\ +\frac{f^{abd}(k^-\epsilon^{\ast \mu}_{b_k}\epsilon^{\ast d}_{\mu_{p_5-k}}+({p_5-k})^-\epsilon^{\ast \mu}_{b_{p_5-k}}\epsilon^{\ast d}_{\mu_k})\Bar{u}_{p_3}\gamma^+T^a u_{p_3+p_5}}{((p_5-k)^+-k^+)^2} \Biggr]\frac{\Bar{u}_{p_3+p_5} \slashed{\epsilon}_{p_1+p_2}v_{p_4}\Bar{v}_{p_2}\slashed{\epsilon}^{\ast}_{p_1+p_2}u_{p_1}}{(p_1^-+p_2^--(p_1+p_2)^-)}\Biggr\}
\end{multline}

\subsection{Amplitudes involving $V_1W_4$ vertices}

\begin{figure}[h]
    \centering
    \includegraphics[width=0.23\linewidth]{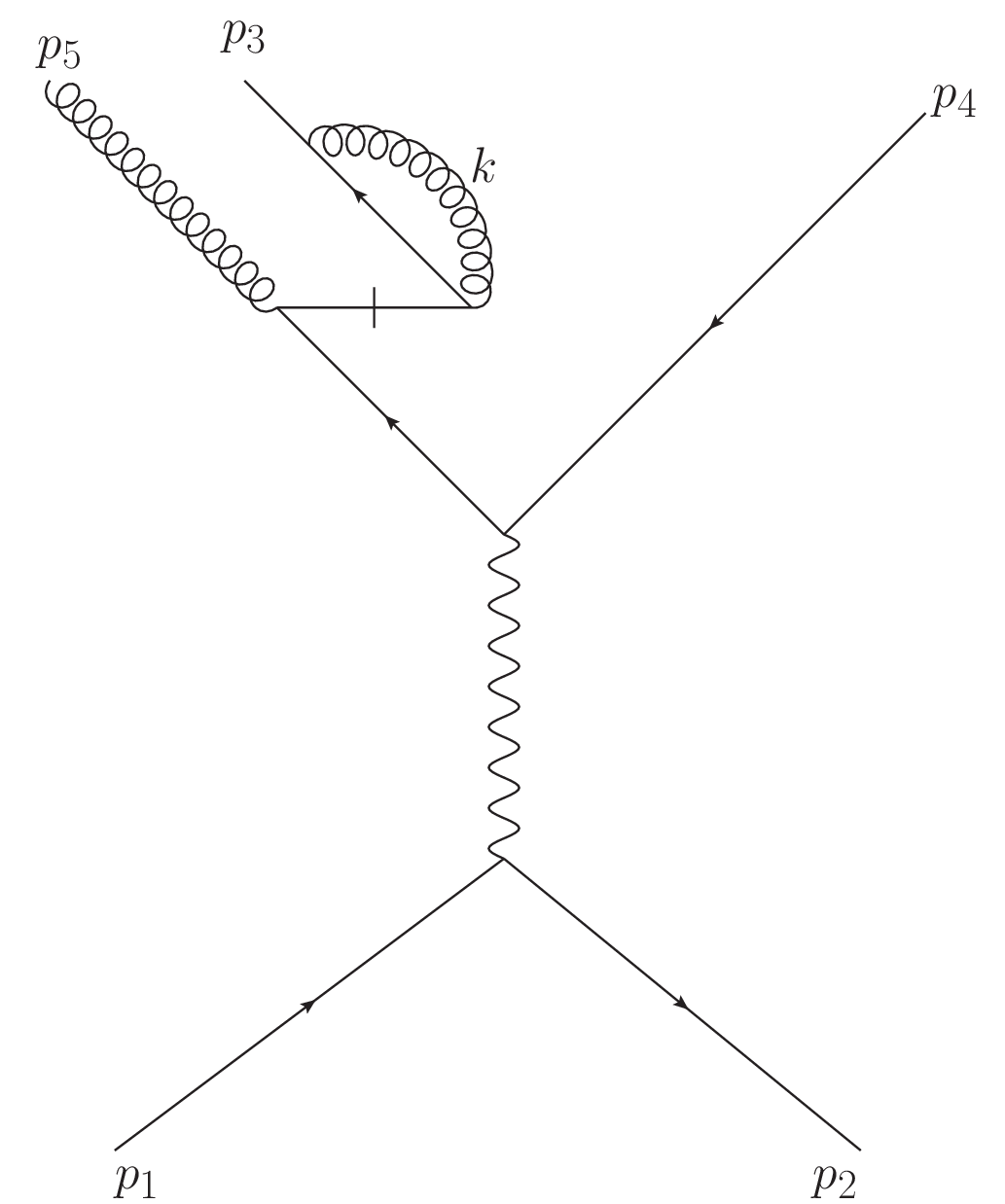} \includegraphics[width=0.23\linewidth]{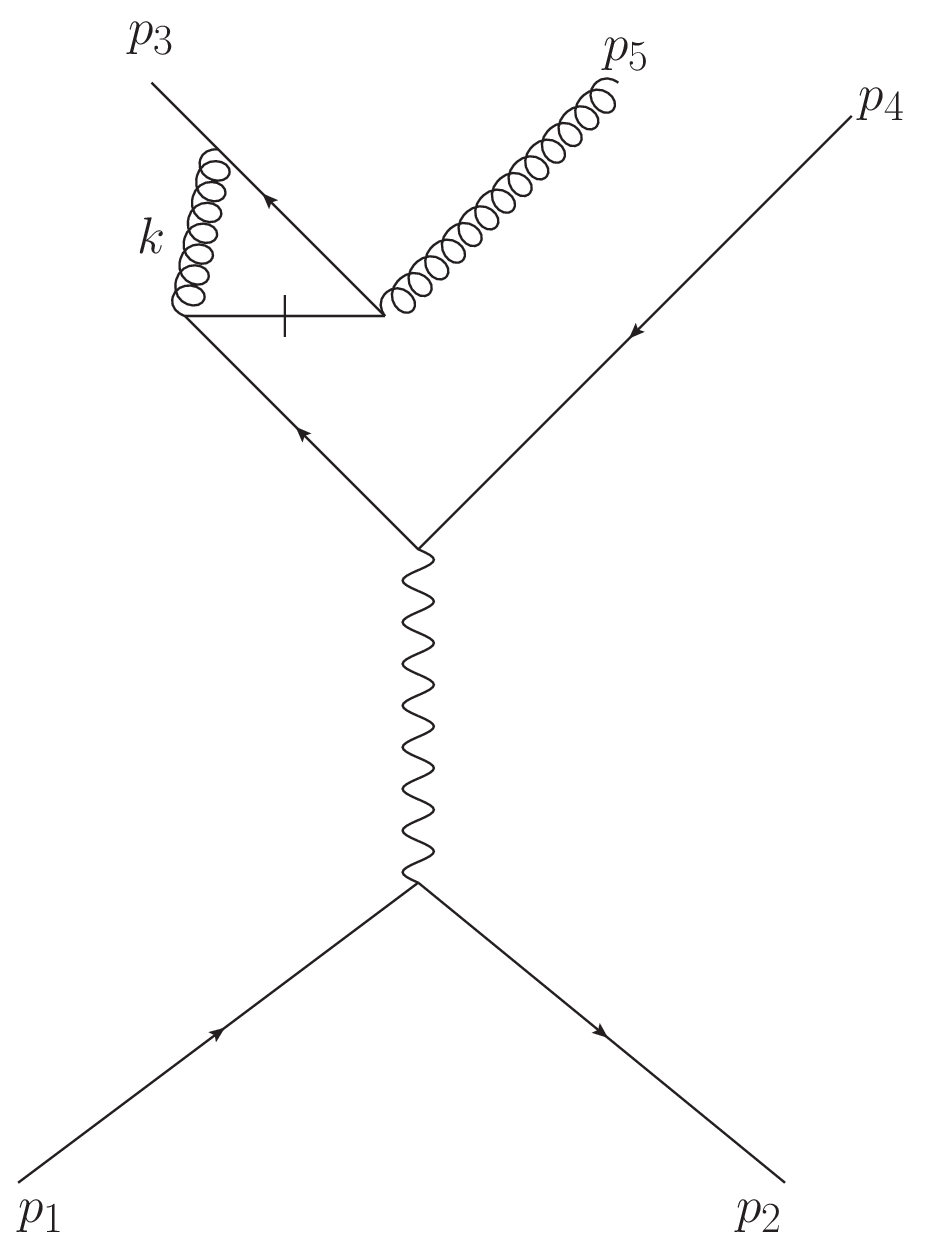} \includegraphics[width=0.23\linewidth]{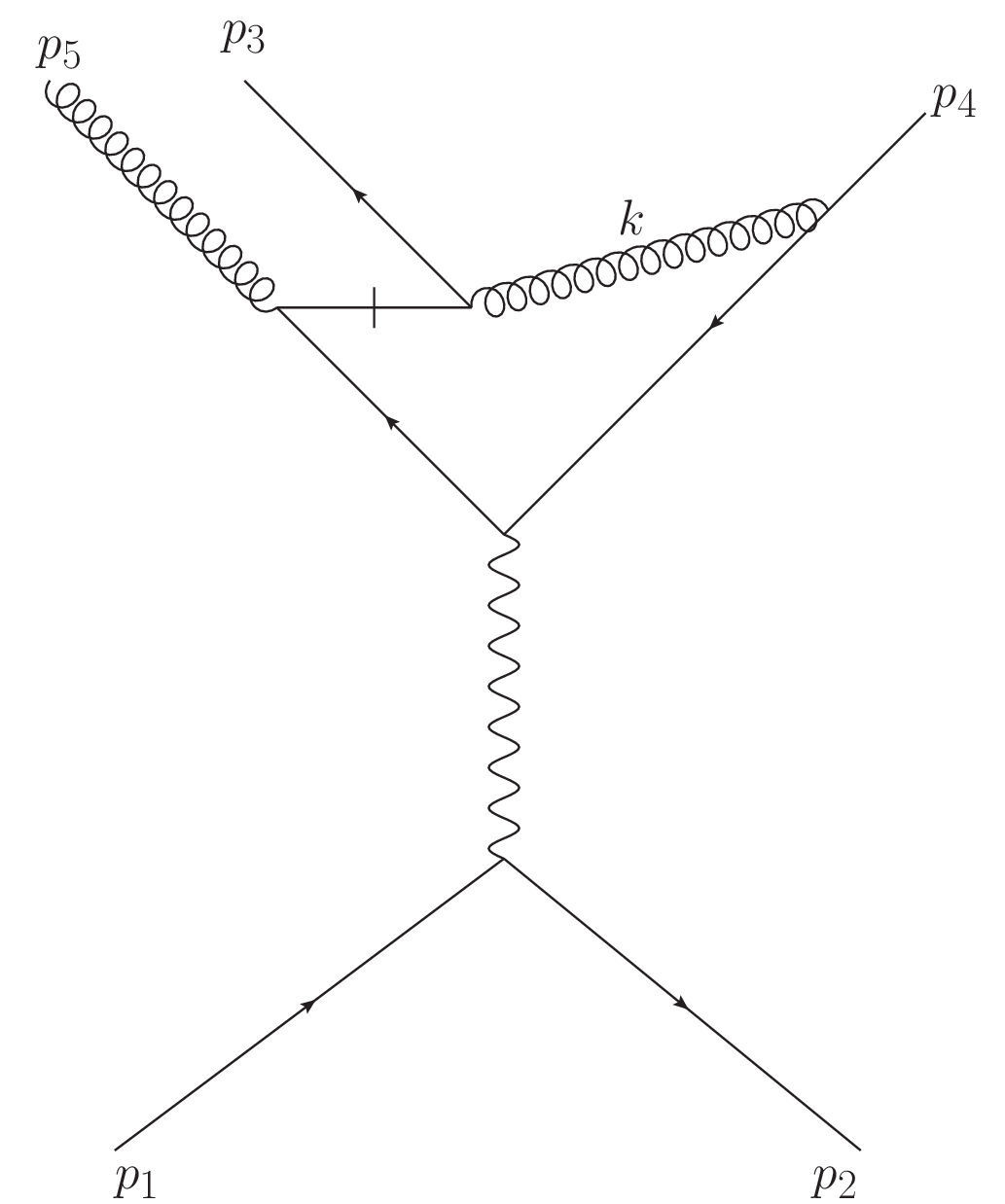}
    \includegraphics[width=0.23\linewidth]{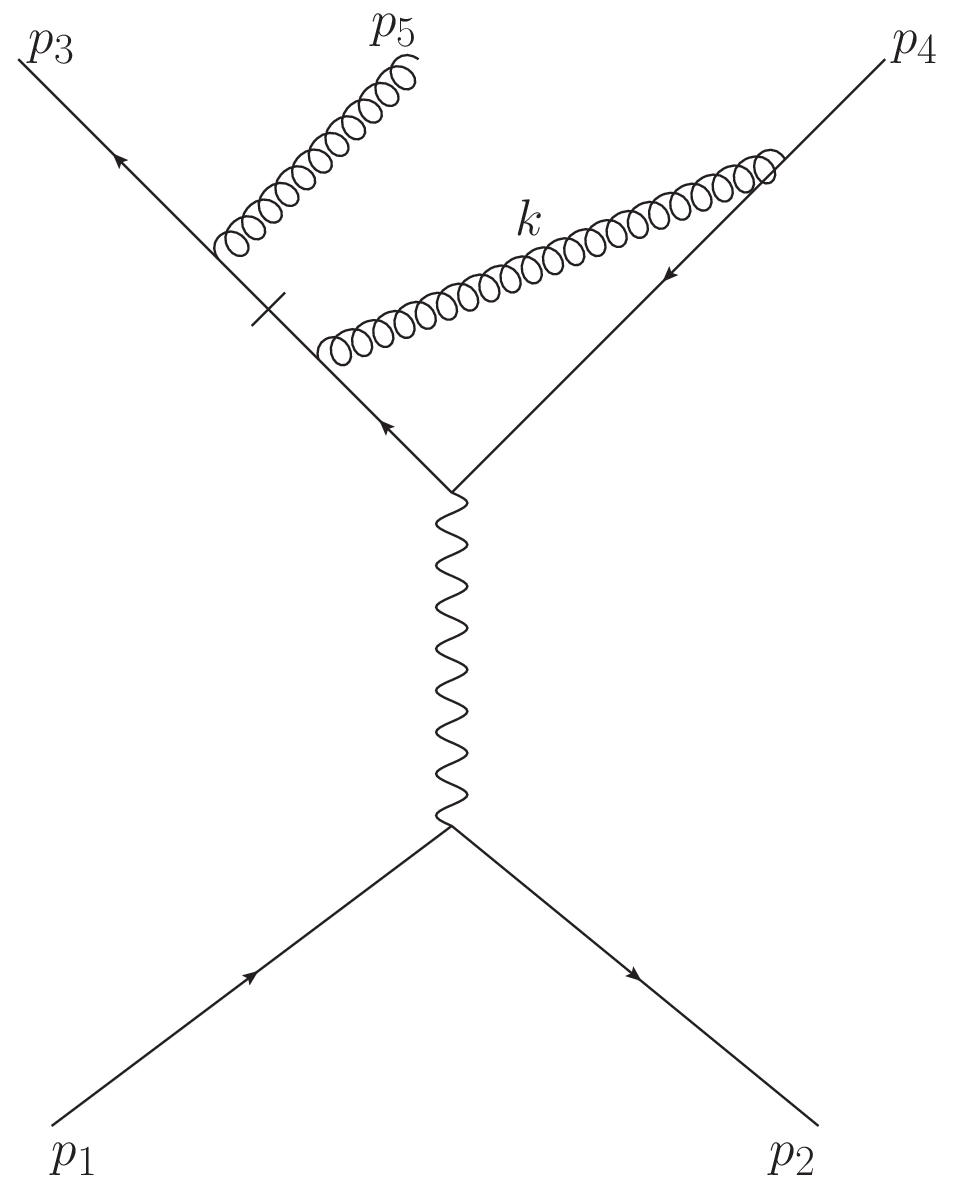}\\
    \hspace{0.02\linewidth} $T_{7a}$ \hspace{0.21\linewidth} $T_{7b}$\hspace{0.21\linewidth}$T_{7c}$\hspace{0.21\linewidth}$T_{7d}$
    \caption{Diagrams involving $V_1W_4$ combination of interaction vertices}
    \label{fig:T7}
\end{figure}

Diagrams involving $V_1W_4$ vertices and contributing to the transition amplitude at $\mathcal{O}(g^3)$ are illustrated in Fig. \ref{fig:T7}.

The amplitudes corresponding to diagrams $T_{7a}$ and $T_{7b}$ are presented below. 
\begin{multline}
    T_{7a}=\frac{-e^2g^3}{2(2\pi)^{21/2}\prod_i\sqrt{2p_i^+}}\int [k]\Biggl[\frac{\Bar{u}_{p_3}\slashed{\epsilon}^{c}_{k}T^c{u}_{p_3-k}\Bar{u}_{p_3-k}T^a\slashed{\epsilon}^{\ast a}_{k}\gamma^+\slashed{\epsilon}^{\ast b}_{p_5}T^b{u}_{p_3+p_5}}{(2(p_3+p_5)^+)(2(p_3-k)^+)(2k^+)(p_5^+-(p_3^++p_5^+))(p_3^--(p_3-k)^--k^-)}\\ \times \frac{\Bar{u}_{p_3+p_5}\slashed{\epsilon}_{p_1+p_2}v_{p_4}\Bar{v}_{p_2}\slashed{\epsilon}^{\ast}_{p_1+p_2}u_{p_1}}{(2(p_1+p_2)^+)(p_1^-+p_2^--(p_1+p_2)^-)(p_3^-+p_5^--(p_3+p_5)^-)}\Biggr]
\end{multline}

\begin{multline}
    T_{7b}=\frac{e^2g^3}{2(2\pi)^{21/2}\prod_i\sqrt{2p_i^+}}\int [k]\Biggl[\frac{\bar{v}_{p_4-k}\slashed{\epsilon}^{c}_{k}T^c{v}_{p_4}\Bar{u}_{p_3}T^a\slashed{\epsilon}^{\ast a}_{k}\gamma^+\slashed{\epsilon}^{\ast b}_{p_5}T^b{u}_{p_3+p_5-k}}{(2(p_3+p_5+k)^+)(2(p_4-k)^+)(2k^+)(p_5^+-(p_3^+p_5^+k^+)(p_4^--(p_4-k)^--k^-)}\\ \times \frac{\Bar{u}_{p_3+p_5+k}\slashed{\epsilon}_{p_1+p_2}v_{p_4-k }\Bar{v}_{p_2}\slashed{\epsilon}^{\ast}_{p_1+p_2}u_{p_1}}{(2(p_1+p_2)^+)(p_1^-+p_2^--(p_1+p_2)^-)(p_1^-+p_2^--(p_4-k)^--(p_3+p_5+k)^-)}\Biggr]
\end{multline}

The amplitudes for the other diagrams can be found in Appendix \ref{app:A}.

$T_{7a}$ and $T_{7b}$ have IR divergences as $(p_3^--(p_3-k)^--k^-)\rightarrow0$ and $(p_4^--(p_4-k)^--k^-)\rightarrow0$ respectively in the limit $k\rightarrow0$.


\section{Transition amplitude in the coherent state basis}\label{sec:coh components}

Transition amplitude in the coherent state basis can be computed by expanding the asymptotic M\o{}ller operator in the coherent state expression upto the desired order and using the resulting Fock state expansion to calculate the transition matrix element. Thus, in addition to the amplitude contributions due to the Fock basis, calculated in sec.\ref{sec:Fock diag}, there will be additional contributions arising from the higher Fock states in the expansion of the coherent state as can be seen from Eq. (\ref{cohstate})-(\ref{eq:transition-amplitude}). 

For example, in coherent state basis, $\mathcal{O}(g^3)$ contributions can also result due to overlap of the $\mathcal{O}(g)$ term in expansion of the coherent state with the final state of $\mathcal{O}(g)$ correction to the tree level process due to real soft/collinear emission, and also due to overlap of the $\mathcal{O}(g^2)$ term in expansion of coherent state with the final state of the tree level process. We will represent these soft/collinear emissions with a dashed line (as is the convention used in Ref.\cite{jaimisrafeynmangauge}) to distinguish these diagrams from the diagrams of the process $e^+ e^- \rightarrow 4 $ jets, for example. We use the same convention for soft/collinear quarks and gluons as the identity of the soft parton will be evident from the vertices attached to it. 

For the present case, it suffices to consider the $\mathcal{O}(g)$ terms in $\Omega_{\Delta(V_1)}\vert q {\bar q} g \rangle $ and $\Omega_{\Delta(V_2)}\vert q {\bar q} g \rangle$ and the $\mathcal{O}(g^2)$ terms in  $\Omega_{\Delta(W_4)}\vert q {\bar q} g \rangle$ in the coherent state expansion.

In the following discussion, we present expressions and diagrams for the additional contributions to the transition amplitude of the process $e^{+}e^{-}\rightarrow q\bar{q}g$ which arise  due to $\mathcal{O}(g)$ and $\mathcal{O}(g^2)$ terms in expansion of the coherent state. As in sec.\ref{sec:Fock diag}, we classify the full set of diagrams according to the vertices involved. The  diagrams fall into following categories:


\subsection{Amplitudes involving $\Omega_{\Delta(V_1)}V_1V_1$ }

\begin{figure}[h]
    \centering
    \includegraphics[width=0.23\linewidth]{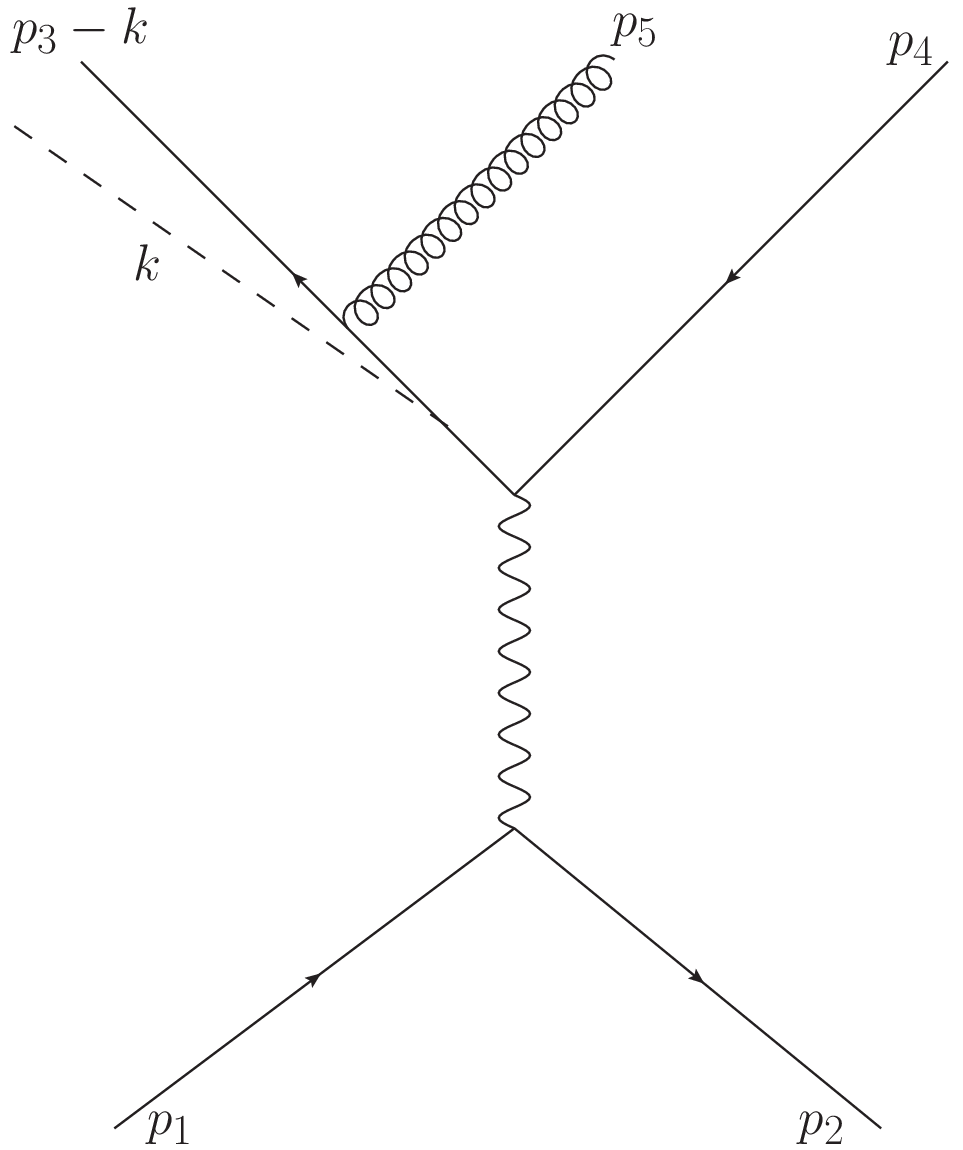} \includegraphics[width=0.23\linewidth]{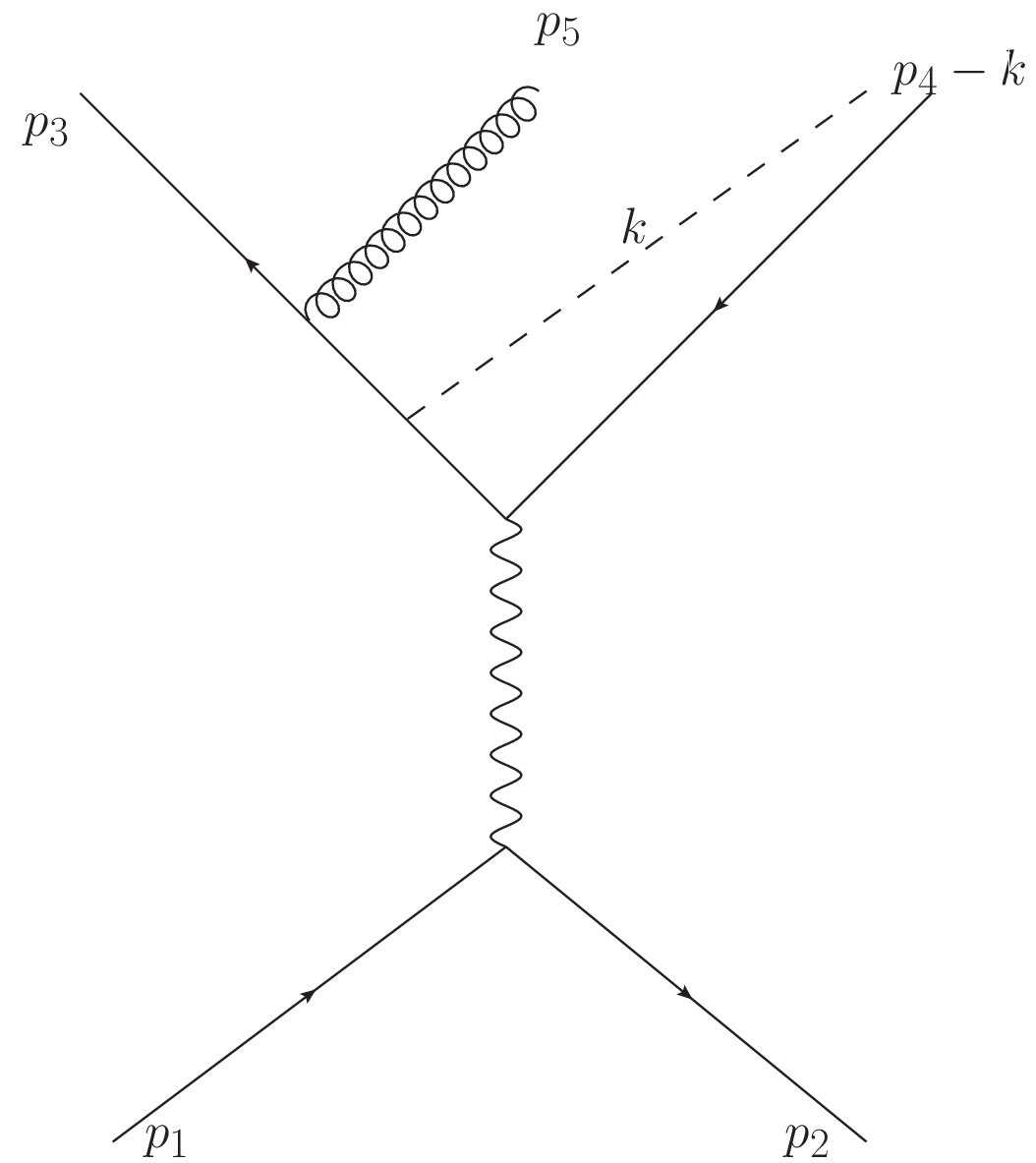} \includegraphics[width=0.23\linewidth]{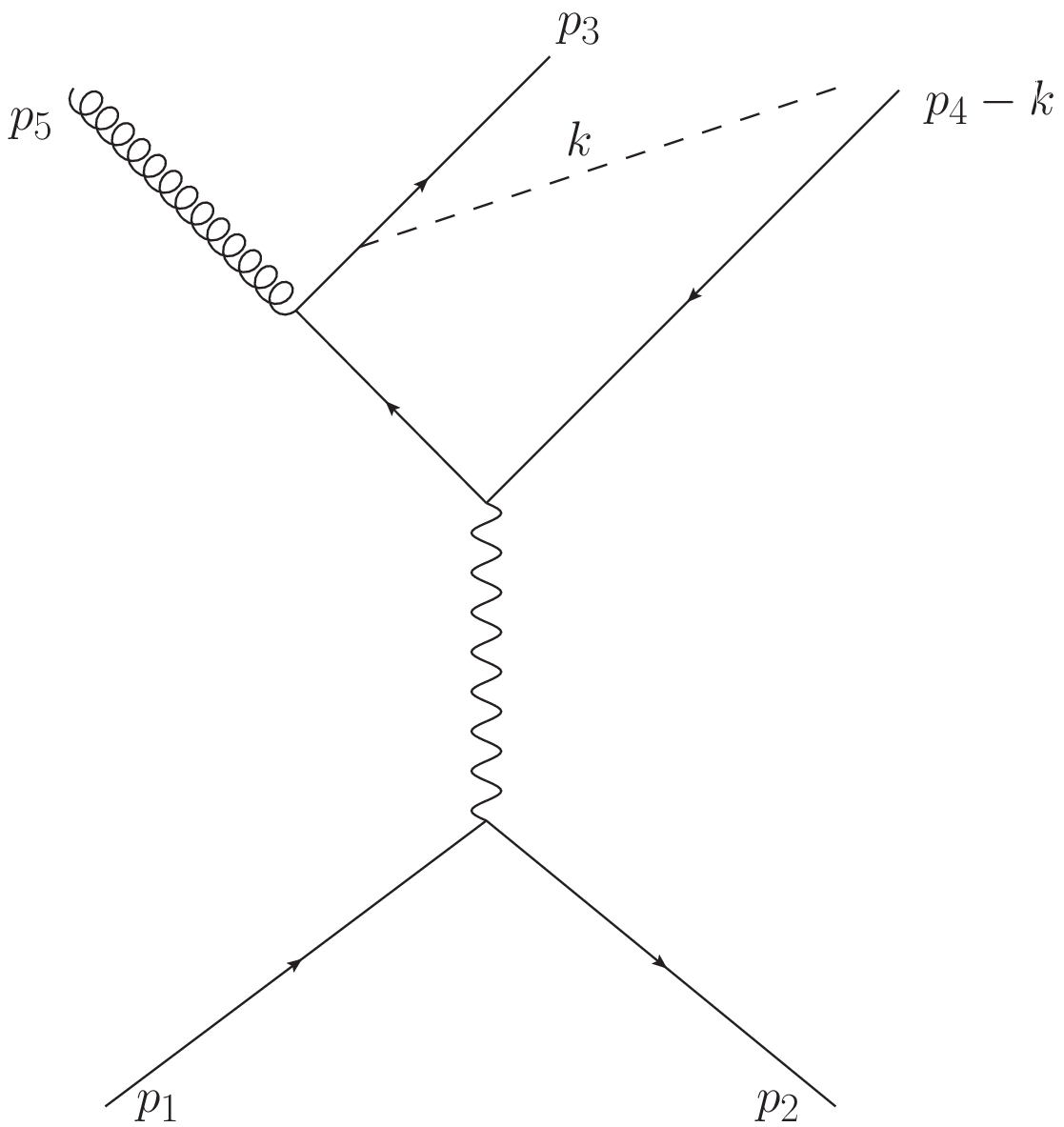}
    \includegraphics[width=0.23\linewidth]{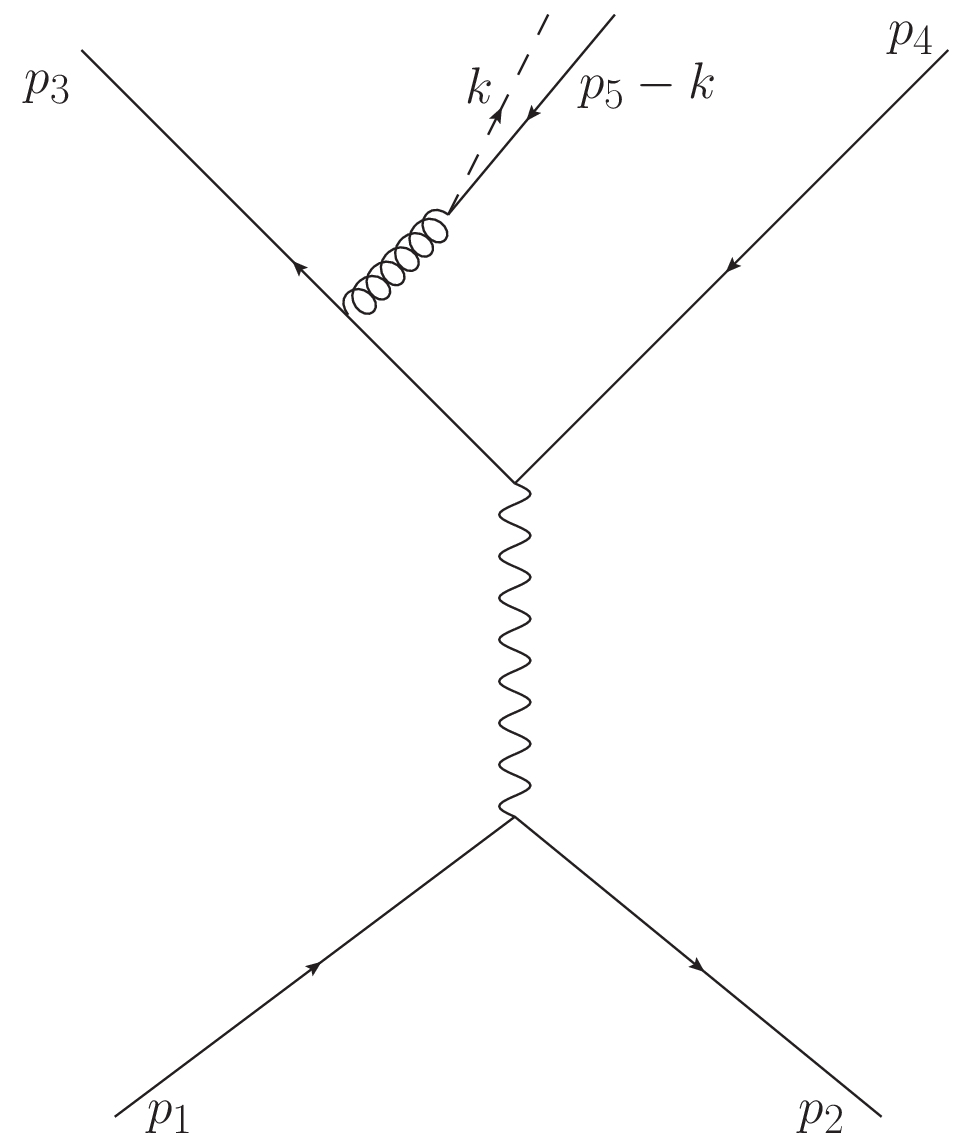}\\
    \hspace{0.02\linewidth} $T_{1a}^{coh}$ \hspace{0.21\linewidth} $T_{1b}^{coh}$\hspace{0.21\linewidth}$T_{1c}^{coh}$\hspace{0.21\linewidth}$T_{1d}^{coh}$\\
    \includegraphics[width=0.23\linewidth]{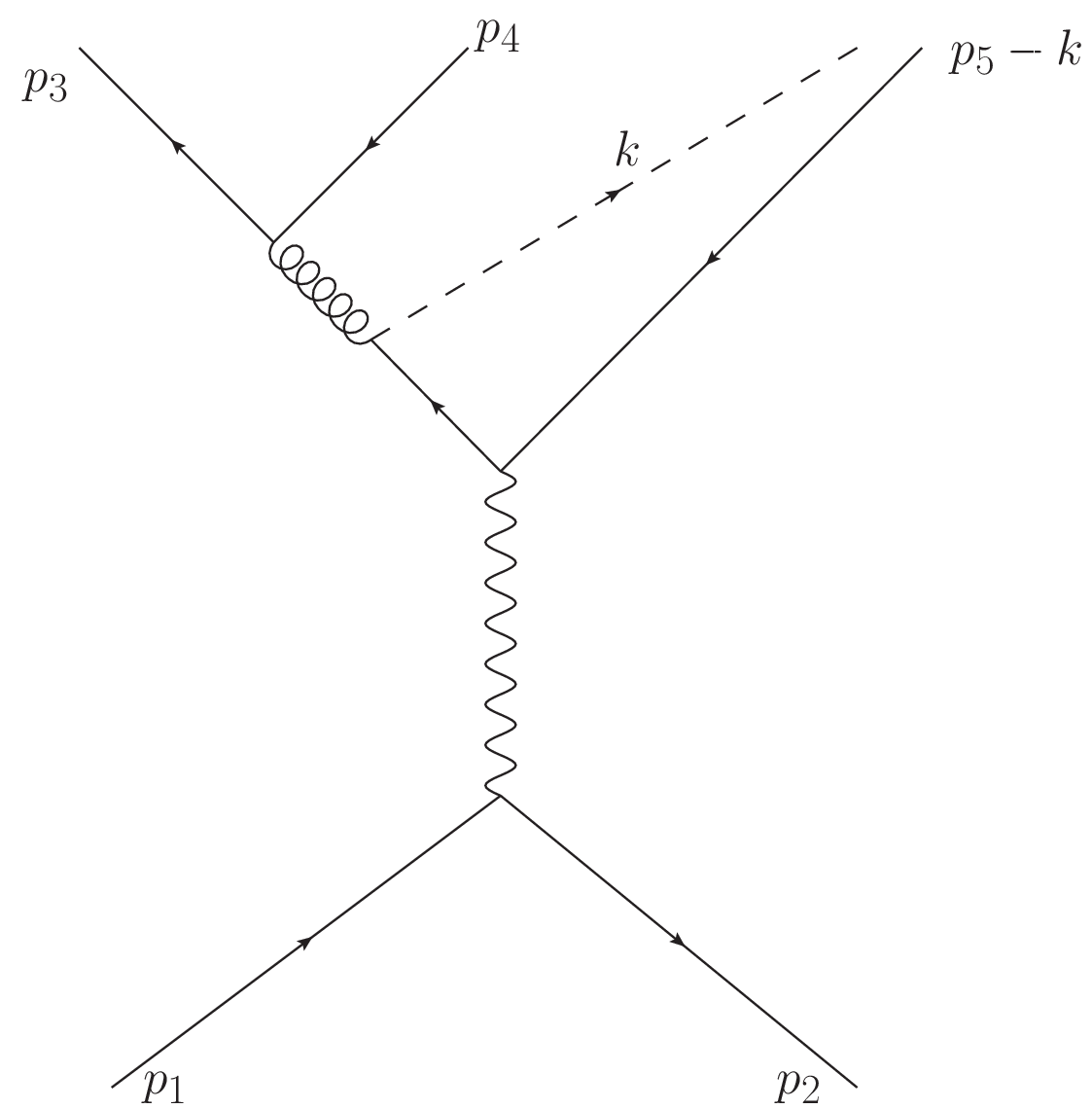}
    \includegraphics[width=0.23\linewidth]{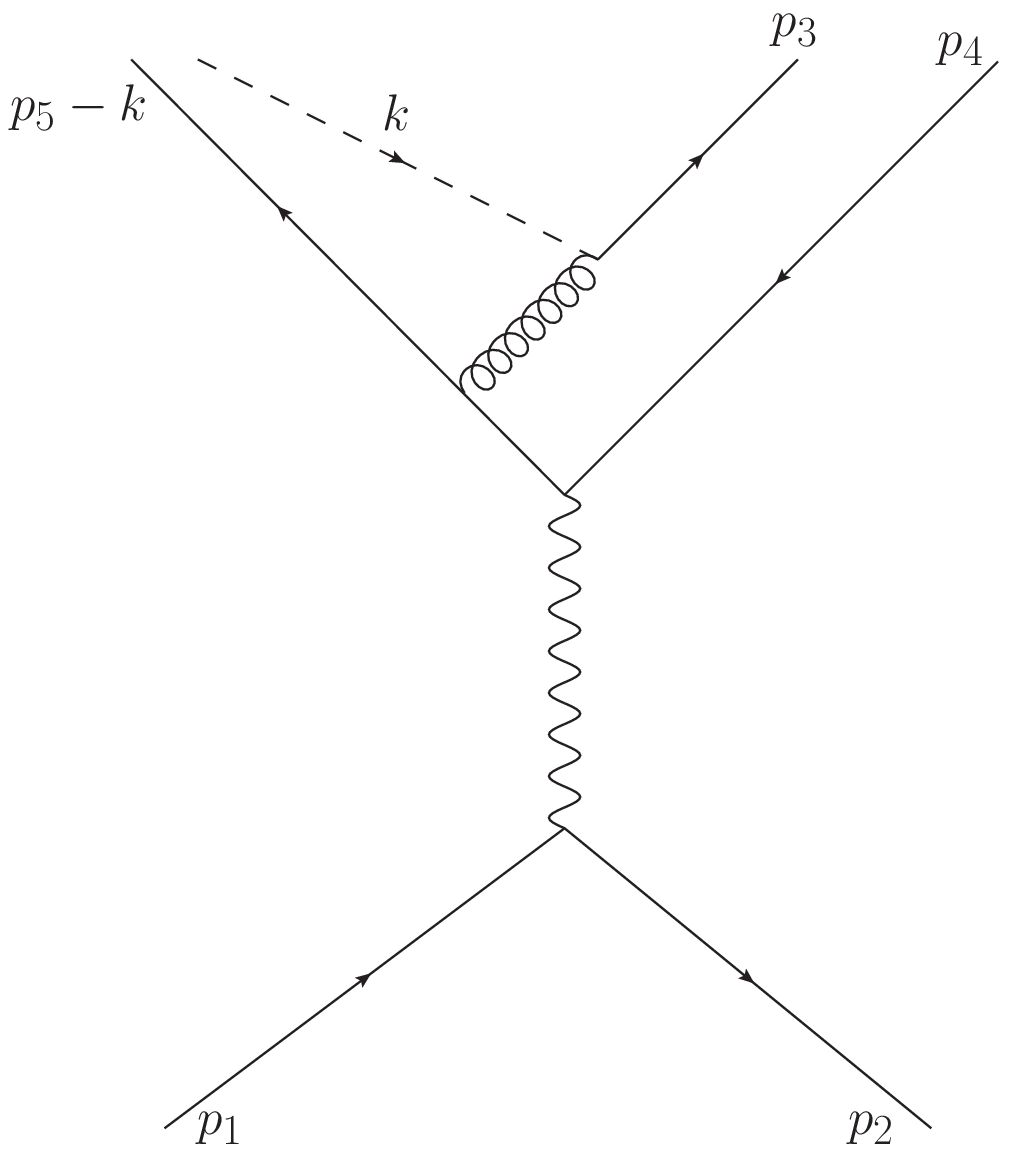}
    \includegraphics[width=0.23\linewidth]{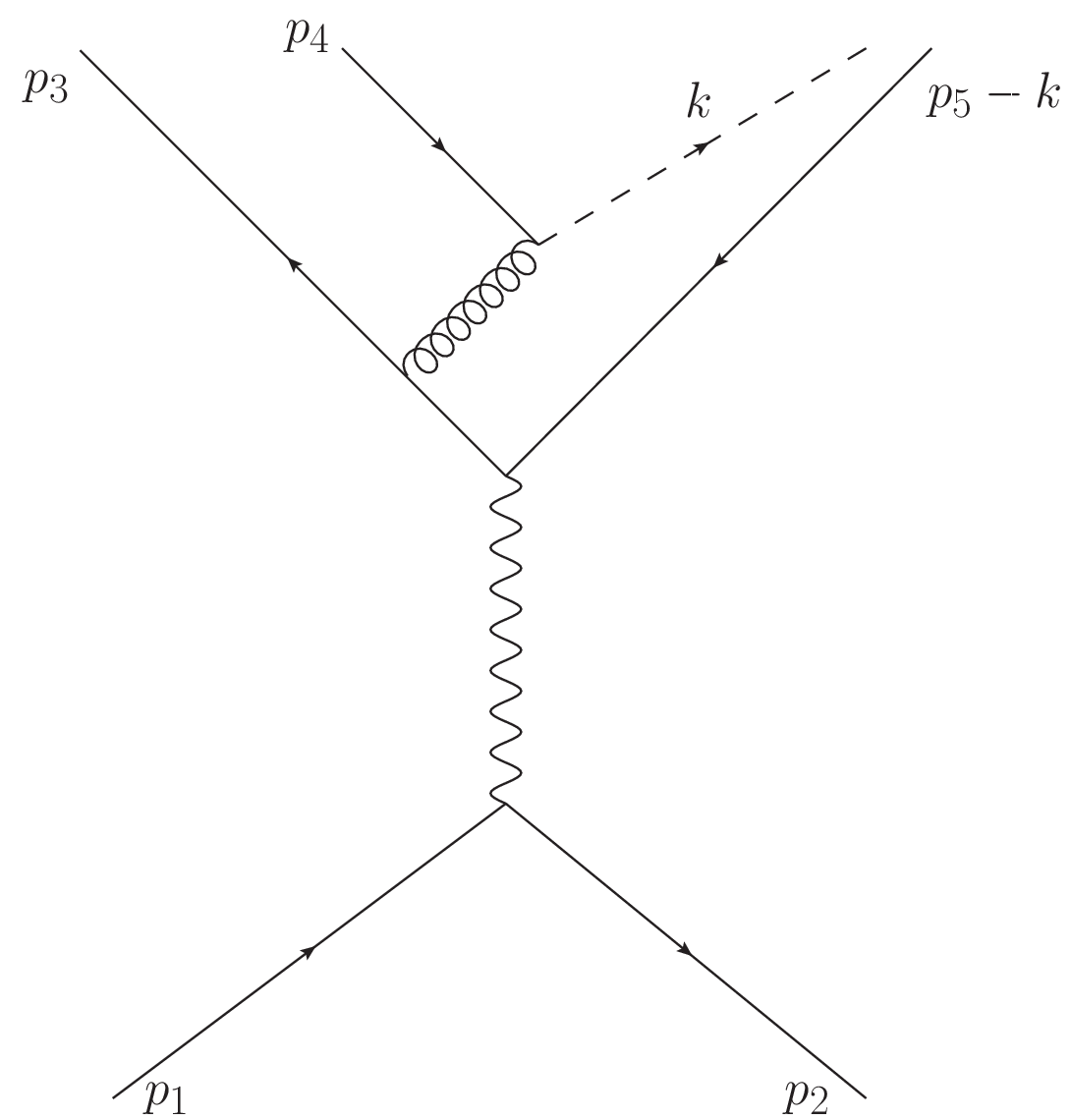}\\
    \hspace{0.02\linewidth} $T_{1e}^{coh}$ \hspace{0.2\linewidth} $T_{1f}^{coh}$\hspace{0.2\linewidth}$T_{1g}^{coh}$\\
    \caption{Diagrams involving $\Omega_{\Delta(V_1)}$ and $V_1V_1$. The dashed lines with arrows represent soft/collinear quarks and the dashed lines without arrows represent soft/collinear gluons}
    \label{fig:T1_coh}
\end{figure}

Amplitudes for the diagrams in Fig. \ref{fig:T1_coh} result from overlap between $\mathcal{O}(g)$ term of $\Omega_{\Delta(V_1)}$ in $\ket{q\Bar{q}g:\text{coh}}$ and the final states of the processes $e^{+}e^{-}\rightarrow q\bar{q}gg$ or $e^{+}e^{-}\rightarrow q\bar{q}q\bar{q}$, where one of the partons in the final state is soft/collinear. These are given by 

\begin{multline}
    T_{1a}^{coh}=\frac{-e^2g^3}{(2\pi)^{15/2}\prod_i\sqrt{2p_i^+}}\int [dk]\Theta_\Delta\Biggl[\frac{\Bar{u}_{p_3}\slashed{\epsilon}^{a}_{k}T^a{u}_{p_3-k}\Bar{u}_{p_3-k}\slashed{\epsilon}^{\ast b}_{p_5}T^b{u}_{p_3+p_5-k}}{(2(p_3-k)^+)(2(p_3+p_5)^+)(2(p_3+p_5-k)^+)(2k^+)(p_3^--(p_3-k)^--k^-)}\\ \times \frac{\Bar{u}_{p_3+p_5-k}\slashed{\epsilon}^{\ast c}_{k}T^c{u}_{p_3+p_5}\Bar{u}_{p_3+p_5}\slashed{\epsilon}_{p_1+p_2}v_{p_4}\Bar{v}_{p_2}\slashed{\epsilon}^{\ast}_{p_1+p_2}u_{p_1}}{(2(p_1+p_2)^+)(p_3^-+p_5^--(p_3^-+p_5)^-)(p_1^-+p_2^--(p_1+p_2)^-)(p_3^-+p_5^--k^--(p_3+p_5-k)^-)}\Biggr]
\end{multline}

where $\Theta_\Delta=\theta(\Delta-k^+)\theta(2k^+\Delta-k_\perp^2)$ limits the phase space to the region where $(p_3^--(p_3-k)^--k^-)<\Delta$

Similarly, in 
\begin{multline}
    T_{1b}^{coh}=\frac{e^2g^3}{(2\pi)^{15/2}\prod_i\sqrt{2p_i^+}}\int [dk]\Theta_\Delta\Biggl[\frac{\Bar{v}_{p_4-k}\slashed{\epsilon}^{a}_{k}T^a{v}_{p_4}\Bar{u}_{p_3}\slashed{\epsilon}^{\ast b}_{p_5}T^b{u}_{p_3+p_5}}{(2(p_4-k)^+)(2(p_3+p_5)^+)(2(p_3+p_5+k)^+)(2k^+)(2(p_1+p_2)^+)(p_4^--(p_4-k)^--k^-)}\\ \times \frac{\Bar{u}_{p_3+p_5}\slashed{\epsilon}^{\ast c}_{k}T^c{u}_{p_3+p_5+k}\Bar{u}_{p_3+p_5+k}\slashed{\epsilon}_{p_1+p_2}v_{p_4-k}\Bar{v}_{p_2}\slashed{\epsilon}^{\ast}_{p_1+p_2}u_{p_1}}{(p_1^-+p_2^--(p_3+p_5+k)^--(p_4^--k)^-)(p_1^-+p_2^--(p_1+p_2)^-)(p_1^-+p_2^--(p_3+p_5)^--(p_4^--k)^--k^-)}\Biggr]
\end{multline}

where $\Theta_\Delta=\theta(\Delta-k^+)\theta(2k^+\Delta-k_\perp^2)$ limits the phase space to the region where $(p_4^--(p_4-k)^--k^-)<\Delta$

A similar result holds for all the following amplitudes. $\Theta_\Delta$ restricts the phase space integration to the region where the energy denominator approaches 0, causing the IR divergence.

\begin{multline}
    T_{1c}^{coh}=\frac{e^2g^3}{(2\pi)^{15/2}\prod_i\sqrt{2p_i^+}}\int [dk]\Theta_\Delta \Biggl[\frac{\Bar{v}_{p_4-k}\slashed{\epsilon}^{a}_{k}T^a{v}_{p_4}\Bar{u}_{p_3}\slashed{\epsilon}^{\ast b}_{k}T^b{u}_{p_3+k}}{(2(p_4-k)^+)(2(p_3+k)^+)(2(p_3+p_5+k)^+)(2k^+)(2(p_1+p_2)^+)(p_4^--(p_4-k)^--k^-)}\\ \times \frac{\Bar{u}_{p_3+k}\slashed{\epsilon}^{\ast c}_{p_5}T^c{u}_{p_3+p_5+k}\Bar{u}_{p_3+p_5+k}\slashed{\epsilon}_{p_1+p_2}v_{p_4-k}\Bar{v}_{p_2}\slashed{\epsilon}^{\ast}_{p_1+p_2}u_{p_1}}{p_1^-+p_2^--(p_1+p_2)^-)(p_1^-+p_2^--(p_3+p_5+k)^--(p_4^--k)^-)((p_3^-+p_4^--(p_3+k)^--(p_4^--k)^-)}\Biggr]
\end{multline}

\begin{multline}
    T_{1d}^{coh}=\frac{-e^2g^3}{(2\pi)^{15/2}\prod_i\sqrt{2p_i^+}}\int [dk]\Theta_\Delta\Biggl[\frac{\Bar{v}_{p_5-k}\slashed{\epsilon}^{\ast a}_{p_5}T^a{u}_{k}\Bar{u}_{k}\slashed{\epsilon}^b_{p_5}T^b{v}_{p_5-k}\Bar{u}_{p_3}\slashed{\epsilon}^{\ast c}_{p_5}T^c{u}_{p_3+p_5}}{(2(p_3+p_5)^+)(2(p_5-k)^+)(2(p_1+p_2)^+)(2p_5^+)(p_5^--(p_5-k)^--k^-)}\\ \times \frac{\Bar{u}_{p_3+p_5}\slashed{\epsilon}_{p_1+p_2}v_{p_4}\Bar{v}_{p_2}\slashed{\epsilon}^\ast_{p_1+p_2}u_{p_1}}{(2k^+)(p_1^-+p_2^--(p_1+p_2)^-)(p_3^-+p_5^--(p_3+p_5)^-)(p_1^-+p_2^--p_3^--p_4^--p_5^-)}\Biggr]
\end{multline}

\begin{multline}
    T_{1e}^{coh}=\frac{-e^2g^3}{(2\pi)^{15/2}\prod_i\sqrt{2p_i^+}}\int [dk]\Theta_\Delta\Biggl[\frac{\Bar{v}_{p_5-k}\slashed{\epsilon}^{\ast a}_{p_5}T^a{u}_{k}\Bar{u}_{p_3}\slashed{\epsilon}^{b}_{p_3+p_4}T^b{v}_{p_4}\Bar{u}_{k}\slashed{\epsilon}^{\ast c}_{p_3+p_4}T^c{u}_{p_3+p_4+k}}{(2(p_3+p_4+k)^+)(2(p_5-k)^+)(2(p_1+p_2)^+)(2(p_3+p_4)^+)(p_5^--(p_5-k)^--k^-)}\\ \times \frac{\Bar{u}_{p_3+p_4+k}\slashed{\epsilon}_{p_1+p_2}v_{p_5-k}\Bar{v}_{p_2}\slashed{\epsilon}^{\ast}_{p_1+p_2}u_{p_1}}{(2k^+)(p_1^-+p_2^--(p_1+p_2)^-)(p_1^-+p_1^--(p_3+p_4)^--(p_5-k)^--k^-)(p_1^-+p_2^--(p_3+p_4+k)^--(p_5-k)^-)}\Biggr]
\end{multline}

\begin{multline}
    T_{1f}^{coh}=\frac{e^2g^3}{(2\pi)^{15/2}\prod_i\sqrt{2p_i^+}}\int [dk]\Theta_\Delta\Biggl[\frac{\Bar{v}_{p_5-k}\slashed{\epsilon}^{\ast a}_{p_5}T^a{u}_{k}\Bar{u}_{p_3}\slashed{\epsilon}^{b}_{p_3+p_5-k}T^b{v}_{p_5-k}\Bar{u}_{k}\slashed{\epsilon}^{\ast c}_{p_3+p_5-k}T^c{u}_{p_3+p_5}}{(2(p_3+p_5)^+)(2(p_5-k)^+)(2(p_1+p_2)^+)(2(p_3+p_5-k)^+)(p_5^--(p_5-k)^--k^-)}\\ \times \frac{\Bar{u}_{p_3+p_5}\slashed{\epsilon}_{p_1+p_2}v_{p_4}\Bar{v}_{p_2}\slashed{\epsilon}^{\ast}_{p_1+p_2}u_{p_1}}{(2k^+)(p_1^-+p_2^--(p_1+p_2)^-)(p_3^-+p_5^--(p_3+p_5)^-)(p_3^-+p_5^--(p_3+p_5-k)^-k^-)}\Biggr]
\end{multline}

\begin{multline}
    T_{1g}^{coh}=\frac{e^2g^3}{(2\pi)^{15/2}\prod_i\sqrt{2p_i^+}}\int [dk]\Theta_\Delta\Biggl[\frac{\Bar{v}_{p_5-k}\slashed{\epsilon}^{\ast a}_{p_5}T^a{u}_{k}\Bar{u}_{k}\slashed{\epsilon}^{b}_{p_4+k}T^b{v}_{p_4}\Bar{u}_{p_3}\slashed{\epsilon}^{\ast c}_{p_4+k}T^c{u}_{p_3+p_4+k}}{(2(p_4+k)^+)(2(p_5-k)^+)(2(p_1+p_2)^+)(2(p_3+p_4+k)^+)(p_5^--(p_5-k)^--k^-)}\\ \times \frac{\Bar{u}_{p_3+p_4+k}\slashed{\epsilon}_{p_1+p_2}v_{p_5-k}\Bar{v}_{p_2}\slashed{\epsilon}^{\ast}_{p_1+p_2}u_{p_1}}{(2k^+)(p_1^-+p_2^--(p_1+p_2)^-)(p_4^-+p_5^--(p_4+k)^--(p_5-k)^-)(p_1^-+p_2^--(p_5-k)^-(p_3+p_4+k)^-)}\Biggr]
\end{multline}

\subsection{Amplitudes involving $\Omega_{\Delta(V_1)}V_2V_1$}

\begin{figure}[h]
    \centering
    \includegraphics[width=0.23\linewidth]{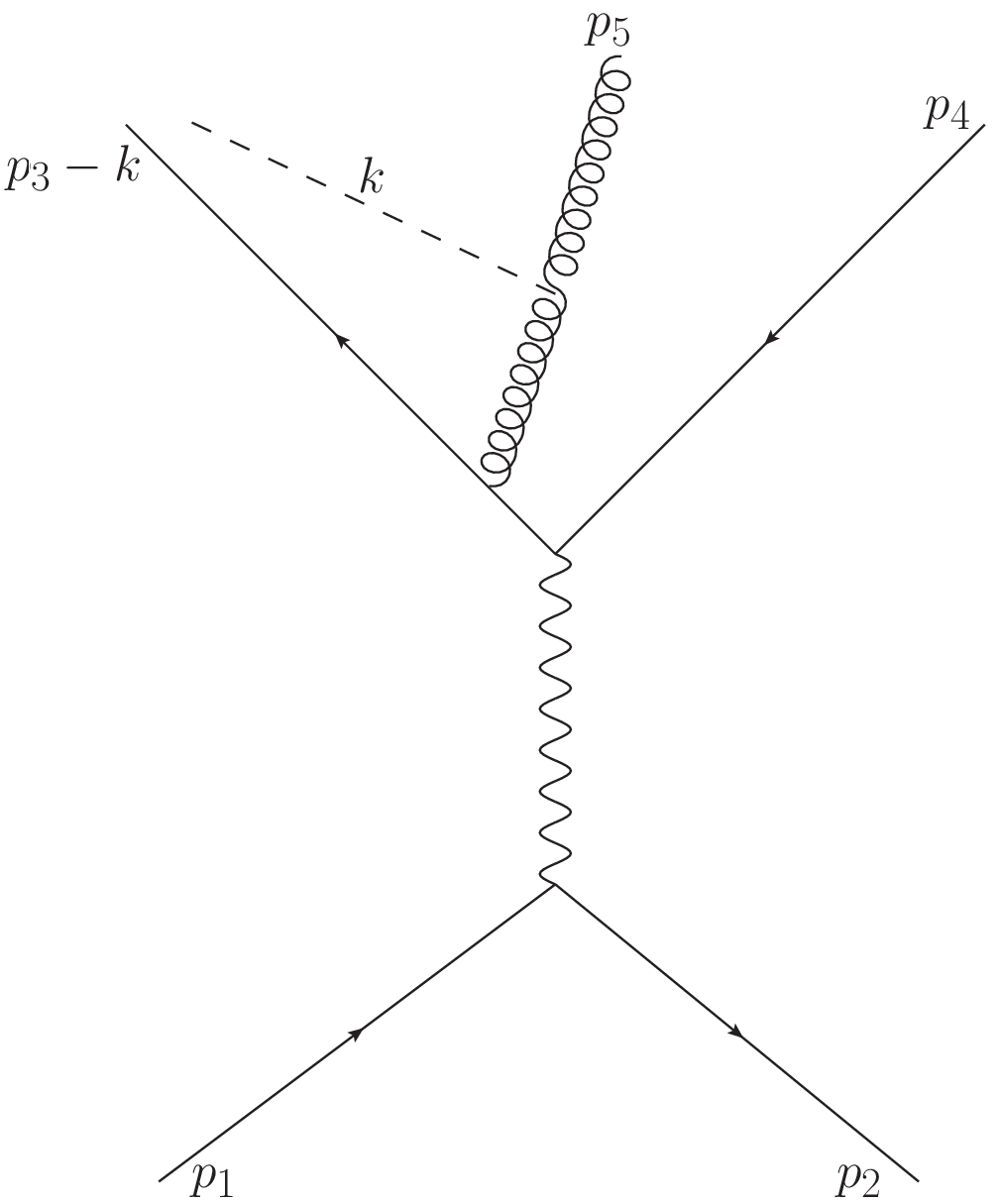} \includegraphics[width=0.255\linewidth]{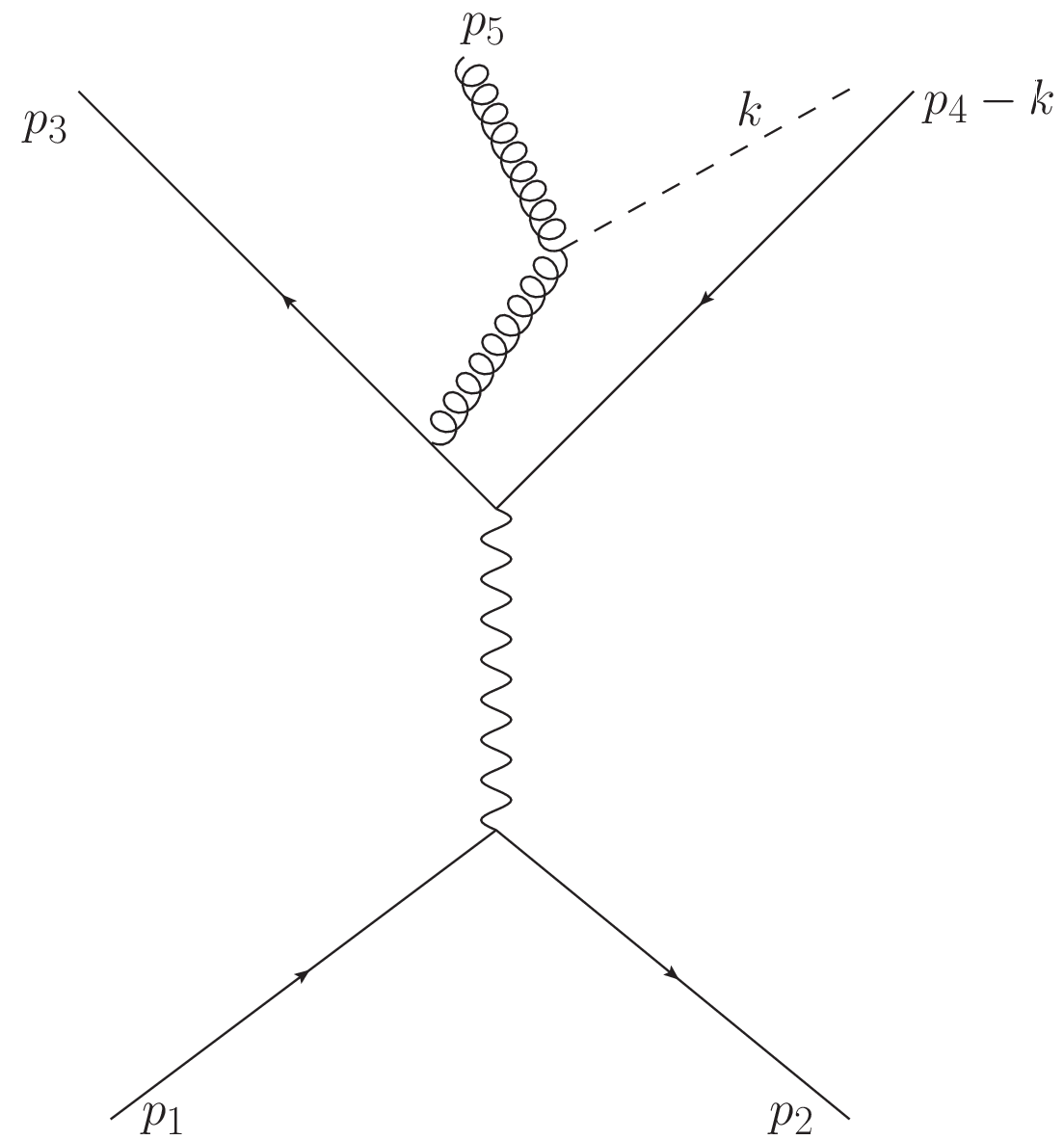}\\
    \hspace{0.02\linewidth} $T_{2a}^{coh}$ \hspace{0.21\linewidth} $T_{2b}^{coh}$\\
    \caption{Diagrams involving $\Omega_{\Delta(V_1)}$ and $V_2V_1$. The dashed lines represent soft/collinear gluons}
    \label{fig:T2_coh}
\end{figure}

Amplitudes for the diagrams in Fig. \ref{fig:T2_coh} result from overlap between $\mathcal{O}(g)$ term of $\Omega_{\Delta(V_1)}$ in $\ket{q\Bar{q}g:\text{coh}}$ and the final states of the process $e^{+}e^{-}\rightarrow q\bar{q}gg$, where one of the partons in the final state is soft/collinear. These are given by

\begin{multline}
    T_{2a}^{coh}=\frac{-e^2g^3}{(2\pi)^{15/2}\prod_i\sqrt{2p_i^+}}\int [dk]\Theta_\Delta\Biggl[\frac{\Bar{u}_{p_3}\slashed{\epsilon}^{e}_{k}T^e{u}_{p_3-k}f^{abd}(ik^\mu \epsilon^{\ast \nu}_{{k}b}(\epsilon^{\ast d}_{\nu p_5}\epsilon^{a}_{\mu p_5+k}+\epsilon^{d}_{\nu p_5+k}\epsilon^{\ast a}_{\mu p_5})+ip_5^\mu \epsilon^{\ast \nu}_{{p_5}b}(\epsilon^{\ast d}_{\nu k}\epsilon^{a}_{\mu p_5+k}}{(2k^+)(2(p_3-k)^+)(2(p_3+p_5)^+)(2(p_1+p_2)^+)(p_1^-+p_2^--(p_1+p_2)^-)}\\  \frac{+\epsilon^{d}_{\nu p_5+k}\epsilon^{\ast a}_{\mu k})-i(p_5+k)^\mu \epsilon^\nu_{{p_5+k}b}(\epsilon^{\ast d}_{\nu p_5}\epsilon^{\ast a}_{\mu k}+\epsilon^{\ast d}_{\nu k}\epsilon^{\ast a}_{\mu p_5}))  \Bar{u}_{p_3-k}\slashed{\epsilon}^{\ast c}_{p_5+k}T^c{u}_{p_3+p_5}\Bar{u}_{p_3+p_5} \slashed{\epsilon}_{p_1+p_2}v_{p_4}\Bar{v}_{p_2}\slashed{\epsilon}^{\ast}_{p_1+p_2}u_{p_1}}{(p_3^-+p_5^--(p_3+p_5)^-)(p_3^-+p_5^--(p_3-k)^--(p_5+k)^-)(p_3^--(p_3-k)^--k^-)}\Biggr]
\end{multline}

\begin{multline}
    T_{2b}^{coh}=\frac{-e^2g^3}{(2\pi)^{15/2}\prod_i\sqrt{2p_i^+}}\int [dk]\Theta_\Delta\Biggl[\frac{\Bar{v}_{p_4-k}\slashed{\epsilon}^{e}_{k}T^e{v}_{p_4}f^{abd}(ik^\mu \epsilon^{\ast \nu}_{{k}b}(\epsilon^{\ast d}_{\nu p_5}\epsilon^{a}_{\mu p_5+k}+\epsilon^{d}_{\nu p_5+k}\epsilon^{\ast a}_{\mu p_5})+ip_5^\mu \epsilon^{\ast \nu}_{{p_5}b}(\epsilon^{\ast d}_{\nu k}\epsilon^{a}_{\mu p_5+k}}{(2k^+)(2(p_3+p_5+k)^+)(2(p_4-k)^+)(2(p_1+p_2)^+)(p_1^-+p_2^--(p_1+p_2)^-)}\\  \frac{+\epsilon^{d}_{\nu p_5+k}\epsilon^{\ast a}_{\mu k})-i(p_5+k)^\mu \epsilon^{\nu}_{{p_5+k}b}(\epsilon^{\ast d}_{\nu p_5}\epsilon^{\ast a}_{\mu k}+\epsilon^{\ast d}_{\nu k}\epsilon^{\ast a}_{\mu p_5}))  \Bar{u}_{p_3}\slashed{\epsilon}^{\ast c}_{p_5+k}T^c{u}_{p_3+p_5+k}\Bar{u}_{p_3+p_5+k} \slashed{\epsilon}_{p_1+p_2}v_{p_4-k}\Bar{v}_{p_2}\slashed{\epsilon}^{\ast}_{p_1+p_2}u_{p_1}}{(p_1^-+p_2^--(p_3+p_5+k)^--(p_4-k)^-)(p_4^-+p_5^--(p_4-k)^--(p_5+k)^-)(p_4^--(p_4-k)^--k^-)}\Biggr]
\end{multline}

\subsection{Amplitudes involving $\Omega_{\Delta(V_2)}V_1V_1$}

\begin{figure}[h]
    \centering
    \includegraphics[width=0.23\linewidth]{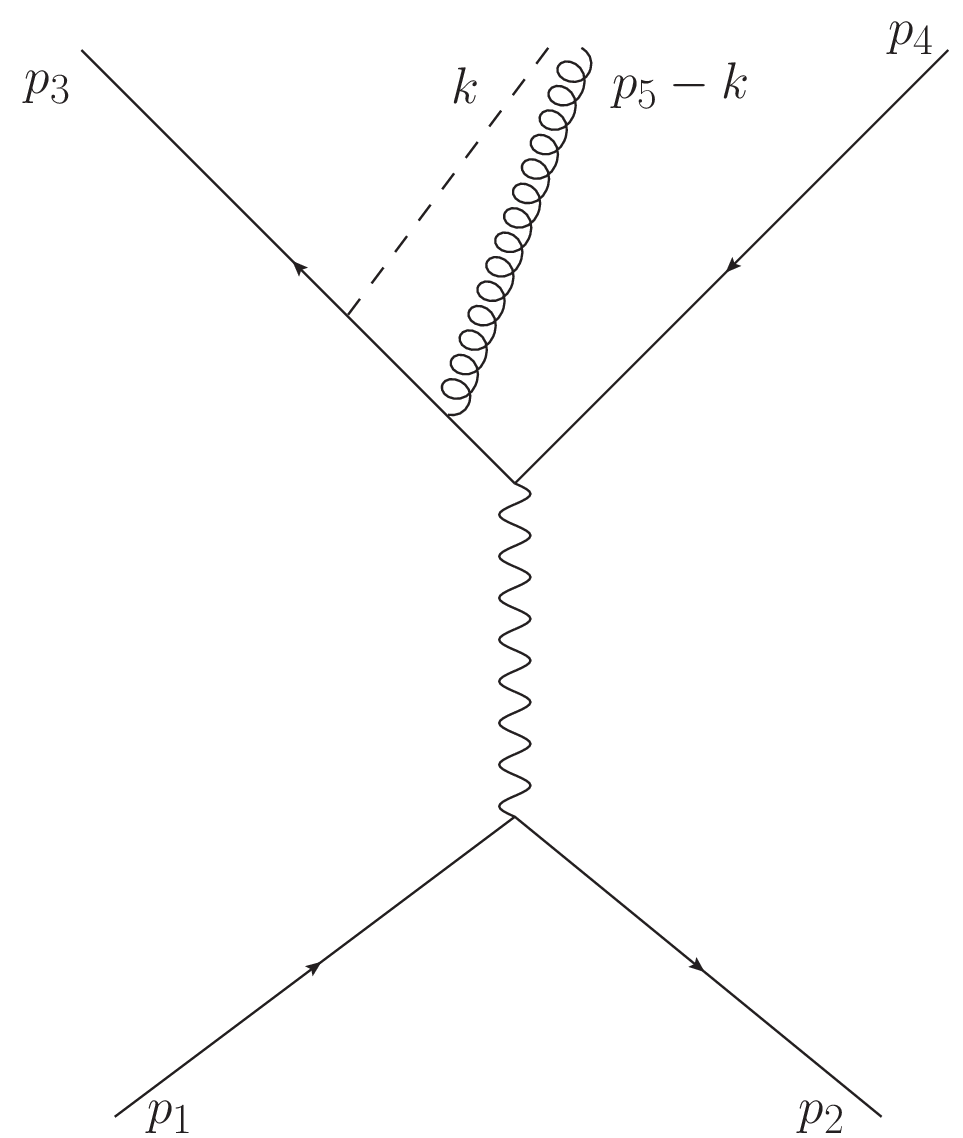} \includegraphics[width=0.23\linewidth]{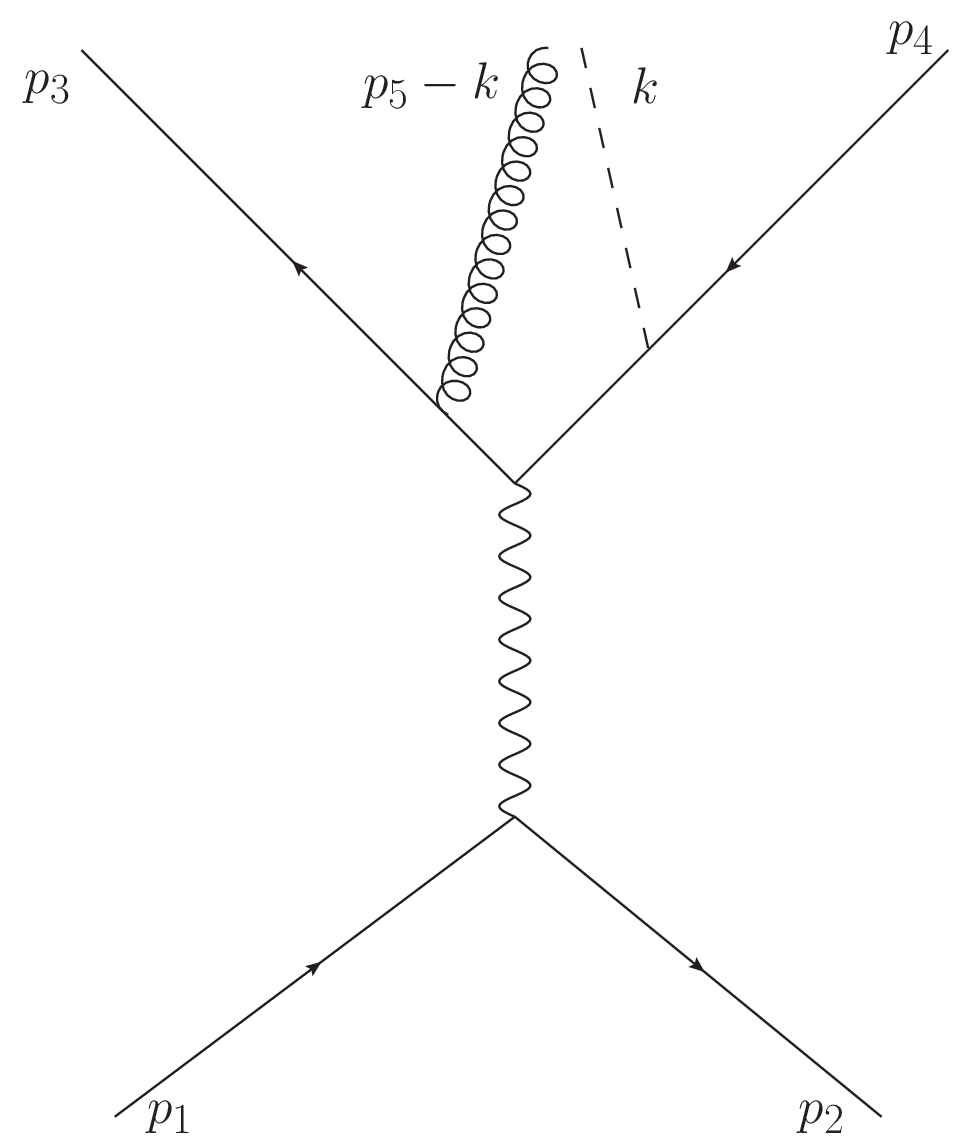}\\
    \hspace{0.02\linewidth} $T_{3a}^{coh}$ \hspace{0.18\linewidth} $T_{3b}^{coh}$\\
    \caption{Diagrams involving $\Omega_{\Delta(V_2)}$ and $V_1V_1$. The dashed lines represent soft/collinear gluons}
    \label{fig:T6_coh}
\end{figure}

Amplitudes for the processes in Fig. \ref{fig:T6_coh} result from overlap between $\mathcal{O}(g)$ term of $\Omega_{\Delta(V_2)}$ in $\ket{q\Bar{q}g:\text{coh}}$ and the final states of the process $e^{+}e^{-}\rightarrow q\bar{q}gg$, where one of the partons in the final state is soft/collinear.
These amplitudes are given by
\begin{multline}
    T_{3a}^{coh}=\frac{-2e^2g^3}{(2\pi)^{15/2}\prod_i\sqrt{2p_i^+}}\int [dk]\Theta_\Delta\Biggl[\frac{f^{abd}(-ik^\mu \epsilon^{\nu}_{b{k}}(\epsilon^{\ast d}_{\nu p_5}\epsilon^{a}_{\mu p_5-k}+\epsilon^{d}_{\nu p_5-k}\epsilon^{\ast a}_{\mu p_5})+ip_5^\mu \epsilon^{\ast \nu}_{b{p_5}}(\epsilon^{d}_{\nu k}\epsilon^{a}_{\mu p_5-k}+\epsilon^{d}_{\nu p_5-k}\epsilon^{a}_{\mu k})}{(2k^+)(2(p_3+p_5-k)^+)(2(p_3+p_5)^+)(2(p_1+p_2)^+)(p_1^-+p_2^--(p_1+p_2)^-)}\\  \frac{-i(p_5-k)^\mu \epsilon^{\nu}_{b{p_5-k}}(\epsilon^{\ast d}_{\nu p_5}\epsilon^{a}_{\mu k}+\epsilon^{d}_{\nu k}\epsilon^{\ast a}_{\mu p_5}))\Bar{u}_{p_3}\slashed{\epsilon}^{\ast e}_{p_5-k}T^e{u}_{p_3+p_5-k}\Bar{u}_{p_3+p_5-k}\slashed{\epsilon}^{\ast c}_{k}T^c{u}_{p_3+p_5}\Bar{u}_{p_3+p_5} \slashed{\epsilon}_{p_1+p_2}v_{p_4}\Bar{v}_{p_2}\slashed{\epsilon}^{\ast}_{p_1+p_2}u_{p_1}}{(p_3^-+p_5^--(p_3+p_5)^-)(p_3^-+p_5^--(p_3+p_5-k)^--k^-)(p_5^--(p_5-k)^--k^-)}\Biggr]
\end{multline}

\begin{multline}
    T_{3b}^{coh}=\frac{-2e^2g^3}{(2\pi)^{15/2}\prod_i\sqrt{2p_i^+}}\int [dk]\Theta_\Delta\Biggl[\frac{f^{abd}(-ik^\mu \epsilon^{\nu}_{b{k}}(\epsilon^{\ast d}_{\nu p_5}\epsilon^{a}_{\mu p_5-k}+\epsilon^{d}_{\nu p_5-k}\epsilon^{\ast a}_{\mu p_5})+ip_5^\mu \epsilon^{\ast \nu}_{b{p_5}}(\epsilon^{d}_{\nu k}\epsilon^{a}_{\mu p_5-k}+\epsilon^{d}_{\nu p_5-k}\epsilon^{a}_{\mu k})}{(2k^+)(2(p_3+p_5-k)^+)(2(p_4+k)^+)(2(p_1+p_2)^+)(p_1^-+p_2^--(p_1+p_2)^-)}\\  \frac{-i(p_5-k)^\mu \epsilon^{\nu}_{b{p_5-k}}(\epsilon^{\ast d}_{\nu p_5}\epsilon^{a}_{\mu k}+\epsilon^{d}_{\nu k}\epsilon^{\ast a}_{\mu p_5}))\Bar{v}_{p_4+k}\slashed{\epsilon}^{\ast e}_{k}T^e{v}_{p_4}\Bar{u}_{p_3}\slashed{\epsilon}^{\ast c}_{p_5-k}T^c{u}_{p_3+p_5-k}\Bar{u}_{p_3+p_5-k}\slashed{\epsilon}_{p_1+p_2}v_{p_4+k}\Bar{v}_{p_2}\slashed{\epsilon}^{\ast}_{p_1+p_2}u_{p_1}}{(p_1^-+p_2^--(p_4+k)^--(p_3+p_5-k)^-)(p_4^-+p_5^--(p_4+k)^--(p_5-k)^-)(p_5^--(p_5-k)^--k^-)}\Biggr]
\end{multline}

\subsection{Amplitudes involving $\Omega_{\Delta(V_1)}W_1$}

\begin{figure}[h]
    \centering
    \includegraphics[width=0.22\linewidth]{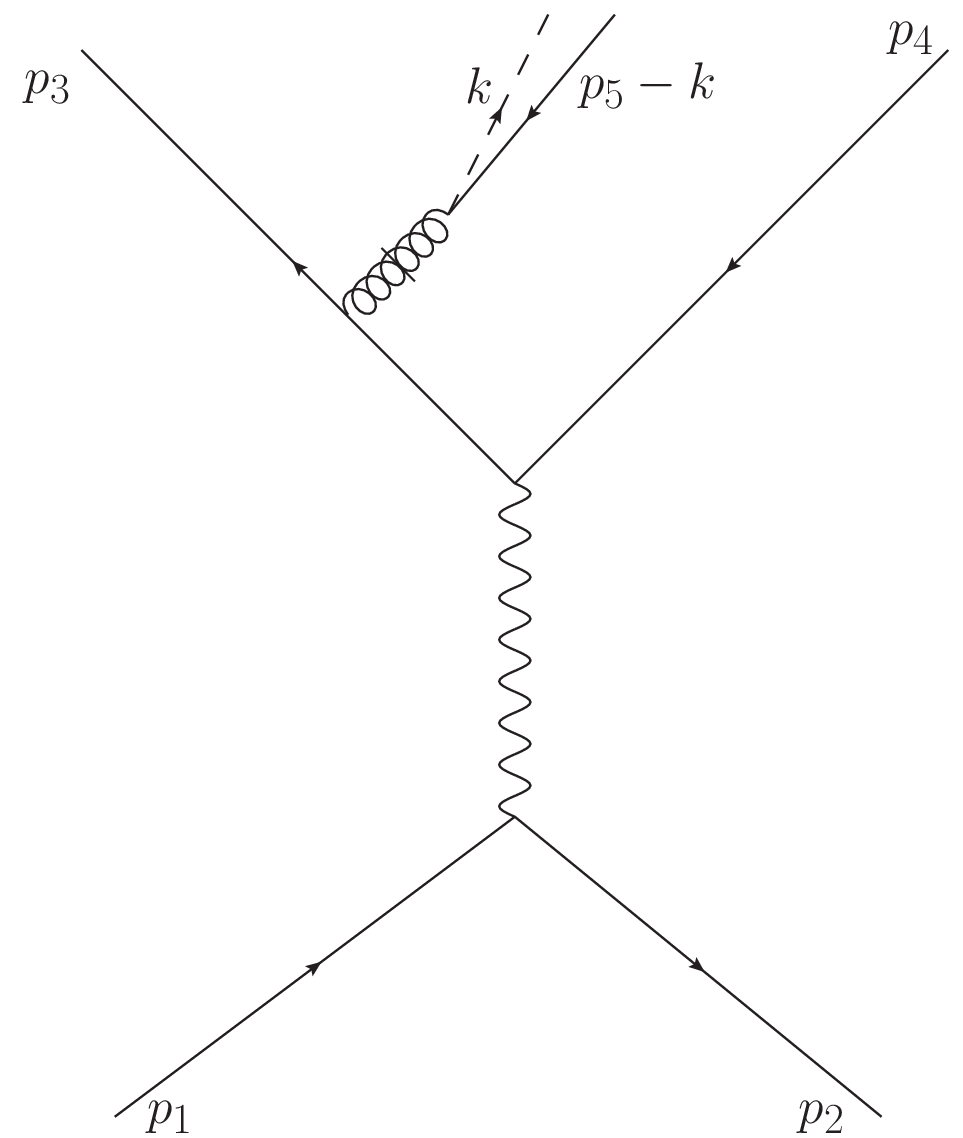} \includegraphics[width=0.255\linewidth]{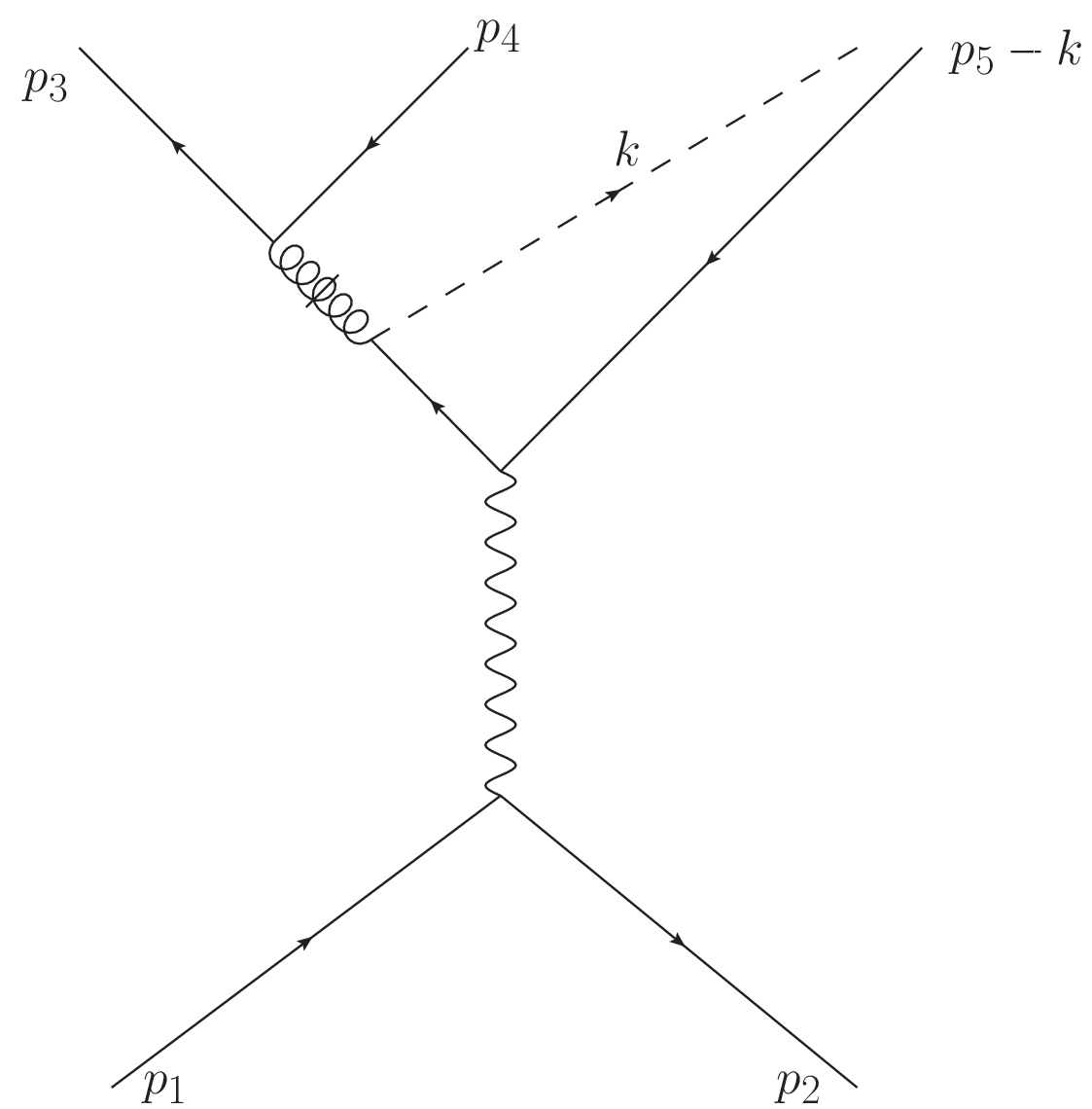} \includegraphics[width=0.23\linewidth]{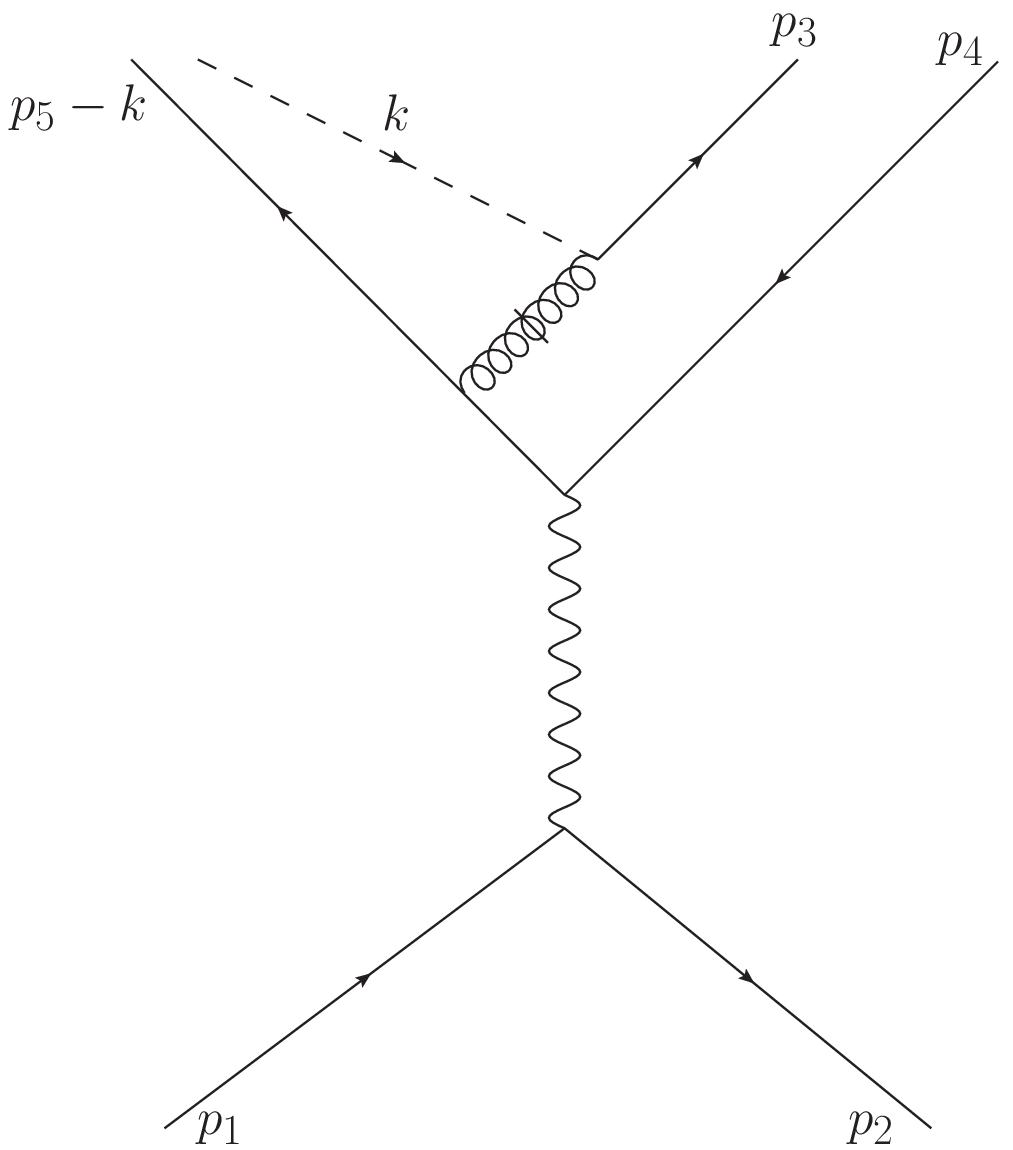} \includegraphics[width=0.255\linewidth]{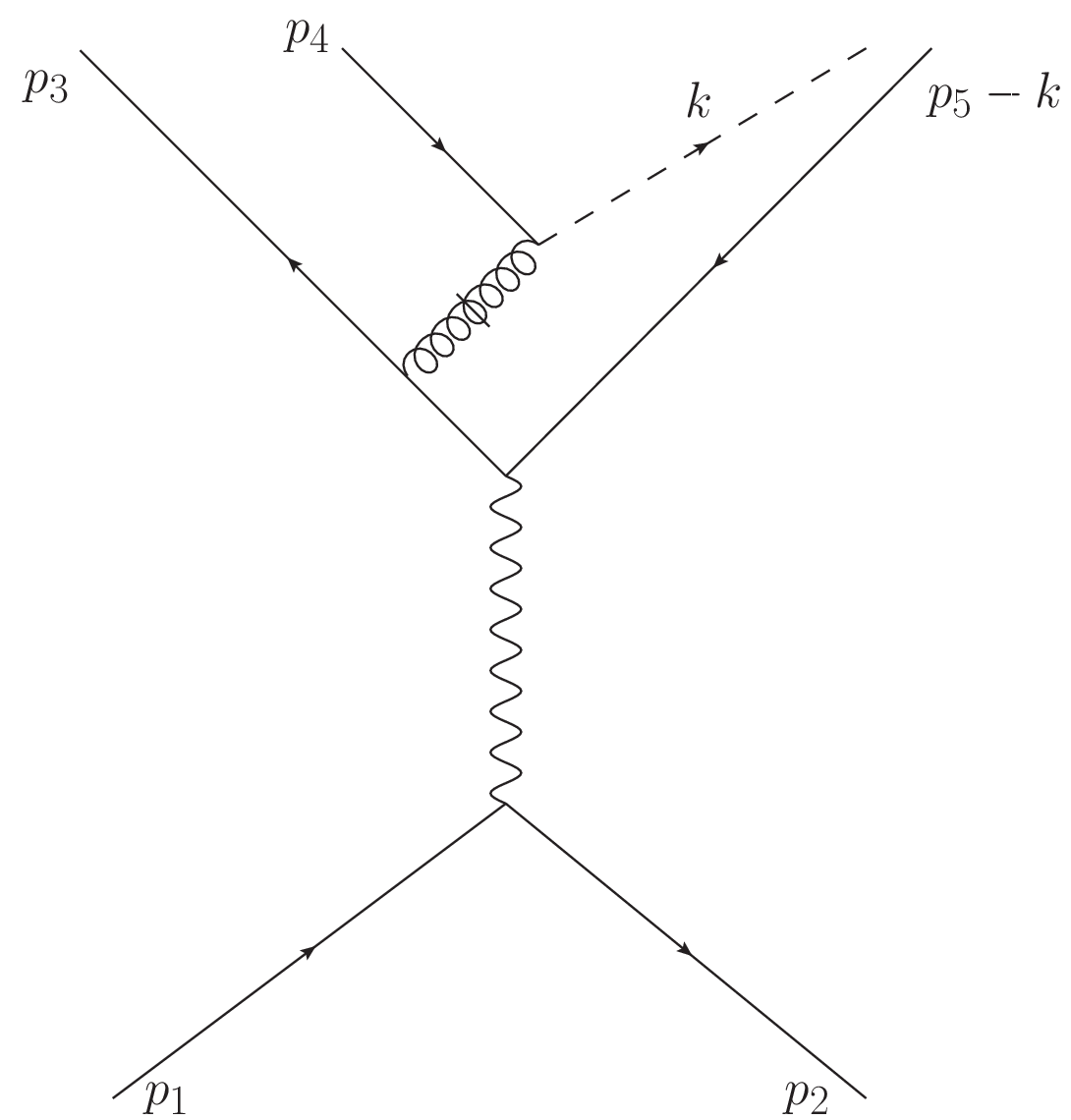}\\
    \hspace{0\linewidth} $T_{4a}^{coh}$ \hspace{0.17\linewidth} $T_{4b}^{coh}$\hspace{0.19\linewidth}$T_{4c}^{coh}$\hspace{0.18\linewidth}$T_{4d}^{coh}$\\
    \caption{Diagrams involving $\Omega_{\Delta(V_1)}$ and $W_1$. The dashed lines represent soft/collinear quarks}
    \label{fig:T3_coh}
\end{figure}

Amplitudes for the diagrams in Fig. \ref{fig:T3_coh} result from overlap between $\mathcal{O}(g)$ term of $\Omega_{\Delta(V_1)}$ in $\ket{q\Bar{q}g:\text{coh}}$ and the final states of the process $e^{+}e^{-}\rightarrow q\bar{q}q\bar{q}$, where one of the partons in the final state is soft/collinear. These are given by

\begin{multline}
    T_{4a}^{coh}=\frac{2e^2g^3}{(2\pi)^{21/2}\prod_i\sqrt{2p_i^+}}\int [dk]\Theta_\Delta\frac{\Bar{v}_{p_5-k}\slashed{\epsilon }^{\ast c}_{p_5}T^c{u}_{k}\Bar{u}_{p_3}\gamma^+T^a{v}_{p_5-k}\Bar{u}_{k}\gamma^+T^a{u}_{p_3+p_5}}{(2(p_5-k)^+)(2(p_3+p_5)^+)(k^+ - (p_3+p_5)^+)^2(p_5^--(p_3-k)^--k^-)}\\  \times\frac{\Bar{u}_{p_3+p_5} \slashed{\epsilon}_{p_1+p_2}v_{p_4}\Bar{v}_{p_2}\slashed{\epsilon}^{\ast}_{p_1+p_2}u_{p_1}}{(2k^+)(2(p_1+p_2)^+)(p_1^-+p_2^--(p_1+p_2)^-)(p_3^-+p_5^--(p_3+p_5)^-)}
\end{multline}

\begin{multline}
    T_{4b}^{coh}=\frac{2e^2g^3}{(2\pi)^{21/2}\prod_i\sqrt{2p_i^+}}\int [dk]\Theta_\Delta\frac{\Bar{v}_{p_5-k}\slashed{\epsilon }^{\ast c}_{p_5}T^c{u}_{k}\Bar{u}_{k}\gamma^+T^a{v}_{p_4}\Bar{u}_{p_3}\gamma^+T^a{u}_{p_3+p_4+k}}{(2(p_5-k)^+)(2(p_3+p_4+k)^+)(p_3^+ - (p_3+p_4+k)^+)^2(p_5^--(p_5-k)^--k^-)}\\  \times\frac{\Bar{u}_{p_3+p_4+k} \slashed{\epsilon}_{p_1+p_2}v_{p_5-k}\Bar{v}_{p_2}\slashed{\epsilon}^{\ast}_{p_1+p_2}u_{p_1}}{(2k^+)(2(p_1+p_2)^+)(p_1^-+p_2^--(p_1+p_2)^-)(p_1^-+p_2^--(p_3+p_4+k)^--(p_5-k)^-)}
\end{multline}

\begin{multline}
    T_{4c}^{coh}=\frac{-2e^2g^3}{(2\pi)^{21/2}\prod_i\sqrt{2p_i^+}}\int [dk]\Theta_\Delta\frac{\Bar{v}_{p_5-k}\slashed{\epsilon }^{\ast c}_{p_5}T^c{u}_{k}\Bar{u}_{k}\gamma^+T^a{v}_{p_5-k}\Bar{u}_{p_3}\gamma^+T^a{u}_{p_3+p_5}}{(2(p_5-k)^+)(2(p_3+p_5)^+)(p_3^+ - (p_3+p_5)^+)^2(p_5^--(p_5-k)^--k^-)}\\  \times\frac{\Bar{u}_{p_3+p_5} \slashed{\epsilon}_{p_1+p_2}v_{p_4}\Bar{v}_{p_2}\slashed{\epsilon}^{\ast}_{p_1+p_2}u_{p_1}}{(2k^+)(2(p_1+p_2)^+)(p_1^-+p_2^--(p_1+p_2)^-)(p_3^-+p_5^--(p_3+p_5)^-)}
\end{multline}

\begin{multline}
    T_{4d}^{coh}=\frac{-2e^2g^3}{(2\pi)^{21/2}\prod_i\sqrt{2p_i^+}}\int [dk]\Theta_\Delta\frac{\Bar{v}_{p_5-k}\slashed{\epsilon}^{\ast c}_{p_5}T^c{u}_{k}\Bar{u}_{p_3}\gamma^+T^a{v}_{p_4}\Bar{u}_{k}\gamma^+T^a{u}_{p_3+p_4+k}}{(2(p_5-k)^+)(2(p_3+p_4+k)^+)(k^+ - (p_3+p_4+k)^+)^2(p_5^--(p_5-k)^--k^-)}\\  \times\frac{\Bar{u}_{p_3+p_4+k} \slashed{\epsilon}_{p_1+p_2}v_{p_5-k}\Bar{v}_{p_2}\slashed{\epsilon}^{\ast}_{p_1+p_2}u_{p_1}}{(2k^+)(2(p_1+p_2)^+)(p_1^-+p_2^--(p_1+p_2)^-)(p_1^-+p_2^--(p_3+p_4+k)^--(p_5-k)^-)}
\end{multline}

\subsection{Amplitudes involving $\Omega_{\Delta(V_1)}W_2$}

\begin{figure}[h]
    \centering
    \includegraphics[width=0.24\linewidth]{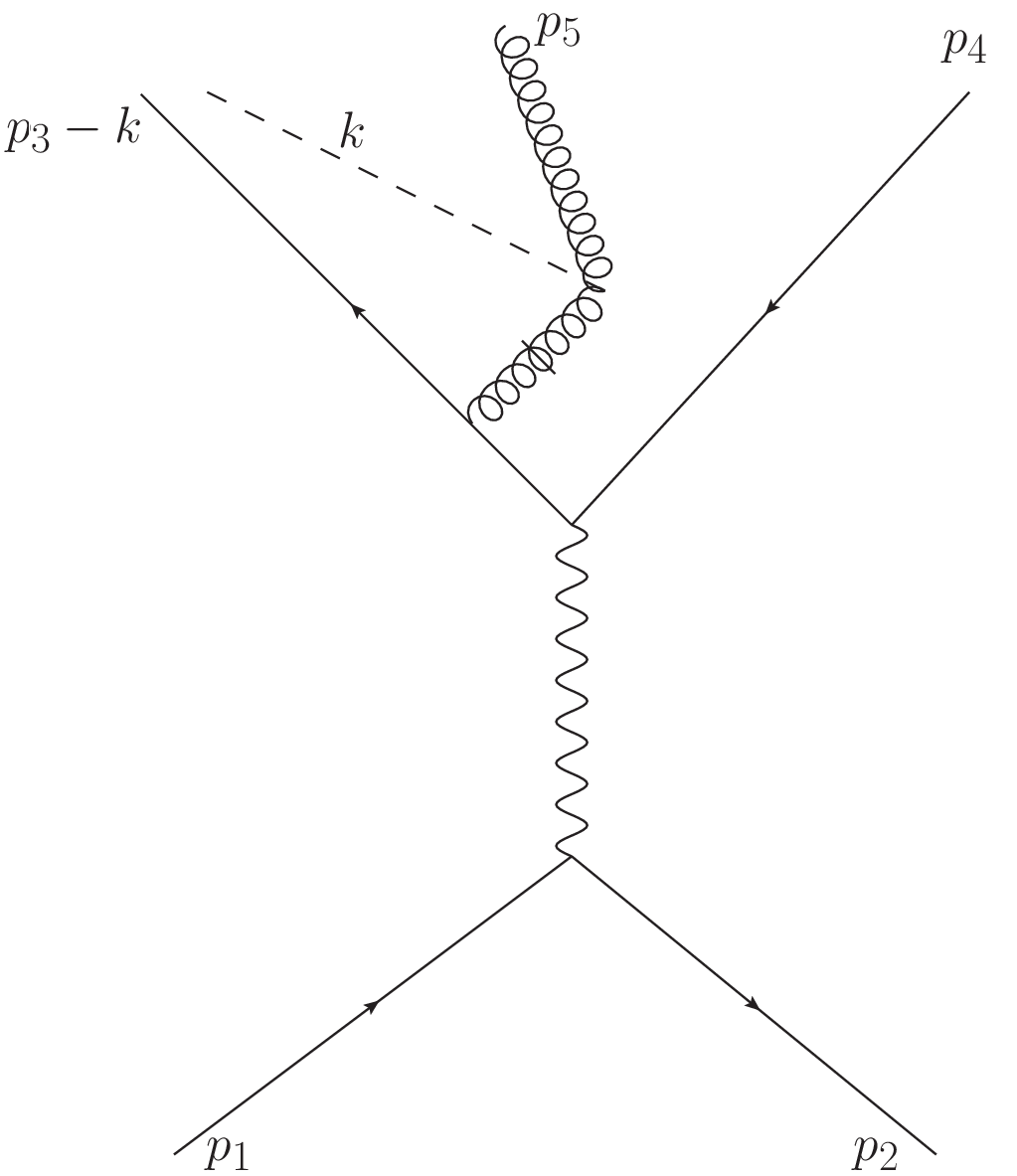} \includegraphics[width=0.255\linewidth]{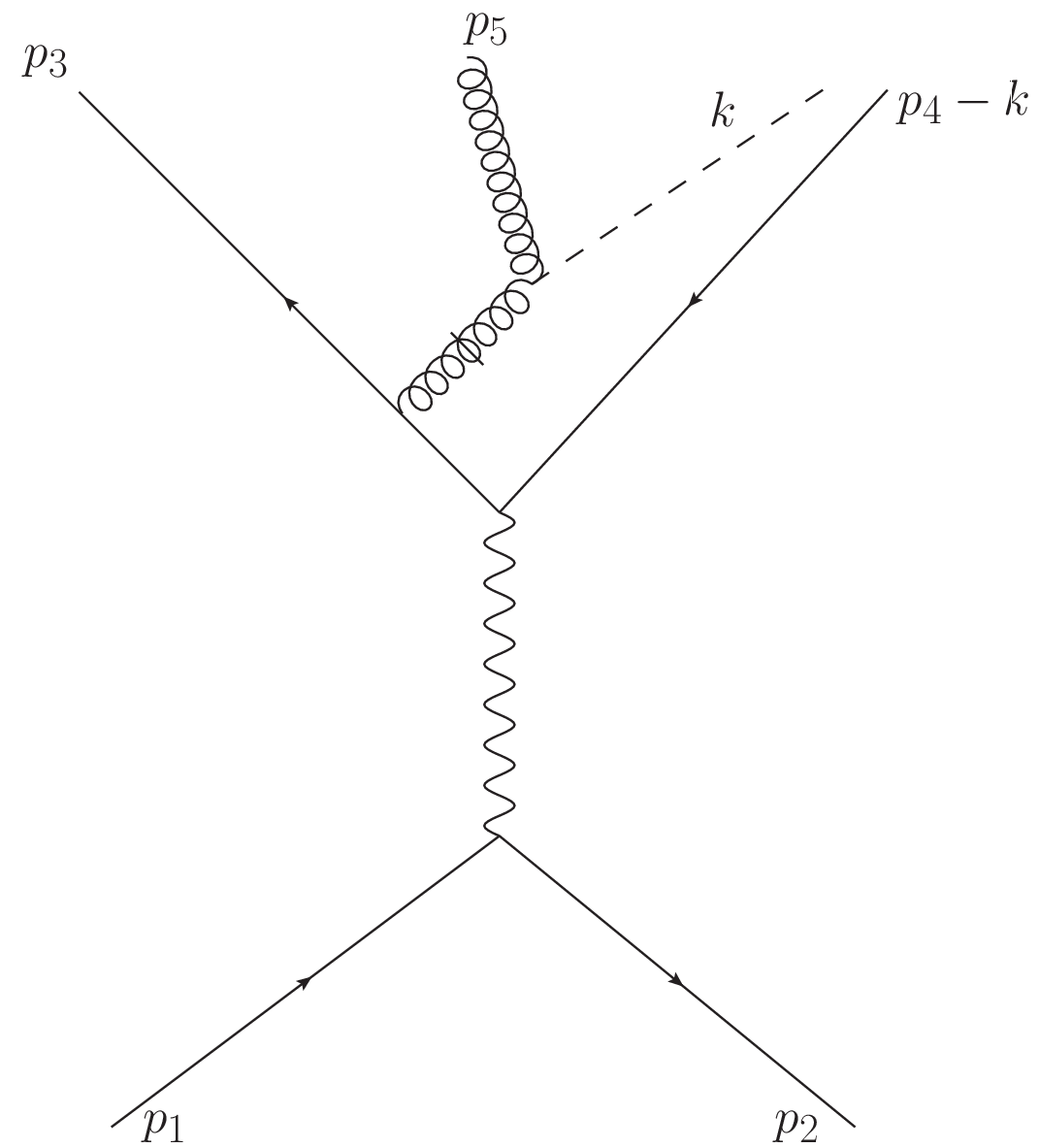}\\
    \hspace{0.02\linewidth} $T_{5a}^{coh}$ \hspace{0.21\linewidth} $T_{5b}^{coh}$\\
    \caption{Diagrams involving $\Omega_{\Delta(V_1)}$ and $W_2$. The dashed lines represent soft/collinear gluons.}
    \label{fig:T4_coh}
\end{figure}

Amplitudes for the diagrams in Fig. \ref{fig:T4_coh} result from overlap between $\mathcal{O}(g)$ term of $\Omega_{\Delta(V_1)}$ in $\ket{q\Bar{q}g:\text{coh}}$ and the final states of the process $e^{+}e^{-}\rightarrow q\bar{q}gg$, where one of the partons in the final state is soft/collinear. These are given by

\begin{multline}
    T_{5a}^{coh}=\frac{e^2g^3}{(2\pi)^{21/2}\prod_i\sqrt{2p_i^+}}\int [dk]\Theta_\Delta\frac{\Bar{u}_{p_3}\slashed{\epsilon}^{e}_{k}T^e{u}_{p_3-k}}{(2(p_1+p_2)^+)(p_1^-+p_2^--(p_1+p_2)^-)}\Biggl[\frac{\Bar{u}_{p_3-k}\gamma^+T^a_{c'c''}{u}_{p_3+p_5}f^{abd}(p_5^-\epsilon^{\ast \mu}_{p_5}\epsilon^{\ast}_{\mu k}+k^-\epsilon^{\ast \mu}_{k}\epsilon^{\ast}_{\mu p_5})}{(2(p_3-k)^+)(2(p_3+p_5)^+)(p_5^+-k^+)^2}\\  +\frac{f^{abd}(p_5^-\epsilon^{\ast \mu}_{p_5}\epsilon^{\ast}_{\mu k}+k^-\epsilon^{\ast \mu}_{k}\epsilon^{\ast}_{\mu p_5})\Bar{u}_{p_3-k}\gamma^+T^a_{c'c''}{u}_{p_3+p_5}}{(2(p_3-k)^+)(2(p_3+p_5)^+)((p_3-k)^+-(p_3+p_5)^+)^2}\Biggr]\frac{\Bar{u}_{p_3+p_5} \slashed{\epsilon}_{p_1+p_2}v_{p_4}\Bar{v}_{p_2}\slashed{\epsilon}^{\ast}_{p_1+p_2}u_{p_1}}{(2k^+)(p_3^-+p_5^--(p_3+p_5)^-)(p_3^--(p_3-k)^--k^-)}
\end{multline}

\begin{multline}
    T_{5b}^{coh}=\frac{-e^2g^3}{(2\pi)^{21/2}\prod_i\sqrt{2p_i^+}}\int [dk]\Theta_\Delta\frac{\Bar{v}_{p_4-k}\slashed{\epsilon}^{c}_{k}T^c{v}_{p_4}}{(2(p_1+p_2)^+)(p_1^-+p_2^--(p_1+p_2)^-)}\Biggl[\frac{\Bar{u}_{p_3}\gamma^+T^a_{c'c''}{u}_{p_3+p_5+k}f^{abd}(p_5^-\epsilon^{\ast \mu}_{p_5}\epsilon^{\ast}_{\mu k}+k^-\epsilon^{\ast \mu}_{k}\epsilon^{\ast}_{\mu p_5})}{(2(p_4-k)^+)(2(p_3+p_5+k)^+)(p_5^+-k^+)^2}\\  +\frac{f^{abd}(p_5^-\epsilon^{\ast \mu}_{p_5}\epsilon^{\ast}_{\mu k}+k^-\epsilon^{\ast \mu}_{k}\epsilon^{\ast}_{\mu p_5})\Bar{u}_{p_3}\gamma^+T^a_{c'c''}{u}_{p_3+p_5+k}}{(2(p_4-k)^+)(2(p_3+p_5+k)^+)(p_3^+-(p_3+p_5+k)^+)^2}\Biggr]\frac{\Bar{u}_{p_3+p_5+k} \slashed{\epsilon}_{p_1+p_2}v_{p_4-k}\Bar{v}_{p_2}\slashed{\epsilon}^{\ast}_{p_1+p_2}u_{p_1}}{(2k^+)(p_1^-+p_2^--(p_3+p_5+k)^--(p_4-k)^-)(p_4^--(p_4-k)^--k^-)}
\end{multline}

\subsection{Amplitudes involving $\Omega_{\Delta(V_2)}W_2$}

\begin{figure}[h]
    \centering
    \includegraphics[width=0.22\linewidth]{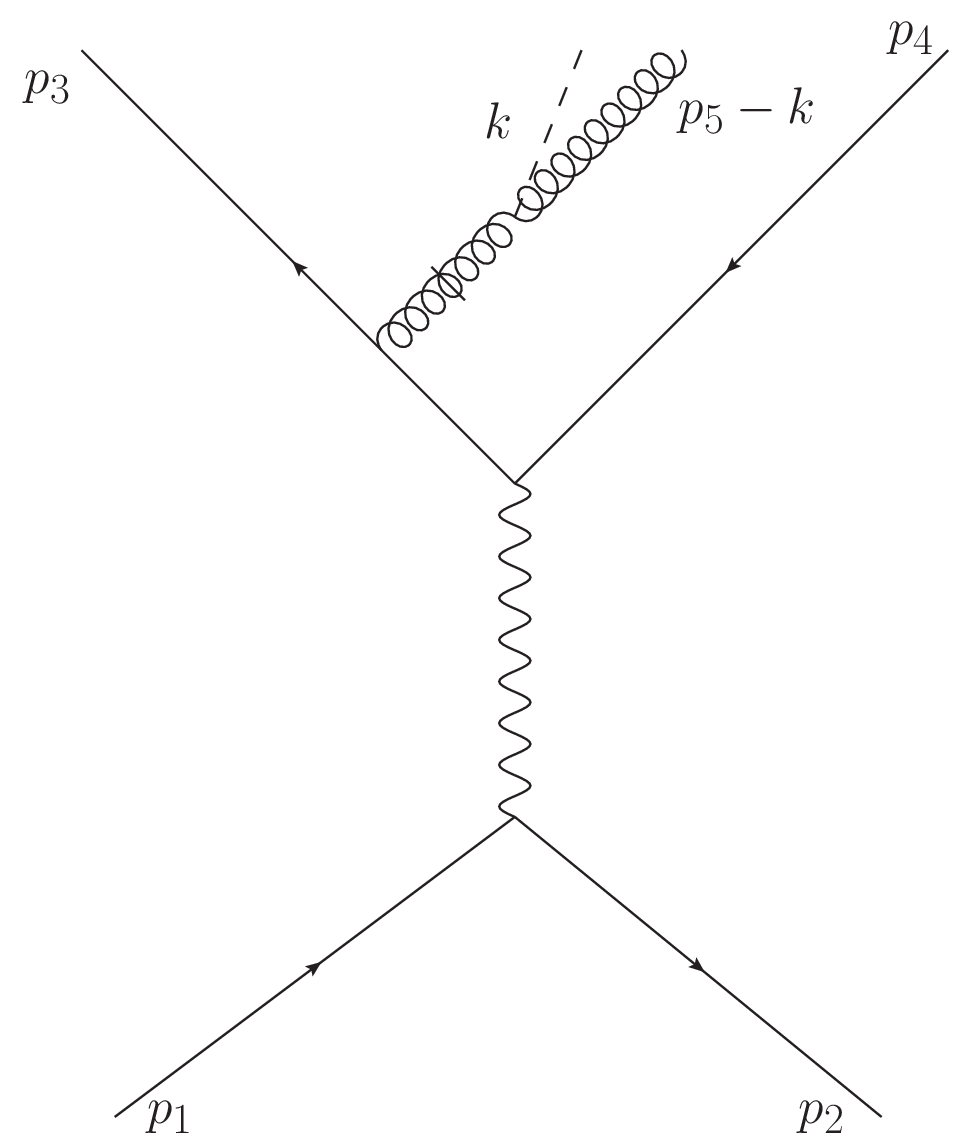}\\
    \hspace{0.02\linewidth} $T_{6a}^{coh}$\\
    \caption{Diagrams involving $\Omega_{\Delta(V_2)}$ and $W_2$. The dashed lines represent soft/collinear gluons}
    \label{fig:T7_coh}
\end{figure}

Amplitudes for the processes in Fig. \ref{fig:T7_coh} result from overlap between $\mathcal{O}(g)$ term of $\Omega_{\Delta(V_2)}$ in $\ket{q\Bar{q}g:\text{coh}}$ and the final states of the process $e^{+}e^{-}\rightarrow q\bar{q}gg$, where one of the partons in the final state is soft/collinear.
This amplitude is given by
\begin{multline}
    T_{6a}^{coh}=\frac{-e^2g^3}{2(2\pi)^{21/2}\prod_i\sqrt{2p_i^+}}\int [dk]\Theta_\Delta\Biggl\{\frac{f^{abd}(p_5^\mu \epsilon^{\ast \nu}_{{p_5}b}(\epsilon^{d}_{\nu k}\epsilon^{a}_{\mu p_5-k}+\epsilon^{d}_{\nu p_5-k}\epsilon^{a}_{\mu k})-k^\mu \epsilon^{\nu}_{b{k}}(\epsilon^{\ast d}_{\nu p_5}\epsilon^{a}_{\mu p_5-k}+\epsilon^{d}_{\nu p_5-k}\epsilon^{\ast a}_{\mu p_5})}{(2k^+)(2(p_5-k)^+)(2(p_3+p_5)^+)(2(p_1+p_2)^+)(p_1^-+p_2^--(p_1+p_2)^-)}\\  \frac{-(p_5-k)^\mu \epsilon^{\nu}_{b{p_5-k}}(\epsilon^{\ast d}_{\nu p_5}\epsilon^{a}_{\mu k}+\epsilon^{d}_{\nu k}\epsilon^{\ast a}_{\mu p_5}))}{(p_3^-+p_5^--(p_3+p_5)^-)(p_5^--(p_5-k)^--k^-)}\Biggl[\frac{\Bar{u}_{p_3}\gamma^+T^a u_{p_3+p_5}f^{abd}(k^-\epsilon^{\ast \mu}_{b_k}\epsilon^{\ast d}_{\mu_{p_5-k}}+({p_5-k})^-\epsilon^{\ast \mu}_{b_{p_5-k}}\epsilon^{\ast d}_{\mu_k})}{((p_4+p_5)^+-p_4^+)^2}\\ +\frac{f^{abd}(k^-\epsilon^{\ast \mu}_{b_k}\epsilon^{\ast d}_{\mu_{p_5-k}}+({p_5-k})^-\epsilon^{\ast \mu}_{b_{p_5-k}}\epsilon^{\ast d}_{\mu_k})\Bar{u}_{p_3}\gamma^+T^a u_{p_3+p_5}}{((p_5-k)^+-k^+)^2} \Biggr]\frac{\Bar{u}_{p_3+p_5} \slashed{\epsilon}_{p_1+p_2}v_{p_4}\Bar{v}_{p_2}\slashed{\epsilon}^{\ast}_{p_1+p_2}u_{p_1}}{(p_1^-+p_2^--(p_1+p_2)^-)}\Biggr\}
\end{multline}


\subsection{Amplitudes involving $\Omega_{\Delta(V_1)}W_4$}

\begin{figure}[h]
    \centering
    \includegraphics[width=0.22\linewidth]{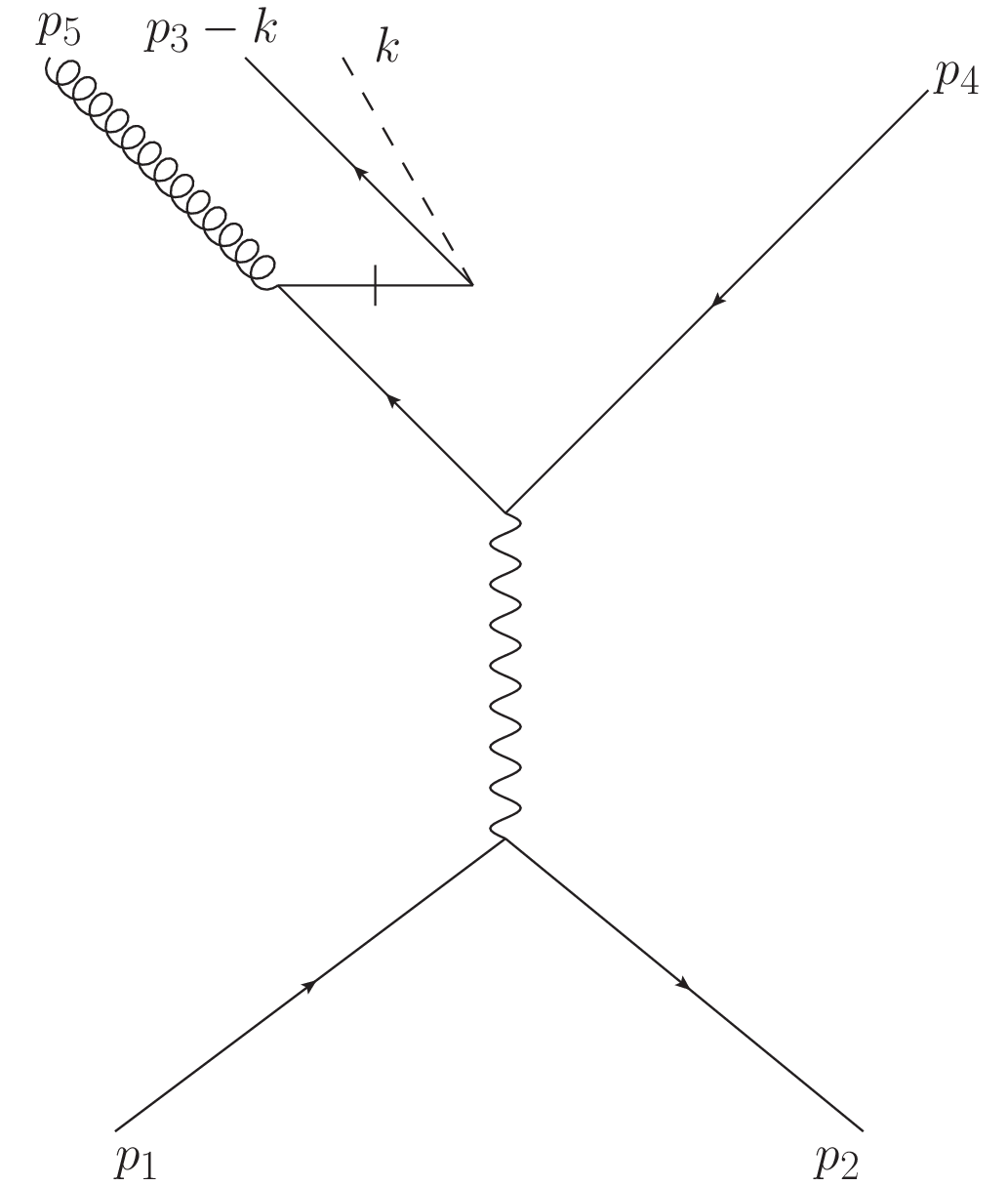} \includegraphics[width=0.2\linewidth]{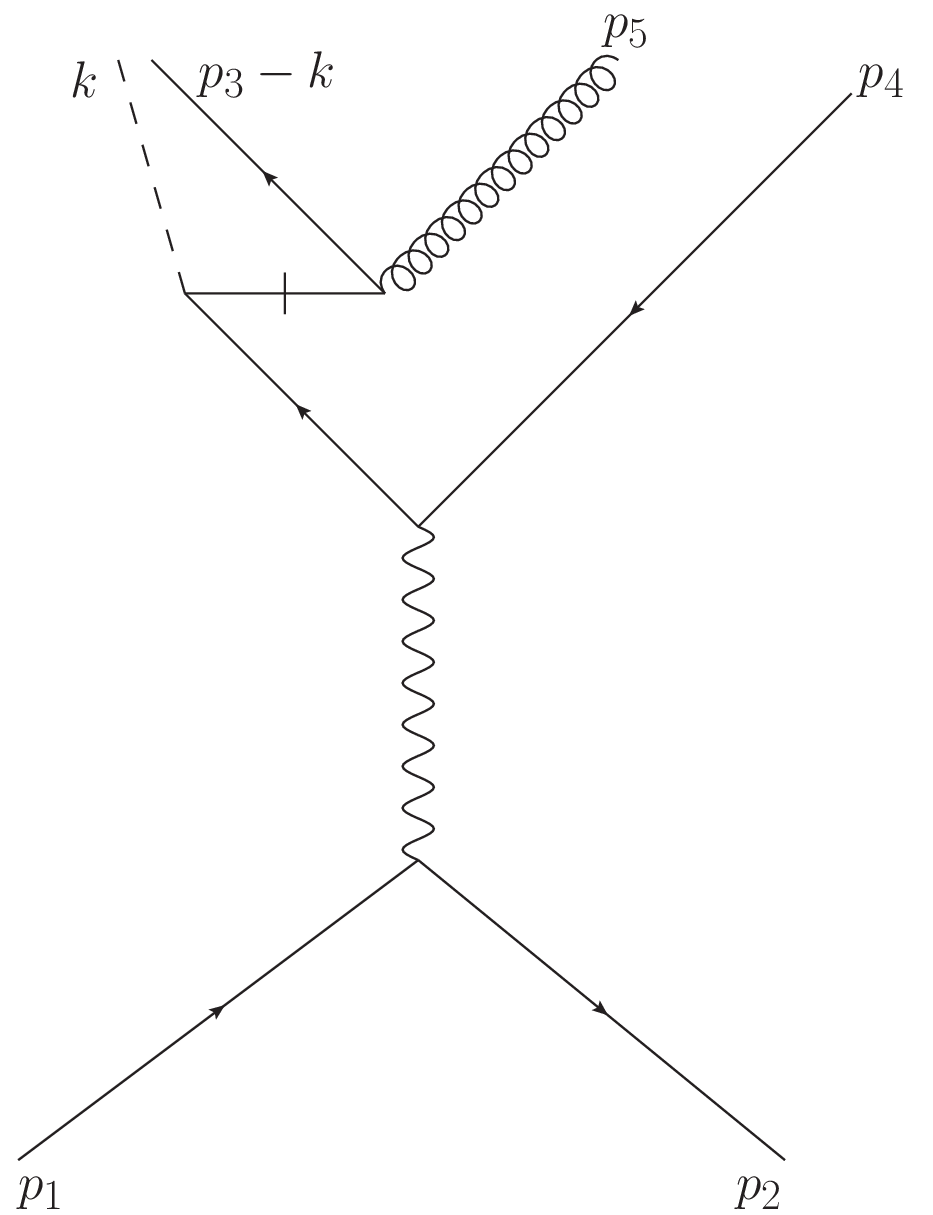} \includegraphics[width=0.23\linewidth]{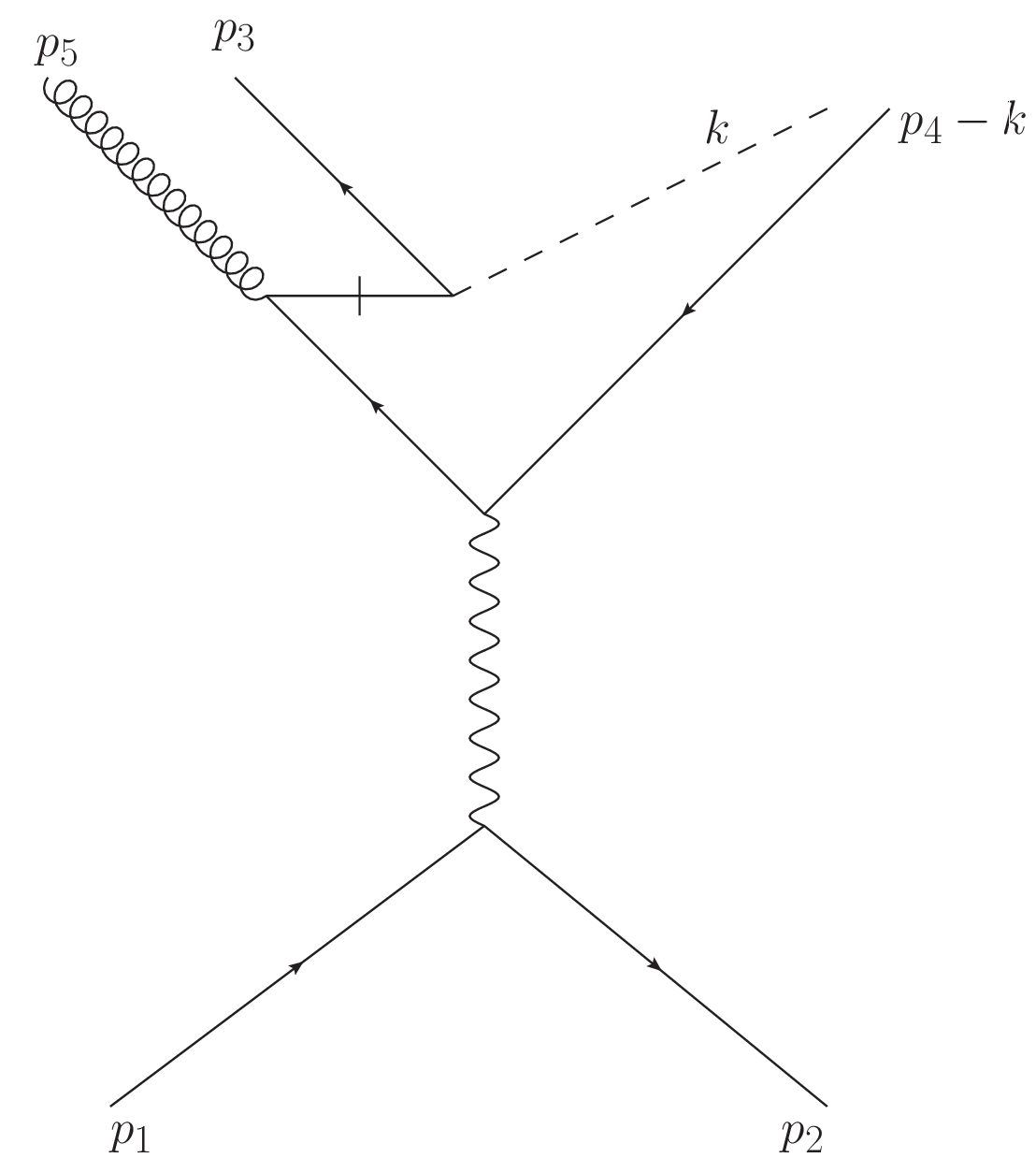} \includegraphics[width=0.22\linewidth]{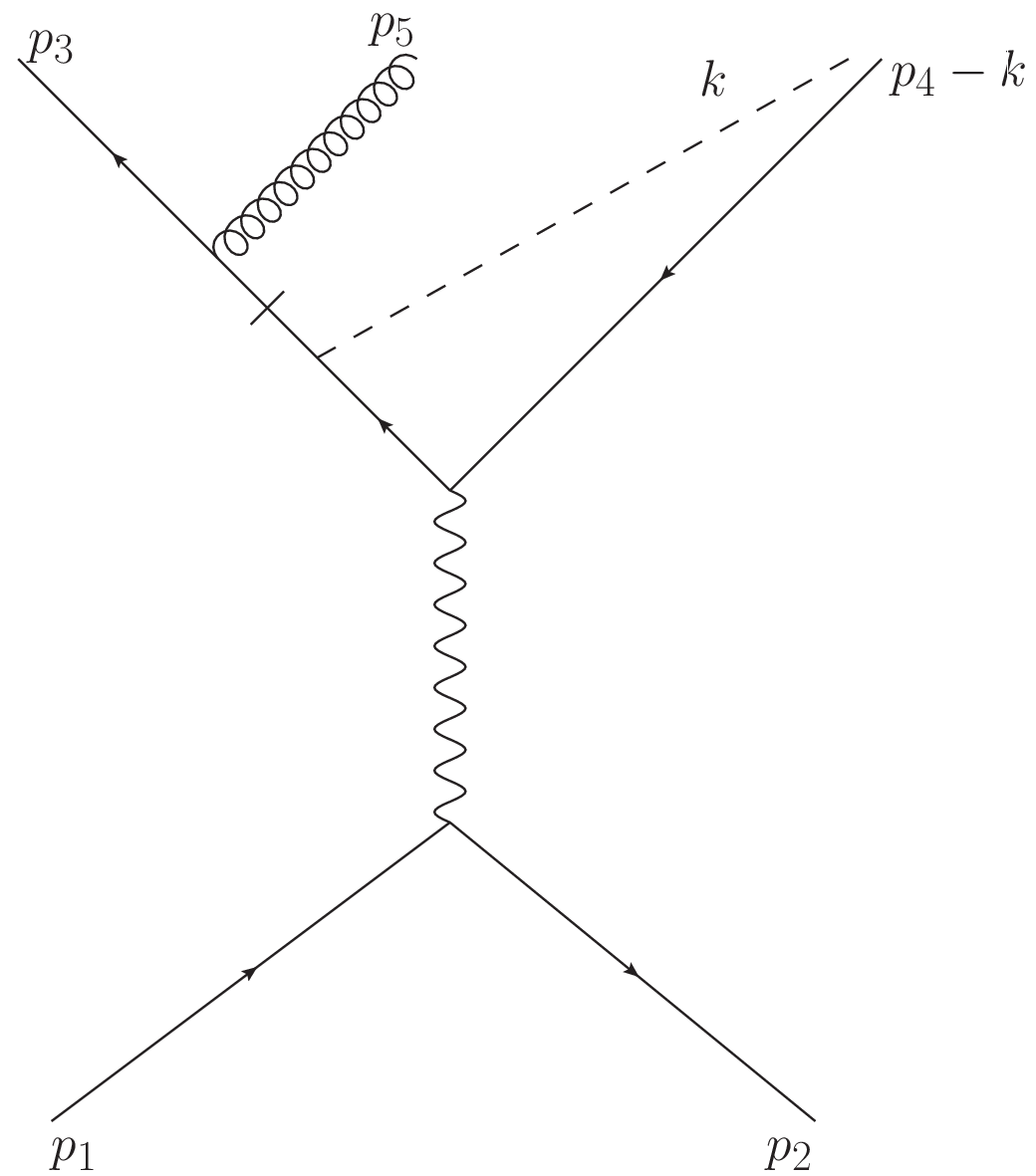}\\
    \hspace{0\linewidth} $T_{7a}^{coh}$ \hspace{0.17\linewidth} $T_{7b}^{coh}$\hspace{0.18\linewidth}$T_{7c}^{coh}$\hspace{0.18\linewidth}$T_{7d}^{coh}$\\
    \caption{Diagrams involving $\Omega_{\Delta(V_1)}$ and $W_4$. The dashed lines represent soft/collinear gluons}
    \label{fig:T5_coh}
\end{figure}

Amplitudes for the processes in Fig. \ref{fig:T5_coh} result from overlap between $\mathcal{O}(g)$ term of $\Omega_{\Delta(V_1)}$ in $\ket{q\Bar{q}g:\text{coh}}$ and the final states of the processes $e^{+}e^{-}\rightarrow q\bar{q}q\bar{q}$ or $e^{+}e^{-}\rightarrow q\bar{q}gg$, where one of the partons in the final state is soft/collinear. These are given by

\begin{multline}
    T_{7a}^{coh}=\frac{e^2g^3}{2(2\pi)^{21/2}\prod_i\sqrt{2p_i^+}}\int [k]\Theta_\Delta\Biggl[\frac{\Bar{u}_{p_3}\slashed{\epsilon}^{c}_{k}T^c{u}_{p_3-k}\Bar{u}_{p_3-k}T^a\slashed{\epsilon}^{\ast a}_{k}\gamma^+\slashed{\epsilon}^{\ast b}_{p_5}T^b{u}_{p_3+p_5}}{(2(p_3+p_5)^+)(2(p_3-k)^+)(2k^+)(p_5^+-(p_3^++p_5^+))(p_3^--(p_3-k)^--k^-)}\\ \times \frac{\Bar{u}_{p_3+p_5}\slashed{\epsilon}_{p_1+p_2}v_{p_4}\Bar{v}_{p_2}\slashed{\epsilon}^{\ast}_{p_1+p_2}u_{p_1}}{(2(p_1+p_2)^+)(p_1^-+p_2^--(p_1+p_2)^-)(p_3^-+p_5^--(p_3+p_5)^-)}\Biggr]
\end{multline}

\begin{multline}
    T_{7b}^{coh}=\frac{-e^2g^3}{2(2\pi)^{21/2}\prod_i\sqrt{2p_i^+}}\int [k]\Theta_\Delta\Biggl[\frac{\bar{v}_{p_4-k}\slashed{\epsilon}^{c}_{k}T^c{v}_{p_4}\Bar{u}_{p_3}T^a\slashed{\epsilon}^{\ast a}_{k}\gamma^+\slashed{\epsilon}^{\ast b}_{p_5}T^b{u}_{p_3+p_5-k}}{(2(p_3+p_5+k)^+)(2(p_4-k)^+)(2k^+)(p_5^+-(p_3^+p_5^+k^+)(p_4^--(p_4-k)^--k^-)}\\ \times \frac{\Bar{u}_{p_3+p_5+k}\slashed{\epsilon}_{p_1+p_2}v_{p_4-k }\Bar{v}_{p_2}\slashed{\epsilon}^{\ast}_{p_1+p_2}u_{p_1}}{(2(p_1+p_2)^+)(p_1^-+p_2^--(p_1+p_2)^-)(p_1^-+p_2^--(p_4-k)^--(p_3+p_5+k)^-)}\Biggr]
\end{multline}

\begin{multline}
    T_{7c}^{coh}=\frac{e^2g^3}{2(2\pi)^{21/2}\prod_i\sqrt{2p_i^+}}\int [k]\Theta_\Delta\Biggl[\frac{\Bar{u}_{p_3}\slashed{\epsilon}^{c}_{k}T^c{u}_{p_3-k}\Bar{u}_{p_3-k}T^a\slashed{\epsilon}^{\ast a}_{p_5}\gamma^+\slashed{\epsilon}^{\ast b}_{k}T^b{u}_{p_3+p_5}}{(2(p_3+p_5)^+)(2(p_3-k)^+)(2k^+)(k^+-(p_3^++p_5^+))(p_3^--(p_3-k)^--k^-)}\\ \times \frac{\Bar{u}_{p_3+p_5}\slashed{\epsilon}_{p_1+p_2}v_{p_4}\Bar{v}_{p_2}\slashed{\epsilon}^{\ast}_{p_1+p_2}u_{p_1}}{(2(p_1+p_2)^+)(p_1^-+p_2^--(p_1+p_2)^-)(p_3^-+p_5^--(p_3+p_5)^-)}\Biggr]
\end{multline}

\begin{multline}
    T_{7d}^{coh}=\frac{-e^2g^3}{2(2\pi)^{21/2}\prod_i\sqrt{2p_i^+}}\int [k]\Theta_\Delta\Biggl[\frac{\bar{v}_{p_4-k}\slashed{\epsilon}^{c}_{k}T^c{v}_{p_4}\Bar{u}_{p_3}T^a\slashed{\epsilon}^{\ast a}_{p_5}\gamma^+\slashed{\epsilon}^{\ast b}_{k}T^b{u}_{p_3+p_5-k}}{(2(p_3+p_5+k)^+)(2(p_4-k)^+)(2k^+)(k+^+-(p_3^+p_5^+k^+)(p_4^--(p_4-k)^--k^-)}\\ \times \frac{\Bar{u}_{p_3+p_5+k}\slashed{\epsilon}_{p_1+p_2}v_{p_4-k }\Bar{v}_{p_2}\slashed{\epsilon}^{\ast}_{p_1+p_2}u_{p_1}}{(2(p_1+p_2)^+)(p_1^-+p_2^--(p_1+p_2)^-)(p_1^-+p_2^--(p_4-k)^--(p_3+p_5+k)^-)}\Biggr]
\end{multline}

\section{Amplitude-level cancellation of divergences}\label{sec:cancel}

In Sections \ref{sec:Fock diag} and \ref{sec:coh components}, we have now presented all the terms in the transition amplitude for the process $e^{+}e^{-}\rightarrow q\bar{q}g$ up to $\mathcal{O}(g^3)$ in the coherent state basis. The total amplitude $T_{fi}$ is
\begin{equation}
    T_{fi}=\sum_{i=1}^7 \sum_\alpha (T_{i\alpha} + T^{coh}_{i\alpha})
\end{equation}
where $\alpha\in \{a,b, \dots\}$. Recall that in the coherent state expansion, the $\mathcal{O}(g)$ and $\mathcal{O}(g^2)$ terms are restricted due to the Heaviside function $\Theta_\Delta$ to only the IR divergent regions of the phase space. Crucially, these $T^{coh}_{i\alpha}$ terms have exactly opposite signs to the $T_{i\alpha}$ terms.

More explicitly, in the region where $\Theta_\Delta=1$,
\begin{equation} \label{eq: opp sign amplitudes}
T_{i\alpha}=-T^{coh}_{i\alpha}
\end{equation}
where $i \in \{1,2,...,7\};\; \alpha \in \{a,b, \dots\}$. As is evident from Eq. (\ref{eq: opp sign amplitudes}), to a given order in perturbation theory, each IR-divergent diagram in the Fock basis has a corresponding diagram arising from the higher order contribution of the coherent state basis. The amplitude of this additional diagram is equal and opposite to that of the Fock state diagram. However, this is true only in the region $\Theta_\Delta=1$. Thus, the finite part of the Fock state amplitude, which lies in the phase space where $\Theta_\Delta=0$, remains unaffected.

Thus, the IR divergences in the Fock basis are cancelled exactly by the additional terms arising in the coherent state basis leading to an IR-finite amplitude. 


\section{Conclusion and Outlook}\label{sec:conclusion}

We have constructed coherent states for LFQCD using the KF method of asymptotic dynamics and have shown that the transition amplitude for the process $e^{+}e^{-}\rightarrow q\bar{q}g$ in this coherent state basis is IR-finite up to $\mathcal{O}(g^3)$ in LF-Hamiltonian perturbation theory. 

In particular, we show that when we calculate the scattering amplitude in the coherent state basis, in addition to the contributions from the Fock basis, there are also contributions arising from the Fock state expansion of the asymptotic states. These additional terms cancel the IR divergences that are present in the amplitude calculated in the Fock basis due to vanishing energy denominators.

A natural extension of this work would be to investigate the possibility of an all-order result. An all-order proof of IR finiteness has been constructed in LFQED in the case of fermion self-energy \cite{jaimisraallorder} and it would be instructive to explore a scattering process like processes of interest in LFQCD with this aim. Methods of handling IR divergences in QCD at the amplitude level may be relevant in the quest for all-order amplitude level parton shower algorithms discussed recently in literature \citep{martinez}.

Another important avenue yet to be explored is the calculation of scattering cross-sections in the coherent-state basis and comparing them against experimental data.

\section*{Acknowledgement}

S.G. would like to thank National Initiative for Undergraduate Sciences, HBCSE, TIFR for facilitating and supporting this work. A.M. would like to thank the Department of Atomic Energy, Govt. of India for the award of the Raja Ramanna Fellowship. D.B. acknowledges the support received from Conselho Nacional
de Desenvolvimento Cient\'{i}fico e Tecnol\'{o}gico (CNPq), Brasil, Process no. 152348/2024-7 for carrying out this
work.

\appendix
\section{}\label{app:A}

We now list all the transition amplitudes that were omitted in Sec.\ref{sec:Fock diag}.

The amplitudes corresponding to the diagrams in Fig. \ref{fig:T1} are given below.

\begin{multline}
    T_{1b}=\frac{-e^2g^3}{(2\pi)^{15/2}\prod_i\sqrt{2p_i^+}}\int [dk]\Biggl[\frac{\Bar{v}_{p_4-k}\slashed{\epsilon}^{a}_{k}T^a{v}_{p_4}\Bar{u}_{p_3}\slashed{\epsilon}^{\ast b}_{p_5}T^b{u}_{p_3+p_5}}{(2(p_4-k)^+)(2(p_3+p_5)^+)(2(p_3+p_5+k)^+)(2k^+)(2(p_1+p_2)^+)(p_4^--(p_4-k)^--k^-)}\\ \times \frac{\Bar{u}_{p_3+p_5}\slashed{\epsilon}^{\ast c}_{k}T^c{u}_{p_3+p_5+k}\Bar{u}_{p_3+p_5+k}\slashed{\epsilon}_{p_1+p_2}v_{p_4-k}\Bar{v}_{p_2}\slashed{\epsilon}^{\ast}_{p_1+p_2}u_{p_1}}{(p_1^-+p_2^--(p_3+p_5+k)^--(p_4^--k)^-)(p_1^-+p_2^--(p_1+p_2)^-)(p_1^-+p_2^--(p_3+p_5)^--(p_4^--k)^--k^-)}\Biggr]
\end{multline}

\begin{multline}
    T_{1c}=\frac{-e^2g^3}{(2\pi)^{15/2}\prod_i\sqrt{2p_i^+}}\int [dk]\Biggl[\frac{\Bar{v}_{p_4-k}\slashed{\epsilon}^{a}_{k}T^a{v}_{p_4}\Bar{u}_{p_3}\slashed{\epsilon}^{\ast b}_{k}T^b{u}_{p_3+k}}{(2(p_4-k)^+)(2(p_3+k)^+)(2(p_3+p_5+k)^+)(2k^+)(2(p_1+p_2)^+)(p_4^--(p_4-k)^--k^-)}\\ \times \frac{\Bar{u}_{p_3+k}\slashed{\epsilon}^{\ast c}_{p_5}T^c{u}_{p_3+p_5+k}\Bar{u}_{p_3+p_5+k}\slashed{\epsilon}_{p_1+p_2}v_{p_4-k}\Bar{v}_{p_2}\slashed{\epsilon}^{\ast}_{p_1+p_2}u_{p_1}}{p_1^-+p_2^--(p_1+p_2)^-)(p_1^-+p_2^--(p_3+p_5+k)^--(p_4^--k)^-)((p_3^-+p_4^--(p_3+k)^--(p_4^--k)^-)}\Biggr]
\end{multline}

\begin{multline}
    T_{1d}=\frac{e^2g^3}{(2\pi)^{15/2}\prod_i\sqrt{2p_i^+}}\int [dk]\Biggl[\frac{\Bar{v}_{p_5-k}\slashed{\epsilon}^{\ast a}_{p_5}T^a{u}_{k}\Bar{u}_{k}\slashed{\epsilon}^{b}_{p_5}T^b{v}_{p_5-k}\Bar{u}_{p_3}\slashed{\epsilon}^{\ast c}_{p_5}T^c{u}_{p_3+p_5}}{(2(p_3+p_5)^+)(2(p_5-k)^+)(2(p_1+p_2)^+)(2p_5^+)(p_5^--(p_5-k)^--k^-)}\\ \times \frac{\Bar{u}_{p_3+p_5}\slashed{\epsilon}_{p_1+p_2}v_{p_4}\Bar{v}_{p_2}\slashed{\epsilon}^{\ast}_{p_1+p_2}u_{p_1}}{(2k^+)(p_1^-+p_2^--(p_1+p_2)^-)(p_3^-+p_5^--(p_3+p_5)^-)(p_1^-+p_2^--p_3^--p_4^--p_5^-)}\Biggr]
\end{multline}

\begin{multline}
    T_{1f}=\frac{-e^2g^3}{(2\pi)^{15/2}\prod_i\sqrt{2p_i^+}}\int [dk]\Biggl[\frac{\Bar{v}_{p_5-k}\slashed{\epsilon}^{\ast a}_{p_5}T^a{u}_{k}\Bar{u}_{p_3}\slashed{\epsilon}^{b}_{p_3+p_5-k}T^b{v}_{p_5-k}\Bar{u}_{k}\slashed{\epsilon}^{\ast c}_{p_3+p_5-k}T^c{u}_{p_3+p_5}}{(2(p_3+p_5)^+)(2(p_5-k)^+)(2(p_1+p_2)^+)(2(p_3+p_5-k)^+)(p_5^--(p_5-k)^--k^-)}\\ \times \frac{\Bar{u}_{p_3+p_5}\slashed{\epsilon}_{p_1+p_2}v_{p_4}\Bar{v}_{p_2}\slashed{\epsilon}^{\ast}_{p_1+p_2}u_{p_1}}{(2k^+)(p_1^-+p_2^--(p_1+p_2)^-)(p_3^-+p_5^--(p_3+p_5)^-)(p_3^-+p_5^--(p_3+p_5-k)^-k^-)}\Biggr]
\end{multline}

\begin{multline}
    T_{1g}=\frac{-e^2g^3}{(2\pi)^{15/2}\prod_i\sqrt{2p_i^+}}\int [dk]\Biggl[\frac{\Bar{v}_{p_5-k}\slashed{\epsilon}^{\ast a}_{p_5}T^a{u}_{k}\Bar{u}_{k}\slashed{\epsilon}^{b}_{p_4+k}T^b{v}_{p_4}\Bar{u}_{p_3}\slashed{\epsilon}^{\ast c}_{p_4+k}T^c{u}_{p_3+p_4+k}}{(2(p_4+k)^+)(2(p_5-k)^+)(2(p_1+p_2)^+)(2(p_3+p_4+k)^+)(p_5^--(p_5-k)^--k^-)}\\ \times \frac{\Bar{u}_{p_3+p_4+k}\slashed{\epsilon}_{p_1+p_2}v_{p_5-k}\Bar{v}_{p_2}\slashed{\epsilon}^{\ast}_{p_1+p_2}u_{p_1}}{(2k^+)(p_1^-+p_2^--(p_1+p_2)^-)(p_4^-+p_5^--(p_4+k)^--(p_5-k)^-)(p_1^-+p_2^--(p_5-k)^-(p_3+p_4+k)^-)}\Biggr]
\end{multline}

The amplitudes corresponding to the diagrams in Fig. \ref{fig:T4} are given below.

\begin{multline}
    T_{4c}=\frac{2e^2g^3}{(2\pi)^{21/2}\prod_i\sqrt{2p_i^+}}\int [dk]\frac{\Bar{v}_{p_5-k}\slashed{\epsilon }^{\ast c}_{p_5}T^c{u}_{k}\Bar{u}_{k}\gamma^+T^a{v}_{p_5-k}\Bar{u}_{p_3}\gamma^+T^a{u}_{p_3+p_5}}{(2(p_5-k)^+)(2(p_3+p_5)^+)(p_3^+ - (p_3+p_5)^+)^2(p_5^--(p_5-k)^--k^-)}\\  \times\frac{\Bar{u}_{p_3+p_5} \slashed{\epsilon}_{p_1+p_2}v_{p_4}\Bar{v}_{p_2}\slashed{\epsilon}^{\ast}_{p_1+p_2}u_{p_1}}{(2k^+)(2(p_1+p_2)^+)(p_1^-+p_2^--(p_1+p_2)^-)(p_3^-+p_5^--(p_3+p_5)^-)}
\end{multline}

\begin{multline}
    T_{4d}=\frac{2e^2g^3}{(2\pi)^{21/2}\prod_i\sqrt{2p_i^+}}\int [dk]\frac{\Bar{v}_{p_5-k}\slashed{\epsilon}^{\ast c}_{p_5}T^c{u}_{k}\Bar{u}_{p_3}\gamma^+T^a{v}_{p_4}\Bar{u}_{k}\gamma^+T^a{u}_{p_3+p_4+k}}{(2(p_5-k)^+)(2(p_3+p_4+k)^+)(k^+ - (p_3+p_4+k)^+)^2(p_5^--(p_5-k)^--k^-)}\\  \times\frac{\Bar{u}_{p_3+p_4+k} \slashed{\epsilon}_{p_1+p_2}v_{p_5-k}\Bar{v}_{p_2}\slashed{\epsilon}^{\ast}_{p_1+p_2}u_{p_1}}{(2k^+)(2(p_1+p_2)^+)(p_1^-+p_2^--(p_1+p_2)^-)(p_1^-+p_2^--(p_3+p_4+k)^--(p_5-k)^-)}
\end{multline}

The amplitudes corresponding to the diagrams in Fig. \ref{fig:T7} are given below.

\begin{multline}
    T_{7c}=\frac{-e^2g^3}{2(2\pi)^{21/2}\prod_i\sqrt{2p_i^+}}\int [k]\Biggl[\frac{\Bar{u}_{p_3}\slashed{\epsilon}^{c}_{k}T^c{u}_{p_3-k}\Bar{u}_{p_3-k}T^a\slashed{\epsilon}^{\ast a}_{p_5}\gamma^+\slashed{\epsilon}^{\ast b}_{k}T^b{u}_{p_3+p_5}}{(2(p_3+p_5)^+)(2(p_3-k)^+)(2k^+)(k^+-(p_3^++p_5^+))(p_3^--(p_3-k)^--k^-)}\\ \times \frac{\Bar{u}_{p_3+p_5}\slashed{\epsilon}_{p_1+p_2}v_{p_4}\Bar{v}_{p_2}\slashed{\epsilon}^{\ast}_{p_1+p_2}u_{p_1}}{(2(p_1+p_2)^+)(p_1^-+p_2^--(p_1+p_2)^-)(p_3^-+p_5^--(p_3+p_5)^-)}\Biggr]
\end{multline}

\begin{multline}
    T_{7d}=\frac{e^2g^3}{2(2\pi)^{21/2}\prod_i\sqrt{2p_i^+}}\int [k]\Biggl[\frac{\bar{v}_{p_4-k}\slashed{\epsilon}^{c}_{k}T^c{v}_{p_4}\Bar{u}_{p_3}T^a\slashed{\epsilon}^{\ast a}_{p_5}\gamma^+\slashed{\epsilon}^{\ast b}_{k}T^b{u}_{p_3+p_5-k}}{(2(p_3+p_5+k)^+)(2(p_4-k)^+)(2k^+)(k+^+-(p_3^+p_5^+k^+)(p_4^--(p_4-k)^--k^-)}\\ \times \frac{\Bar{u}_{p_3+p_5
    +k}\slashed{\epsilon}_{p_1+p_2}v_{p_4-k }\Bar{v}_{p_2}\slashed{\epsilon}^{\ast}_{p_1+p_2}u_{p_1}}{(2(p_1+p_2)^+)(p_1^-+p_2^--(p_1+p_2)^-)(p_1^-+p_2^--(p_4-k)^--(p_3+p_5+k)^-)}\Biggr]
\end{multline}

\bibliography{references}

 \end{document}